\documentstyle[11pt,epsf]{article}
 1
 1
 1

\def\lsim{\mathrel{\raise.3ex\hbox{$<$\kern-.75em\lower1ex\hbox{$\sim$}}}}
\def\gsim{\mathrel{\raise.3ex\hbox{$>$\kern-.75em\lower1ex\hbox{$\sim$}}}}
\catcode`\@=11
\def\slash{\mathpalette\make@slash}
\def\make@slash#1#2{\setbox\z@\hbox{$#1#2$}%
  \hbox to 0pt{\hss$#1/$\hss\kern-\wd0}\box0}
\catcode`\@=12 % @ signs are no longer letters
\begin{document}
\noindent
\thispagestyle{empty}
\renewcommand{\thefootnote}{\fnsymbol{footnote}}
\begin{flushright}
{\bf UCSD/PTH 98-02}\\
{\bf hep-ph/9803454}\\
{\bf March 1998}\\
\end{flushright}
\vspace{.5cm}
\begin{center}
  \begin{Large}\bf
Bottom Quark Mass from $\Upsilon$ Mesons
  \end{Large}
  \vspace{1.5cm}

\begin{large}
 A.H. Hoang
\end{large}
\begin{center}
\begin{it}
   Department of Physics,
   University of California, San Diego,\\
   La Jolla, CA 92093--0319, USA\\ 
\end{it} 
\end{center}

  \vspace{4cm}
  {\bf Abstract}\\
\vspace{0.3cm}
%\setcounter{footnote}{0}
%\renewcommand{\thefootnote}{\arabic{footnote}}
%\addtocounter{footnote}{-1}
%
\noindent
\begin{minipage}{15.0cm}
\begin{small}
The bottom quark pole mass $M_b$ is determined using a sum rule which
relates the masses and the electronic decay widths of the $\Upsilon$
mesons to large $n$ moments of the vacuum polarization function
calculated from nonrelativistic quantum chromodynamics. The complete
set of next-to-next-to-leading order (i.e.\ ${\cal{O}}(\alpha_s^2,
\alpha_s\,v, v^2)$ where $v$ is the bottom quark c.m.\ velocity)
corrections is calculated and leads to a considerable reduction of
theoretical uncertainties compared to a pure next-to-leading order
analysis. However, the theoretical uncertainties remain much larger
than the experimental ones. For a two parameter fit for $M_b$, and the
strong $\overline{\mbox{MS}}$ coupling $\alpha_s$, and using the
scanning method to estimate theoretical uncertainties, the
next-to-next-to-leading order analysis yields
$4.74$~GeV~$\le M_b\le 4.87$~GeV and 
$0.096 \le \alpha_s(M_z) \le 0.124$ if experimental
uncertainties are included at the $95\%$ confidence level
and if two-loop running for $\alpha_s$ is employed. $M_b$ and
$\alpha_s$ have a sizeable positive correlation. For the running
$\overline{\mbox{MS}}$ bottom quark mass this leads to 
$4.09$~GeV~$\le m_b(M_{\Upsilon(1S)}/2)\le 4.32$~GeV.
If $\alpha_s$ is
taken as an input, the result for the bottom quark pole mass reads
$4.78$~GeV~$\le M_b\le 4.98$~GeV 
($4.08$~GeV~$\le m_b(M_{\Upsilon(1S)}/2)\le 4.28$~GeV)
for $0.114\lsim \alpha_s(M_z)\le 0.122$.
The discrepancies between the results of three previous analyses on the
same subject by Voloshin, Jamin and Pich, and K\"uhn {\it et al.} are
clarified.  A comprehensive review on the calculation of the heavy
quark-antiquark pair production cross section through a vector current
at next-to-next-to leading order in the nonrelativistic expansion is
presented.
\end{small}
\end{minipage}
\end{center}
\setcounter{footnote}{0}
\renewcommand{\thefootnote}{\arabic{footnote}}
\vspace{1.2cm}
%
%
% text
%
\newpage
\noindent
\section{Introduction}
\label{sectionintroduction}
Quantum chromodynamics (QCD) is the established theory of the
strong interactions. The determination of its parameters, the strong
coupling and the quark masses, and continuous tests of its consistency
with experimental measurements belong to the most important tasks
within particle physics.
For the strong coupling an almost countless number of determinations
exists. The most precise determinations now quote uncertainties in
$\alpha_s(M_z)$ of less than $5\%$.\footnote{
Throughout this paper the strong coupling is defined in the
$\overline{\mbox{MS}}$ scheme.
} 
The remarkable feature of the $\alpha_s$ determinations, however,
is their consistency to each other (see e.g.~\cite{Stirling1} for a
review). The situation for the quark masses can certainly be described
as much less coherent. For the bottom quark pole mass, which
represents an important ingredient for the theoretical description of
$B$ mesons decays the determination of the corresponding
Cabibbo-Kobayashi-Maskawa matrix elements, the situation is
particularly confusing. In the past few years there have been
three determinations by Voloshin ($M_b = 4.827\pm
0.007$~GeV)~\cite{Voloshin1}, and later by Jamin and Pich ($M_b =
4.60\pm 0.02$~GeV)\cite{Jamin1} and K\"uhn {\it et al.} ($M_b =
4.75\pm 0.04$~GeV)~\cite{Kuhn1} which, although they have all been
obtained from
the same experimental data on the spectrum and the electronic decay
widths of the $\Upsilon$ mesons, are contradictory to each other if
the quoted uncertainties are taken seriously. Further, the three
analyses~\cite{Voloshin1,Jamin1,Kuhn1} were all based on the same sum
rule which relates
large $n$ moments (i.e.\ large number of derivatives at zero momentum
transfer) of the vacuum correlator of two bottom-antibottom vector
currents to an integral over the total production cross section of
hadrons containing a bottom and an antibottom quark in $e^+e^-$
annihilation. In the limit of large $n$ the moments can be calculated
in a nonrelativistic expansion~\cite{Novikov1,Voloshin2} and higher
order (relativistic)
corrections can be implemented in a systematic way. 
%One might
%certainly argue that the spread of the results obtained in
%Refs.~\cite{Voloshin1,Jamin1,Kuhn1} simply reflects the true
%theoretical uncertainty
%inherent in this sum rule approach. However, we find it quite
%unsatisfactory that the uncertainty in the bottom quark pole mass
%shall be estimated in such a way.  

This paper contains a determination of the bottom quark pole
mass, where theoretical uncertainties are treated in a conservative
way. It is partly motivated by the belief that a carefully performed
analysis of theoretical uncertainties is mandatory in order to see
whether the uncertainties presented in
Refs.~\cite{Voloshin1,Jamin1,Kuhn1} are
realistic. In this work the method of choice is to scan all
theoretical parameters independently over reasonably large windows. We
will show that this method to estimate theoretical uncertainties is
more conservative than the methods used in
Refs.~\cite{Voloshin1,Jamin1,Kuhn1}. In
particular it renders the results obtained by Voloshin and K\"uhn {\it
et al.} consistent to each other. With the scanning method the 
precise results of Voloshin and K\"uhn {\it et al.} can only be
obtained if some model-like assumptions are imposed which are beyond
first principles QCD. The by far bigger part of the
motivation for this work, however, comes from the fact that now the
technical and conceptual tools have been
developed~\cite{Hoang1,Hoang2,Hoang3,Melnikov1} to
include the next-to-next-to-leading order (NNLO)
relativistic corrections to the large $n$ moments into the analysis. A
large fraction of this paper is devoted to a comprehensive
presentation and review of the concepts and calculations necessary to
determine
those NNLO contributions. In particular, we use the
concept of effective field theories formulated in the frame work of
nonrelativistic quantum chromodynamics (NRQCD)~\cite{Caswell1,Bodwin1}
to deal with
the problems of ultraviolet divergences which arise if relativistic
corrections to the expressions in the nonrelativistic limit are
calculated. However, we regard NRQCD merely as a technical tool and
do not spend too much time on formal considerations. Whenever possible we
rely on physical rather than formal arguments and use results from
older literature even if they have not been derived in the framework
of NRQCD. It is the main intention of this work to calculate the NNLO
corrections to the large $n$ moments and to analyze their impact on
the determination of $M_b$. We show that the NNLO corrections
lead to a considerable reduction of theoretical uncertainties in the
determination of $M_b$.
% (although nowhere in this work the precision
%claimed in Refs.~\cite{Voloshin1,Jamin1,Kuhn1} can be reached).

The program of this paper is as follows: In
Section~\ref{sectionbasicidea} we introduce our notation and explain
the ideas and concepts on which our analysis and calculations are
based on. NRQCD is introduced and a recipe for the calculation of the
moments at NNLO is presented. Because the heavy quark-antiquark cross
section in the threshold regime represents an important intermediate
step in the calculation of the moments, Section~\ref{sectionbasicidea}
also contains a comprehensive review on the basic concepts involved in
the calculation of the vector current induces cross section at
NNLO. In Section~\ref{sectioncalculatemoments} all calculations are
carried out explicitly and all relevant formulae are
displayed. Section~\ref{sectionproperties} contains a discussion on
some peculiarities of the large $n$ moments. A detailed
description of the treatment of the experimental data, the fitting
procedure and the scanning method is given in
Section~\ref{sectionfitting}. In Section~\ref{sectionresults} the
numerical results are presented and discussed. Two different
determinations of $M_b$ are carried out. First, $M_b$ and $\alpha_s$
are fitted simultaneously and, second, $M_b$ is fitted while
$\alpha_s$ is taken as in input. In Section~\ref{sectioncomments},
finally, we comment on the three previous analyses in
Refs.~\cite{Voloshin1,Jamin1,Kuhn1}
and Section~\ref{sectionconclusions} contains the conclusions.
Attached to this paper are three appendices which contain material
which we found too detailed to be presented in the main body of the
paper. 

The reader who is mainly interested in the results for the bottom
quark mass can safely skip most of Section~\ref{sectionbasicidea}, and
Sections~\ref{sectioncalculatemoments} and \ref{sectionproperties}
completely. 
\par
\vspace{0.5cm}
\section{The Basic Ideas and Notation}
\label{sectionbasicidea}
{\bf\underline{The Sum Rule}}\\[1mm]
We start our consideration from the correlator of two electromagnetic
currents of bottom quarks at momentum transfer $q$
\begin{equation}
\Pi_{\mu\nu}(q) \, = \,
-\,i \int \!{\rm d}x\,e^{i\,q.x}\,
   \langle\, 0\,|\,T\,j^b_\mu(x)\,j^b_\nu(0)\,|\,0\, \rangle
\,,
\label{vacpoldef}
\end{equation}
where
\begin{equation}
j^b_\mu(x) \, =  \bar b(x)\,\gamma_\mu\,b(x)
\,.
\end{equation}
The symbol $b$ denotes the bottom quark Dirac field. We define the
$n$'th moment $P_n$ of the vacuum polarization function as
\begin{equation}
P_n \, \equiv \,
\frac{4\,\pi^2\,Q_b^2}{n!\,q^2}\,
\bigg(\frac{d}{d q^2}\bigg)^n\,\Pi_\mu^{\,\,\,\mu}(q)\bigg|_{q^2=0}
\,,
\label{momentdef}
\end{equation}
where $Q_b=-1/3$ is the electric charge of the bottom quark.
Due to causality the n-th moment $P_n$ can be written in terms of a
dispersion integration
\begin{equation}
P_n \, = \,
\int \frac{d s}{s^{n+1}}\,R(s)
\,,
\label{momentcrosssectionrelation}
\end{equation}
where
\begin{equation}
R(s) \, = \, \frac{\sigma(e^+e^-\to\gamma^*\to \mbox{``$b\bar b$''})}
{\sigma_{pt}}
\label{Rdefinitioncovariant}
\end{equation}
is the total photon mediated cross section of bottom quark-antiquark
production in $e^+e^-$ annihilation normalized to the point cross
section $\sigma_{pt}=4 \pi \alpha^2/3 s$. Assuming global duality,
$P_n$ can be either calculated from experimental data for the total
cross section in $e^+e^-$ annihilation\footnote{
At the level of precision in this work the Z mediated cross section
can be safely neglected.
}
or theoretically using quantum chromodynamics (QCD). It is the basic
idea of this sum rule to set the moments calculated from experimental
data, $P_n^{ex}$, equal to those determined theoretically from QCD,
$P_n^{th}$, and to use this relation to determine the bottom quark
mass (and the strong coupling) by fitting theoretical and experimental
moments for various values of $n$.~\cite{Novikov1,Voloshin2,Reinders1}
 
At this point it is mandatory to discuss the range of $n$ for
which the theoretical moments can be calculated sufficiently accurate
(using perturbative QCD) 
to allow for a reliable extraction of $M_b$ and $\alpha_s$. From
Eq.~(\ref{momentcrosssectionrelation})  is
obvious that each moment $P_n$ effectively corresponds to a {\it
smearing} of the cross section $R$ over some energy region $\Delta E$
located around the threshold point. Thus, only if the smearing range is
{\it sufficiently} larger than $\Lambda_{\mbox{\tiny
QCD}}\sim{\cal{O}}(200-300~\mbox{MeV})$,
a perturbative calculation of the moments is
feasible~\cite{Poggio1}. [In Ref.~\cite{Poggio1} is was
argued that $\Delta E$ should be larger than $4 M_b\alpha_s$ to avoid
the complications involving a resummation of the Coulomb singularities
$\propto(\alpha_s/v)^m$. Because this resummation is explicitly carried
out at the NNLO level in this work, we have to take
$\Lambda_{\mbox{\tiny QCD}}$, the typical hadronic scale, as the size
of the minimal smearing range.]
We therefore conclude that $n$ is not allowed to be
too large if perturbative QCD shall be employed. We can  derive
an approximate upper bound for the allowed values of $n$ by changing
the integration variable in
relation~(\ref{momentcrosssectionrelation}) to the energy $E\equiv
\sqrt{q^2}-2 M_b$. For $n\gg 1$ only energies $E\ll M_b$ contribute,
which allows us to expand expression~(\ref{momentcrosssectionrelation})
for small $E/M_b$ (while regarding $(E/M_b) n$ of order one)
\begin{equation}
P_n \, \stackrel{n\gg 1}{=} \,
\frac{1}{(4\,M_b^2)^n}\,\int \frac{d E}{M_b}\,
\exp\bigg(-\frac{E}{M_b}\,n\bigg)\,R\Big((2\,M_b+E)^2\Big)\,
\bigg[\, 1+
{\cal{O}}\bigg(\frac{E}{M_b},\frac{E^2}{M_b^2}\,n\bigg)  
\,\bigg]
\,.
\label{momentcrosssectionrelation2}
\end{equation}
From Eq.~(\ref{momentcrosssectionrelation2}) we see that the
size of the smearing range $\Delta E$ for large $n$ is or order $M_b/n$,
\begin{equation}
\Delta E \, \sim \, \frac{M_b}{n}
\,.
\end{equation} 
Demanding that $\Delta E$ is larger than
$\Lambda_{\mbox{\tiny QCD}}$ yields that the values of $n$ for which a
perturbative calculation of the moments can trusted should be
sufficiently smaller than $15-20$. To avoid systematic
theoretical errors as much as possible we take
\begin{equation}
n_{max} \, = \, 10
\end{equation}
as the maximal value for $n$ employed in this work. 
On the other hand, it is also desirable to choose $n$ as large as
possible because the experimental cross section for electron positron
annihilation into $b\bar b$ hadrons is much better known in the
$\Upsilon$ resonance regime $\sqrt{s} \sim 9.5 - 10.5$~GeV than above
the $B\bar B$ threshold. By taking $n$ large the lower lying resonance
contributions in Eq.~(\ref{momentcrosssectionrelation}) are enhanced
relative to the continuum contributions leading effectively to a
suppression of the experimental uncertainties in the continuum cross
section~\cite{Novikov1,Voloshin2,Shifman1}. For our analysis we choose
\begin{equation}
n_{min} \, = \, 4
\end{equation}
as the minimal value for $n$. It is the regime
$4\le n\le 10$ which we will refer to as ``large $n$'' in this work.
It is a very important fact that for $4\le n\le 10$ 
the bottom-antibottom quark dynamics in the theoretical moments
$P_n^{th}$ is already nonrelativistic in nature. This can be seen
by once again examining
relation~(\ref{momentcrosssectionrelation2}). Because for a given
value of $n$ only energies $E\lsim M_b/n$ contribute, the
corresponding bottom quark velocities $v=\sqrt{E/M_b}$ (in the
c.m. frame) are in the range $|v|\lsim 0.5$, i.e.\ they are always
considerably smaller than the speed of light. In particular, the
velocity is already as large as the typical
size of the strong coupling $\alpha_s(M_b v)\approx 0.3$ governing the
exchange of longitudinal polarized gluons (in Coulomb gauge) among the
bottom-antibottom quark pair. This leads to the breakdown of the
conventional multi-loop perturbation expansion because the
exchange of $m$ longitudinal gluons generates singular terms
$\propto (\alpha_s/v)^m$, $m=0,1,2,\ldots$, (Coulomb singularities) in
the cross section for small velocities. These singular terms would
have to be resummed\footnote{
In this context ``resummation'' technically means that one carries out
the resummation of singular terms in the (formal) kinematic regime
$\alpha_s\ll |v|$. The resulting series (uniquely) define analytic
functions which can then be continued to the regime $|v|\lsim\alpha_s$. 
%The resonances are represented by poles of the
%analytic functions. Expression~(\ref{CoulombGreenfunctionregularized})
%just represents the LO ``resummation'' in the nonrelativistic
%expansion. See also~\cite{X}.
} 
to all orders in multi-loop perturbation theory in
order to arrive at a viable
description of the bottom-antibottom quark
dynamics. In other words, the Coulomb interaction
between the bottom and the antibottom quark has to be treated
exactly~\cite{Voloshin2}. Because this is
an impossible talk in the framework of covariant multi-loop
perturbation theory it is mandatory to calculate the cross section and
the theoretical moments in the nonrelativistic approximation
by solving the Schr\"odinger equation supplemented by relativistic
corrections.\\

\noindent
{\bf\underline{Perturbative NRQCD and the Cross Section}}\\[1mm]
In this paper we use NRQCD~\cite{Caswell1,Bodwin1} to set up a
consistent framework in which the corrections to the nonrelativistic
limit (in form of the nonrelativistic Schr\"odinger equation) can be
determined in a systematic manner at NNLO. This corresponds to
corrections up to order $\alpha_s^2$, $\alpha_s\,v$ and $v^2$ to the
expressions in the nonrelativistic limit. We count orders of
$\alpha_s$ as orders of $v$ because we treat the $b\bar b$ system as
Coulombic. In the framework of multi-loop
perturbation theory this would correspond to a resummation of all
terms $\propto \alpha_s^m v^k$ with $m+k=1,2,3$ in the
cross section for the small velocity expansion. NRQCD is an effective
field theory of QCD
designed to handle nonrelativistic heavy-quark-antiquark systems to in
principle arbitrary precision. NRQCD is based on the separation of
long- and short-distance effects by reformulating QCD in terms of an
unrenormalizable Lagrangian containing all possible operators in
accordance to the symmetries in the nonrelativistic limit. Treating
all quarks of the first and second generation as massless and taking
into account only those terms relevant for the NNLO calculation in
this work the NRQCD Lagrangian reads~\cite{Bodwin1}
\begin{eqnarray}
\lefteqn{
{\cal{L}}_{\mbox{\tiny NRQCD}} \, = \,
- \frac{1}{2} \,\mbox{Tr} \, G^{\mu\nu} G_{\mu\nu} 
+ \sum_{q=u,d,s,c} \bar q \, i \slash{D} \, q
}\nonumber\\[2mm] & &
+\, \psi^\dagger\,\bigg[\,
i D_t 
+ a_1\,\frac{{\mbox{\boldmath $D$}}^2}{2\,M_t} 
+ a_2\,\frac{{\mbox{\boldmath $D$}}^4}{8\,M_t^3}
\,\bigg]\,\psi + \ldots 
\nonumber\\[2mm] & &
+ \,\psi^\dagger\,\bigg[\, 
\frac{a_3\,g}{2\,M_t}\,{\mbox{\boldmath $\sigma$}}\cdot
    {\mbox{\boldmath $B$}}
+ \, \frac{a_4\,g}{8\,M_t^2}\,(\,{\mbox{\boldmath $D$}}\cdot 
  {\mbox{\boldmath $E$}}-{\mbox{\boldmath $E$}}\cdot 
  {\mbox{\boldmath $D$}}\,)
+ \frac{a_5\,g}{8\,M_t^2}\,i\,{\mbox{\boldmath $\sigma$}}\,
  (\,{\mbox{\boldmath $D$}}\times 
  {\mbox{\boldmath $E$}}-{\mbox{\boldmath $E$}}\times 
  {\mbox{\boldmath $D$}}\,)
 \,\bigg]\,\psi 
+\ldots
\nonumber\\[2mm] & &
+ \mbox{$\chi\chi^\dagger$ bilinear terms and higher dimensional operators}
\,.
\label{NRQCDLagrangian}
\end{eqnarray}
The gluons and massless quarks are described by the conventional
relativistic Lagrangian, where $G_{\mu\nu}$ is the gluon field
strength tensor, $q$ the Dirac spinor of a massless quark and $D_\mu$ the gauge
covariant derivative. For convenience, all color indices in
Eq.~(\ref{NRQCDLagrangian}) and throughout this work are
suppressed. The nonrelativistic bottom and antibottom quarks are
described by the Pauli spinors $\psi$ and $\chi$, respectively.
$D_t$ and {\boldmath $D$} are the time and space components of the
gauge covariant derivative $D$ and $E^i = G^{0 i}$ and $B^i =
\frac{1}{2}\epsilon^{i j k} G^{j k}$ the electric and magnetic
components of the gluon field strength tensor (in Coulomb gauge).
The straightforward $\chi^\dagger \chi$ bilinear terms are omitted and
can be obtained using charge symmetry. The short-distance coefficients
$a_1,\ldots,a_5$ are normalized to one at the Born level. The actual
form of the higher order contributions to the short-distance
coefficients $a_1,\ldots,a_5$ (and also to $b_1, b_2$ in
Eq.~(\ref{currentexpansion})) is irrelevant for this work, because we
will later use the ``direct matching'' procedure~\cite{Hoang1,Hoang4}
at the level of the final result for the cross section.

Let us first discuss the  cross section $R$ in
the nonrelativistic regime. To formulate $R$ in the nonrelativistic
regime at NNLO in NRQCD we start from the fully covariant
expression for the total cross section
\begin{eqnarray}
R(q^2) & = &
\frac{4\,\pi\,Q_b^2}{q^2}\,\mbox{Im}\,[\,
-i\,\int\,dx\,e^{i\,q.x}\,
  \langle\, 0\,| T\,j^b_\mu(x) \, j^{b\,\mu}(0)\, |\, 0\,\rangle]
\nonumber\\[2mm] & \equiv &
\frac{4\,\pi\,Q_b^2}{q^2}\,\mbox{Im}\,[\,
\langle\, 0\,| T\, \tilde j^b_\mu(q) \,
 \tilde j^{b\,\mu}(-q)\, |\, 0\,\rangle]
\,,
\label{crosssectioncovariant}
\end{eqnarray}
and expand the electromagnetic current (in momentum space)
$\tilde j_\mu(\pm q) = (\tilde{\bar b}\gamma^\mu \tilde b)(\pm
q)$ which produces/annihilates
a $b\bar b$ pair with c.m. energy $\sqrt{q^2}$ in terms of ${}^3\!S_1$
NRQCD currents up to dimension eight ($i=1,2,3$)
\begin{eqnarray}
\tilde j_i(q) & = & b_1\,\Big({\tilde \psi}^\dagger \sigma_i 
\tilde \chi\Big)(q) -
\frac{b_2}{6 M_t^2}\,\Big({\tilde \psi}^\dagger \sigma_i
(\mbox{$-\frac{i}{2}$} 
\stackrel{\leftrightarrow}{\mbox{\boldmath $D$}})^2
 \tilde \chi\Big)(q) + \ldots
\,,
\nonumber\\[2mm]
\tilde j_i(-q) & = & b_1\,\Big({\tilde \chi}^\dagger \sigma_i 
\tilde \psi\Big)(-q) -
\frac{b_2}{6 M_t^2}\,\Big({\tilde \chi}^\dagger \sigma_i
(\mbox{$-\frac{i}{2}$} 
\stackrel{\leftrightarrow}{\mbox{\boldmath $D$}})^2
 \tilde \psi\Big)(-q) + \ldots 
\,,
\label{currentexpansion}
\end{eqnarray}
where the constants $b_1$ and $b_2$ are short-distance coefficients
normalized to one at the Born level. Only the spatial components of
the currents contribute contribute at the NNLO level.
Inserting expansion~(\ref{currentexpansion}) back into 
Eq.~(\ref{crosssectioncovariant}) leads to the nonrelativistic
expansion of the cross section at the NNLO level
\begin{eqnarray}
R_{\mbox{\tiny NNLO}}^{\mbox{\tiny thr}}(\tilde E) & = &
\frac{\pi\,Q_b^2}{M_b^2}\,C_1(\mu_{\rm hard},\mu_{\rm fac})\,
\mbox{Im}\Big[\,
{\cal{A}}_1(E,\mu_{\rm soft},\mu_{\rm fac})
\,\Big]
\nonumber\\[2mm]
& & - \,\frac{4 \, \pi\,Q_b^2}{3 M_b^4}\,
C_2(\mu_{\rm hard},\mu_{\rm fac})\,
\mbox{Im}\Big[\,
{\cal{A}}_2(E,\mu_{\rm soft},\mu_{\rm fac})
\,\Big]
+ \ldots
\,,
\label{crosssectionexpansion}
\end{eqnarray}
where
\begin{eqnarray}
{\cal{A}}_1 & = & \langle \, 0 \, | 
\, ({\tilde\psi}^\dagger \vec\sigma \, \tilde \chi)\,
\, ({\tilde\chi}^\dagger \vec\sigma \, \tilde \psi)\,
| \, 0 \, \rangle
\,,
\label{A1def}
\\[2mm]
{\cal{A}}_2 & = & \mbox{$\frac{1}{2}$}\,\langle \, 0 \, | 
\, ({\tilde\psi}^\dagger \vec\sigma \, \tilde \chi)\,
\, ({\tilde\chi}^\dagger \vec\sigma \, (\mbox{$-\frac{i}{2}$} 
\stackrel{\leftrightarrow}{\mbox{\boldmath $D$}})^2 \tilde \psi)\,
+ \mbox{h.c.}\,
| \, 0 \, \rangle
\,.
\label{A2def}
\end{eqnarray}
The cross section is expanded in terms of a sum of absorptive parts of
nonrelativistic current correlators, each of them multiplied by a
short-distance coefficient. In fact, the right-hand side (RHS) of
Eq.~(\ref{crosssectionexpansion})
just represents an application of the factorization formalism proposed
in~\cite{Bodwin1}. The second term on the RHS of
Eq.~(\ref{crosssectionexpansion}) is suppressed by $v^2$, i.e.\ of
NNLO. This can be seen explicitly by using the equations of motion
from from the NRQCD Lagrangian, which relates the correlator
${\cal{A}}_2$ directly to ${\cal{A}}_1$,
\begin{equation}
{\cal{A}}_2 = M_b\,E\,{\cal{A}}_2
\,.
\label{A2toA1}
\end{equation}
Relation~(\ref{A2toA1}) has also been used to obtain the coefficient
$-4/3$ in front of the second term on the RHS of
Eq.~(\ref{crosssectionexpansion}).
The nonrelativistic current correlators ${\cal{A}}_1$ and
${\cal{A}}_2$ contain the resummation of the singular terms mentioned
in the previous paragraph. They incorporate all the
long-distance\footnote{
In the context of this paper ``long-distance'' is not equivalent to
``nonperturbative''.
}
dynamics governed by soft scales like the relative three momentum
$\sim M_b\alpha_s$ or the binding energy of
the $b\bar b$ system $\sim M_b\alpha_s^2$.\footnote{
It is not clear at all whether there a not even smaller energy scales 
$\sim M_b\,\alpha_s^k$, $k>2$, which might become relevant. However, those
scales can only be produced by higher order effects like the hyperfine
splitting, which should be irrelevant at least for the total cross
section at NNLO.
} 
The constants $C_1$ and $C_2$ (which are also normalized to one at
the Born level), on the other hand, describe short-distance effects
involving hard scales of the order of the bottom quark mass. They only
represent a simple power series in $\alpha_s$ and do not contain any
resummations in $\alpha_s$. Because we consider
the total $b\bar b$ cross section normalized to the point cross
section, Eq.~(\ref{Rdefinitioncovariant}), $C_1$ and $C_2$ are
independent of $q^2$. In Eq.~(\ref{crosssectionexpansion}) we have
also indicated the dependence of the correlators and the
short-distance coefficients on the various renormalization scales: The
factorization scale $\mu_{\rm fac}$ essentially represents the boundary
between hard and soft momenta. The dependence on the factorization scale
becomes explicit because of ultraviolet (UV) divergences contained in
NRQCD. Because, as in any effective field theory, this boundary is not
defined unambiguously, both the
correlators and the short-distance coefficients in general depend on
$\mu_{\rm fac}$. The soft scale $\mu_{\rm soft}$ and the hard scale
$\mu_{\rm hard}$, on the other hand, are inherent to the correlators and
the short-distance constants, respectively, governing their
perturbative expansion. If we would have all orders in $\alpha_s$ and
$v$ at hand, the dependence of the cross section 
$R_{\mbox{\tiny NNLO}}^{\mbox{\tiny thr}}$ on variations of each the
three scales would vanish exactly. (It is important
that the soft and the hard scale, which both originate from the light
degrees of freedom in the NRQCD Lagrangian, and the factorization
scale are each considered as independent. They can each be defined in
different regularization schemes. In this work we will use the
$\overline{\mbox{MS}}$ scheme for the soft and the hard scale and a
cutoff scheme for the factorization scale.)
Unfortunately, we only perform the calculation up to NNLO in
$\alpha_s$ and $v$ which leads to a residual dependence on the three
scales $\mu_{\rm fac}$, $\mu_{\rm soft}$ and $\mu_{\rm hard}$. In
particular, (as we will demonstrate in Section~\ref{sectionproperties})
the dependence on the soft scale $\mu_{\rm soft}$ is
quite strong, clearly because it governs the perturbative
expansion of the correlators where convergence of the perturbation
series can be expected to be worse than for the short-distance
constants. It is therefore necessary to fix a certain window for each
of the renormalization scales for which the perturbative series for
the cross section shall be evaluated. At this point one can basically
only rely on physical intuition, which tells that the renormalization
scales should be of the same order as the physical scales governing the
particular problem. This means that the soft scale
should be the order of the relative momentum of the $b\bar b$
system\footnote{
We will see later that at NNLO all interactions can be treated as
instantaneous. As a consequence scales of the order of the binding
energy $\sim M_b\,\alpha_s^2$ can be ignored.
}
$\sim M_b\,\alpha_s$, and that the hard scale should be of
order $M_b\sim 5$~GeV. The factorization scale, on the other hand,
should cover (at least partly) the soft and and hard regime.
Because there is in our opinion no unique way to make this statement
more quantitative, it is important to choose the corresponding windows
``reasonably large''. In our case the choices are as follows:
\begin{eqnarray}
1.5\,\mbox{GeV}\, \le & \mu_{\rm soft} & \le \, 3.5\,\mbox{GeV}
\,,
\nonumber\\[1mm]
2.5\,\mbox{GeV} \,\le & \mu_{\rm hard} & \le \, 10\,\mbox{GeV}
\,,
\nonumber\\[1mm]
2.5\,\mbox{GeV} \,\le & \mu_{\rm fac} & \le \, 10\,\mbox{GeV}
\,.
\label{choiceofscales}
\end{eqnarray}
We will show in Sections~\ref{sectionfitting} and \ref{sectionresults}
that the dependence of the
theoretical moments $P_n^{th}$ on theses scales represents the
dominant source of the uncertainties in the extraction of $M_b$. Thus,
it is the choice given in Eqs.~(\ref{choiceofscales}) which
determines the size of the uncertainties!\\

\begin{figure}[t!] % figpropagators
\begin{center}
\leavevmode
\epsfxsize=3cm
\epsffile[220 420 420 550]{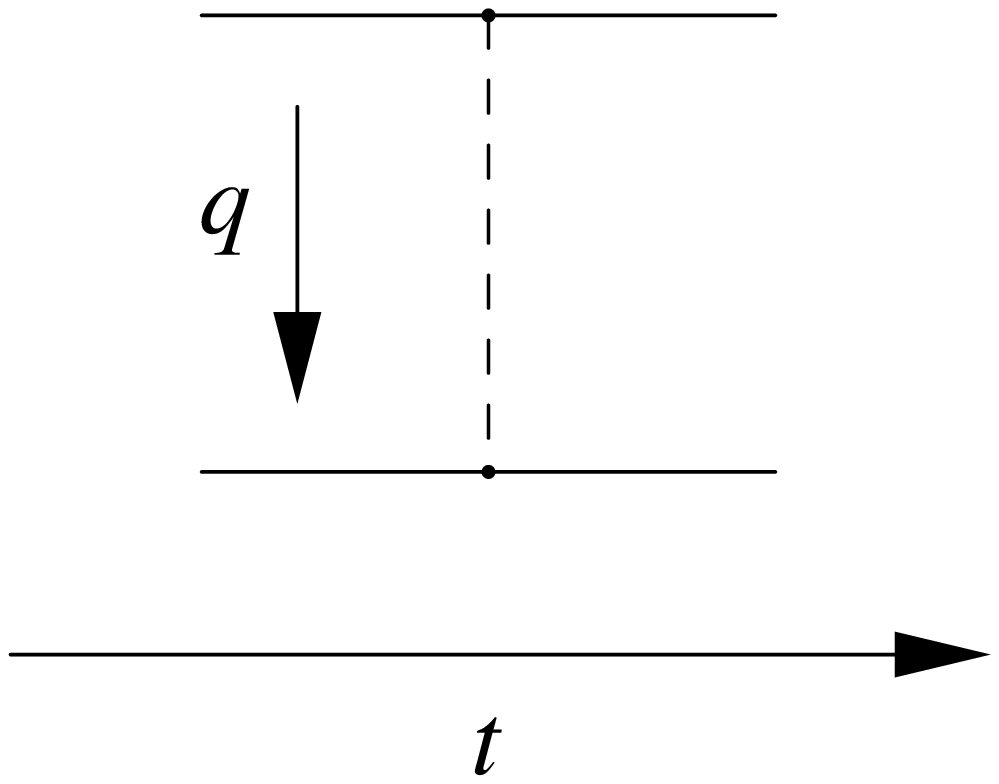}
\mbox{\large $G^{0 0}_{\rm long} = 
\frac{\displaystyle i}{\displaystyle \mbox{}\,\,\vec q^{\,2}\,\,\mbox{}}$} 
\hspace{2cm} 
\epsfxsize=3cm
\leavevmode
\epsffile[220 420 420 550]{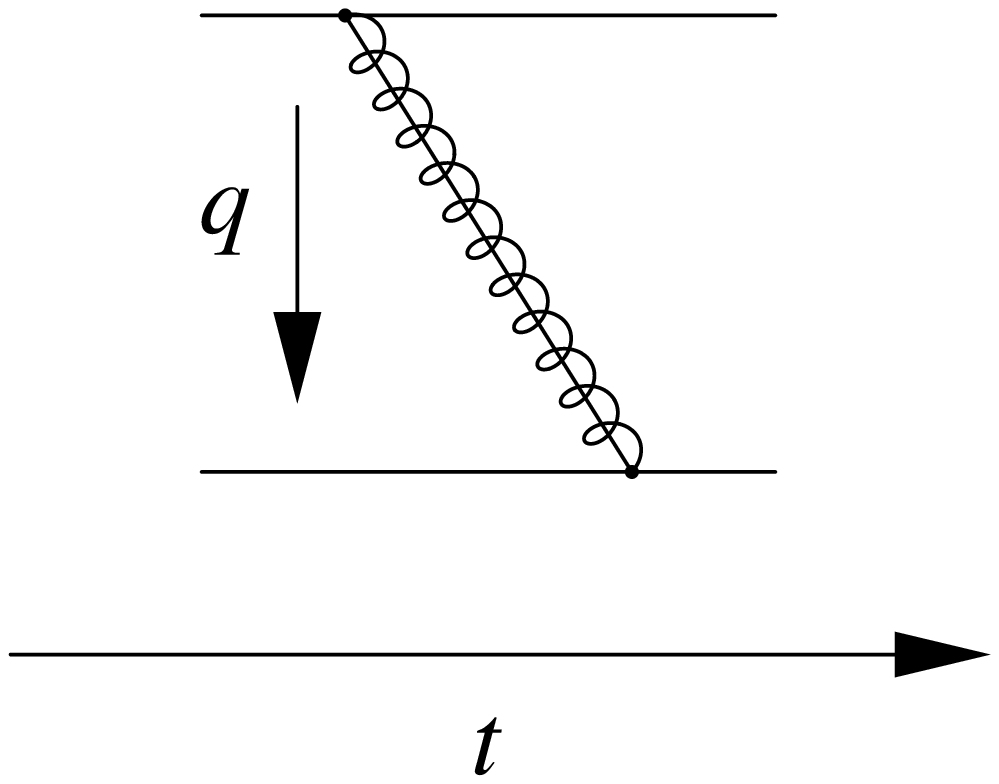}
\mbox{\large $G^{i j}_{\rm trans} = 
\frac{\displaystyle i}{\displaystyle  \mbox{}\,\,q^2\,\,\mbox{}}
\bigg(\delta_{i j}-
\frac{\displaystyle q^i q^j}{\displaystyle \vec q^{\,2}}\bigg)$} 
\vskip  2.0cm
 \caption{\label{figpropagators} 
Graphical representation of the longitudinal  and the transverse
gluon exchange including the corresponding Feynman rules for the
momentum exchange $q=(q^0,\vec q)$. The exchange of a longitudinal
gluon is instantaneous in time because its does not have an energy
dependence. As a consequence the longitudinal exchange can be
described by an instantaneous potential. The exchange of a transverse
gluon, on the other hand, is retarded in time and, in general, cannot
be described in terms of an instantaneous potential.
}
%\label{figpropagators}
 \end{center}
\end{figure}

\noindent
{\bf\underline{Instantaneous Interactions and Retardation
Effects}}\\[1mm]
To calculate the correlators ${\cal{A}}_1$ and ${\cal{A}}_2$
we use methods originally developed for QED bound state calculations
in the framework of NRQED~\cite{Hoang1,Caswell1,Lepage1,Hoang5} and
transfer them (with the appropriate modifications to account for the
non-Abelian effects) to the problem of heavy quark-antiquark production
in the kinematic regime close to the threshold. Because the Coulomb
gauge is the standard gauge in which QED bound state calculations are
carried out we also use the Coulomb gauge for the calculations in this
work. The Coulomb gauge separates the gluon propagator into a
longitudinal and a transverse piece (see Fig.~\ref{figpropagators}). 
The longitudinal propagator does not have an energy dependence and
therefore represents an instantaneous interaction. As a consequence,
in configuration space representation a longitudinal gluon exchange can
be written as an instantaneous potential (which only depends on the
spatial distance). Through the time derivative in the NRQCD Lagrangian
the longitudinal gluon exchange leads to the Coulomb potential which
is the dominant (LO) interaction between the bottom quarks in the
nonrelativistic limit. Through the $1/M_b^2$ couplings of the bottom
quarks to the chromo-electric field the longitudinal exchange also
leads to the Darwin and spin-orbit potential, which contribute at the
NNLO level. Because these potentials are instantaneous their treatment
is straightforward in the framework of a two-body Schr\"odinger
equation.

For the transverse gluon the situation is more
subtle. Because all couplings of the bottom quarks to the
chromo-magnetic field are of order $1/M_b$ the exchange of a
transverse gluon between two bottom quark lines is a NNLO
effect. However, in contrast to the Darwin and the spin-orbit
interaction, the propagation of the transverse gluon energy has an
energy dependence, i.e.\ it is an interaction with a temporal
retardation. Physically this means that the transverse gluon can
travel alongside the $b\bar b$ pair for some time
period~\cite{Voloshin2,Voloshin3}. In this time
period the $b\bar b$ pair is part of a higher order Fock $b\bar
b$-gluon state which, in principle, cannot be treated in terms of a
two-body Schr\"odinger equation. Fortunately, in our case we can
neglect the energy dependence
of the transverse gluon propagator completely. This can be easily
understood by considering a typical diagram describing the exchange of
a transverse gluon between the $b\bar b$ pair in the background of a
continuous Coulomb exchange of longitudinal gluons, see
e.g.\ Fig.~\ref{figtransverse}a. 
\begin{figure}[t] % figladder
\begin{center}
\leavevmode
\epsfxsize=2cm
\epsffile[220 390 420 520]{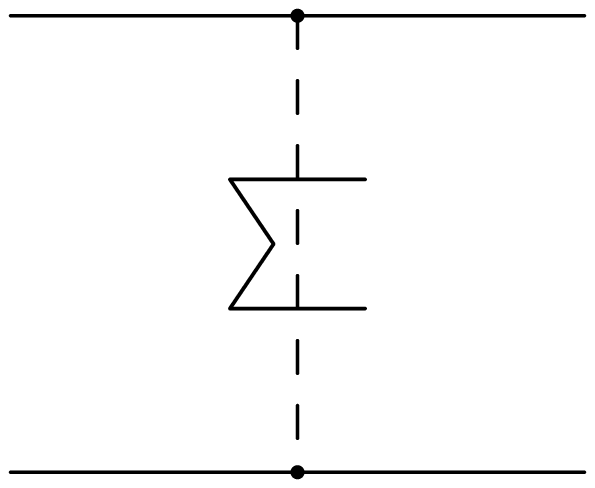}
\mbox{\large\mbox{}\quad = \quad\mbox{}}
\leavevmode
\epsfxsize=2cm
\epsffile[220 390 420 520]{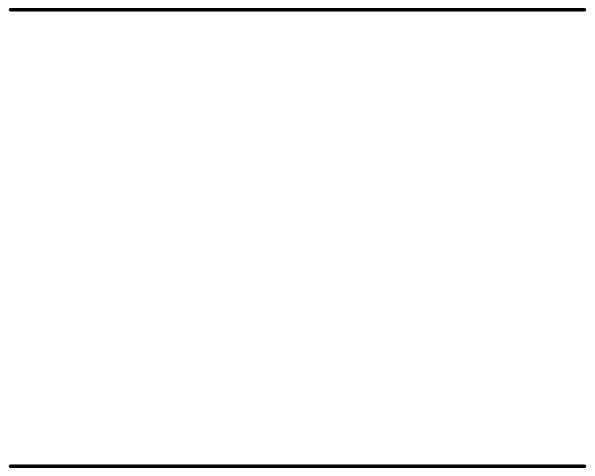}
\mbox{\large + }
\leavevmode
\epsfxsize=2cm
\epsffile[220 390 420 520]{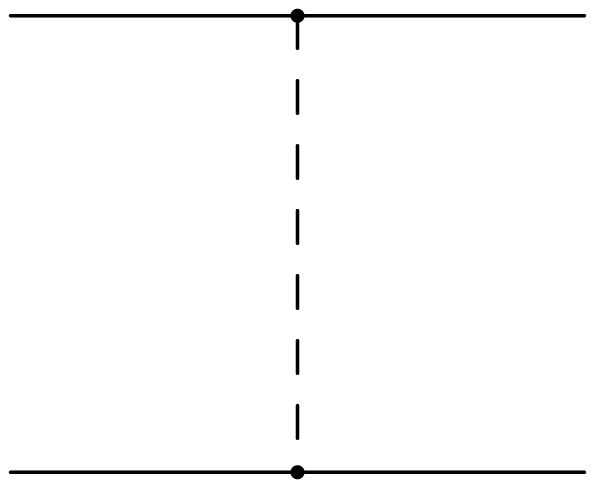}
\mbox{\large + }
\leavevmode
\epsfxsize=2cm
\epsffile[220 390 420 520]{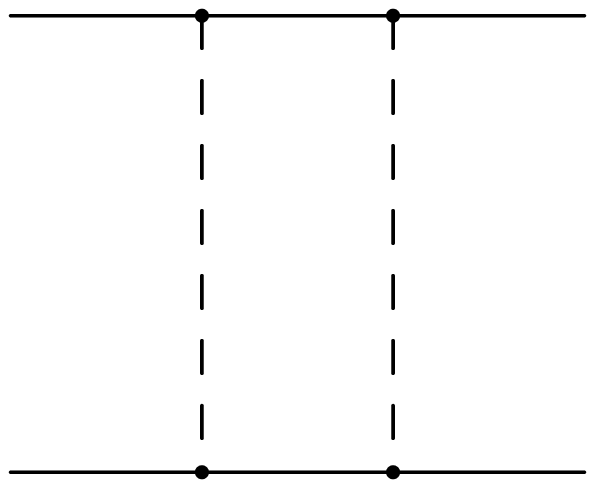}
\mbox{\large + }
\leavevmode
\epsfxsize=2cm
\epsffile[220 390 420 520]{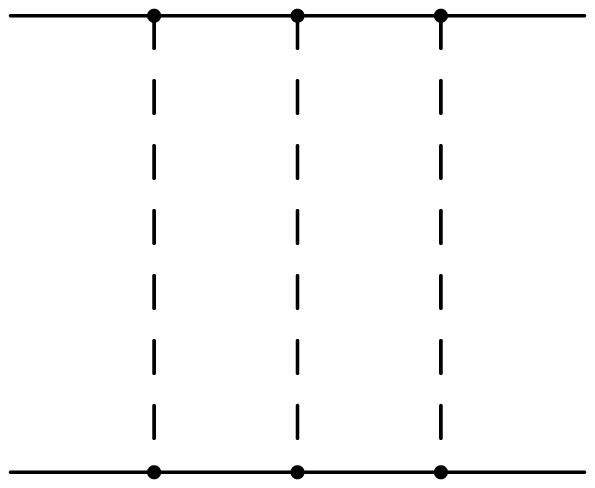}
\mbox{\large + \ldots }
\vskip  1.cm
 \caption{\label{figladder} Graphical representation of the
resummation of Coulomb ladder diagrams to all orders. The
quark-antiquark propagation contains the nonrelativistic kinetic
energy. The resummation is carried out explicitly by calculating the
Green function the nonrelativistic Schr\"odinger equation with the
Coulomb potential at the Born level, see
Eq.~(\ref{CoulombGreenfunctioncomplete}).}
%\label{figladder}
 \end{center}
\end{figure}
\begin{figure}[t] % figtransverse
\begin{center}
\leavevmode
\epsfxsize=2cm
\epsffile[220 420 420 550]{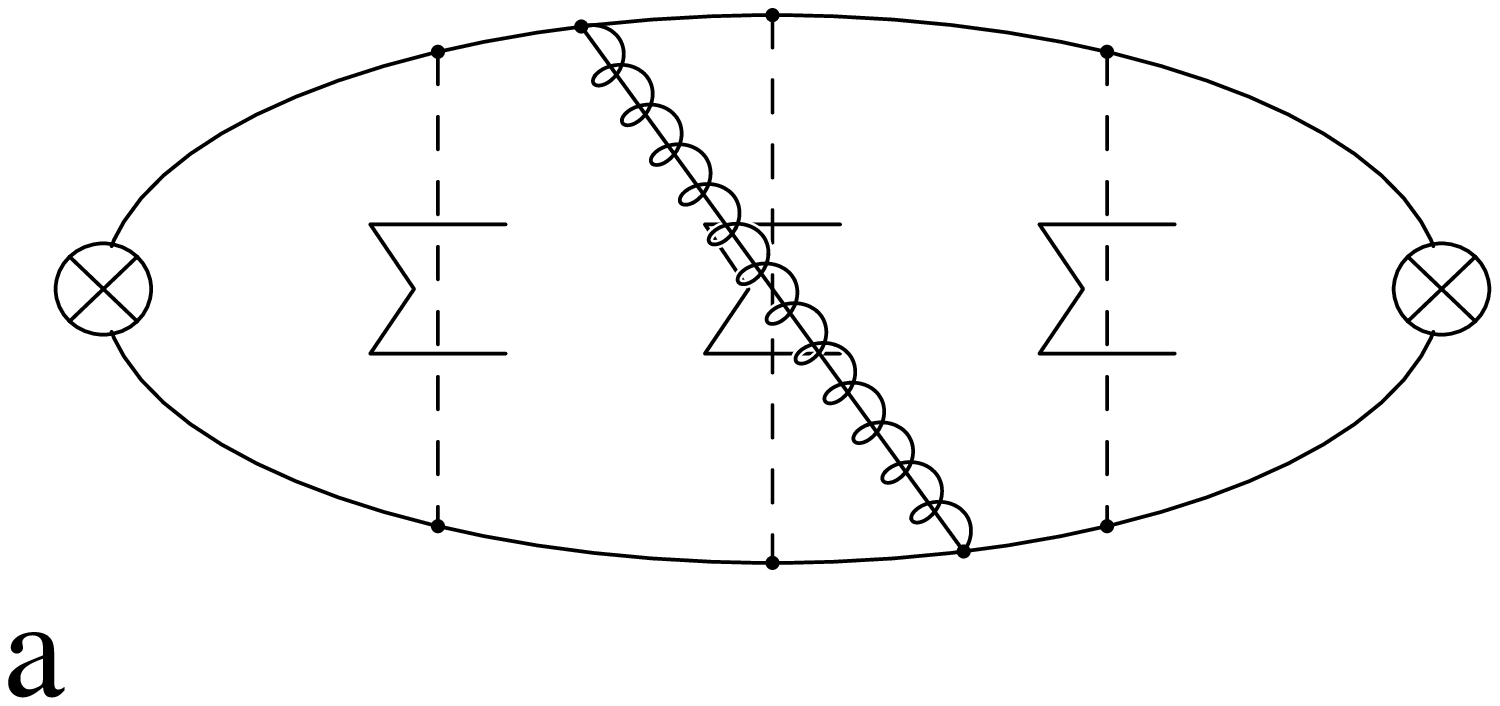}
\hspace{3cm}
\leavevmode
\epsfxsize=2cm
\epsffile[220 420 420 550]{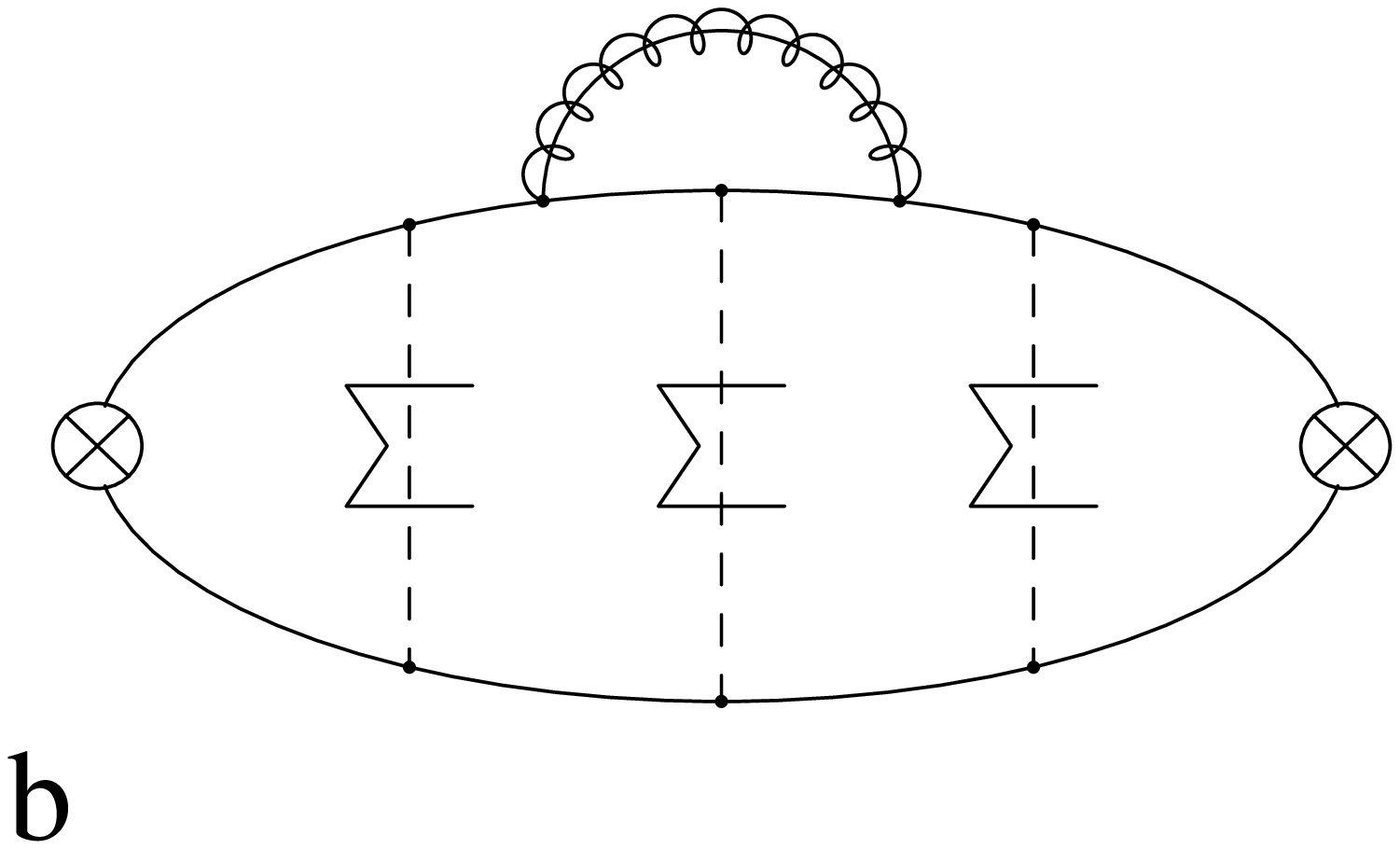}
\hspace{3cm}
\leavevmode
\epsfxsize=2cm
\epsffile[220 420 420 550]{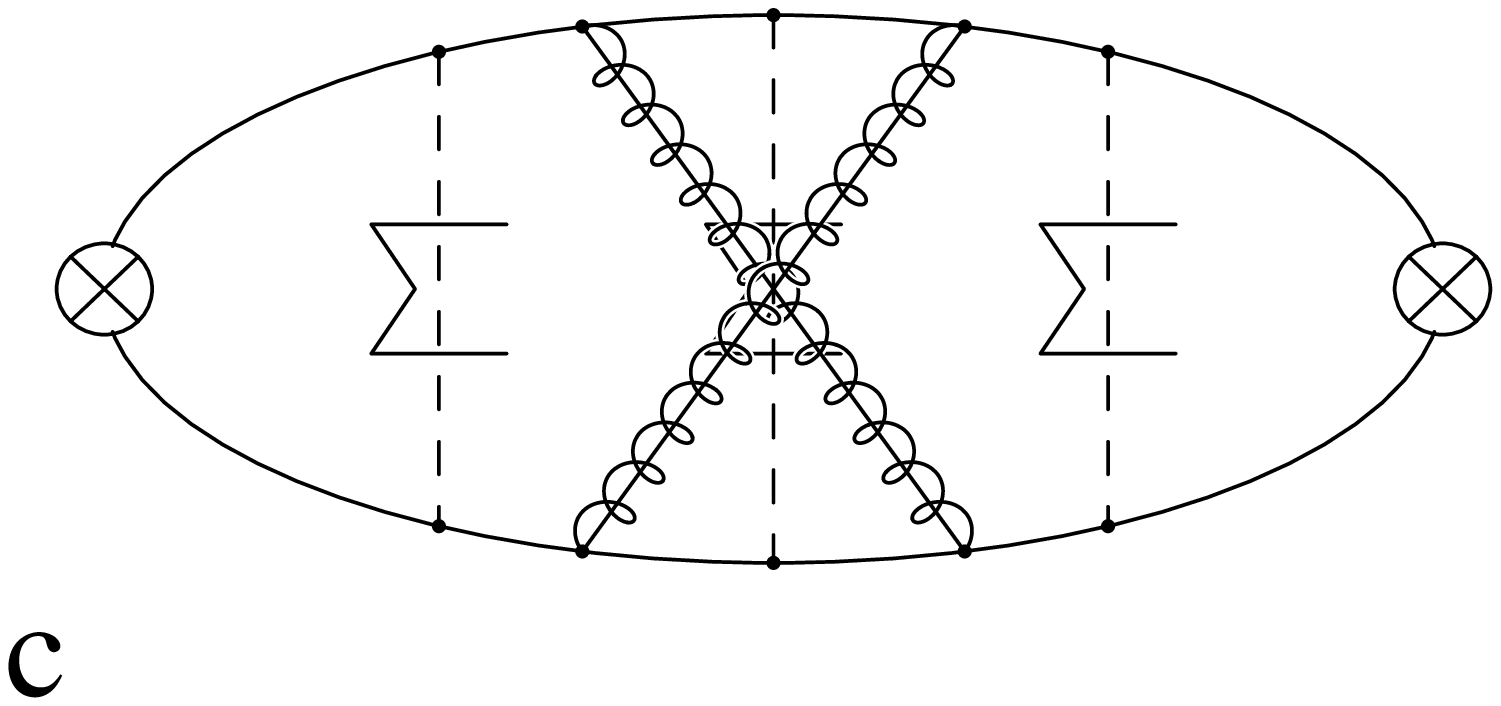}
\vskip  1.5cm
 \caption{\label{figtransverse} Typical diagrams describing the
exchange of a transverse gluons (in Coulomb gauge) in the background of
the Coulomb exchange of longitudinal gluons. Longitudinal lines with a
$\sum$ sign represent the summation of Coulomb ladder diagrams to all
orders, see Fig.~\ref{figladder}.}
%\label{figtransverse}
 \end{center}
\end{figure}
If both ends of the transverse gluon end at bottom quarks the typical
energy carried by the gluon can only be of order $M_b\,v^2$, the c.m.\
kinetic energy of the bottom quarks. The typical three momentum of the
gluon, on the other hand, can either be of order $M_b\,v$, the
relative momentum in the $b\bar b$ system, or also of order
$M_b\,v^2$.
If the three momentum is of order $M_b\,v^2$, the
transverse gluon is essentially real and needs, in addition to the
$v^2$ suppression coming from the couplings to the quarks, another
phase space factor $v$ to exist. Thus, the transverse gluons with this
energy-momentum configuration lead to effects suppressed by $v^3$,
which is beyond the NNLO level. If the three momentum of the transverse
gluon is of order $M_b\,v$, on the other hand, it is far off-shell and
we can neglect the small energy component in a first
approximation. [It should be emphasized that this argument implies the
hierarchy $M_b\alpha_s\gg M_b\alpha_s^2$, which is
conceivable for the $b\bar b$ system where $\alpha_s\sim 0.3$.]
From that one can see that at NNLO the transverse gluon
exchange can, like the longitudinal one, also be treated as an
instantaneous interaction. This means that in
Fig.~\ref{figtransverse}a only those diagrams contribute at NNLO where
the transverse line does not cross any longitudinal line.
The differences between longitudinal and
transverse gluons will only become manifest beyond the NNLO level. 
For the same reason any self-energy or crossed-ladder type diagram
(see Figs.~\ref{figtransverse}b,c for typical examples) can be safely
neglected at the NNLO level. In fact, the situation is in complete
analogy to the hydrogen
atom or the positronium in QED, where it is well known that retardation
effects lead to the ``Lamb-shift'' corrections which are suppressed by
$\alpha^3$ relative to the LO nonrelativistic contributions. [Of
course, the crossed exchange and self-energy type diagrams have to be
taken into account in the two-loop calculation of the cross section in
full QCD needed to determine the ${\cal{O}}(\alpha_s^2)$ contributions
to the short-distance coefficients. Those short-distance constants,
however, describe effect from high momenta of order
$M_b$ which are not contained in the correlators.] 

From the considerations above we can draw the following
conclusions regarding the calculation of the correlators 
${\cal{A}}_1$ and ${\cal{A}}_2$ at NNLO:
\begin{itemize}
\item[1.] We can treat the problem of $b\bar b$ production
close to threshold as a pure two-body problem. This means that the
NRQCD Lagrangian effectively reduces to a two-body Schr\"odinger
equation from which the correlators can be determined.
\item[2.] All interactions between the bottom and the antibottom quark
can be written as time independent, instantaneous potentials,
which means that only ladder-like diagrams have to taken into account.
\item[3.] We can use the well known analytic solutions of the
nonrelativistic Coulomb problem for
positronium~\cite{Wichmann1,Hostler1,Schwinger1} and use
Rayleigh-Schr\"odinger time-independent perturbation theory (TIPT) to
determine the corrections caused by all higher order interactions and
effects. 
\end{itemize} 
However, there is one remark in order: although the effects of the
transverse gluon exchange having a temporal retardation are formally
beyond
the NNLO level, this is not a proof that they are indeed smaller than
the NNLO contributions calculated in this work. It is in fact rather
likely that the retardation effects cannot be calculated at all using
perturbative methods because the characteristic scale of the
coupling governing the emission, absorption or interaction of a gluon
which has energy and momentum of order $M_b\alpha_s^2$ would be
of the order of $0.5 - 1$~GeV. This is already quite close
to the typical hadronization scale $\Lambda_{\mbox{\tiny QCD}}$. From
this point of view it seems that the NNLO analysis presented here
cannot be improved any more, at least not with perturbative
methods. This problem might even cast doubts on the reliability of the
NNLO corrections themselves and underlines the necessity that the
preferred ranges for the renormalization scales,
Eqs.~(\ref{choiceofscales}), are chosen sufficiently large. We will
ignore further implications of this problem for the calculations and
analyses carried out in this work. (See also the paragraph on
nonperturbative effects.)\\

\noindent
{\bf\underline{Instantaneous Potentials}}\\[1mm]
At the Born level all potentials relevant for the nonrelativistic
cross section at NNLO can be obtained directly from the NRQCD
Lagrangian considering (color singlet) $b\bar b\to b\bar b$ single gluon
t-channel exchange scattering diagrams. In configuration space
representation the Born level potentials read 
($r\equiv|\vec r|$, $C_F=4/3$, $C_A=3$, $T=1/2$,
$a_s\equiv\alpha_s(\mu_{\rm soft})$) 
\begin{eqnarray}
V_c^{(0)}(\vec r) & = & -\,\frac{C_F\,a_s}{r}
\,,
\label{Vcoulomb}
\nonumber\\[2mm] 
V_{\mbox{\tiny BF}}(\vec r) & = & 
\frac{C_F\,a_s\,\pi}{M_b^2}\,
\Big[\,
1 + \frac{8}{3}\,\vec S_b\,\vec S_{\bar b}
\,\Big]
\,\delta^{(3)}(\vec r)
+ \frac{C_F\,a_s}{2 M_b^2 r}\,\Big[\,
\vec\nabla^2 + \frac{1}{r^2} \vec r\, (\vec r \, \vec\nabla) \vec\nabla
\,\Big]
\nonumber\\[2mm] & &
- \,\frac{3\,C_F\,a_s}{M_b^2 r^3}\,
\Big[\,
\frac{1}{3}\,\vec S_b \,\vec S_{\bar b} - 
\frac{1}{r^2}\,\Big(\vec S_b\,\vec r\,\Big)
\,\Big(\vec S_{\bar b}\,\vec r\,\Big)
\,\Big]
+ \frac{3\,C_F\,a_s}{2 M_b^2 r^3}\,\vec L\,(\vec S_b+\vec S_{\bar b})
\,,
\label{VBreitFermi}
\end{eqnarray}
where $\vec S_b$ and $\vec S_{\bar b}$ are the bottom and antibottom
quark spin operators and $\vec L$ is the angular momentum operator.
$V_c^{(0)}$ is the well known Coulomb potential. It constitutes the
LO interaction and will (together with the nonrelativistic kinetic
energy) be taken into account exactly.
It arises from the exchange of a longitudinal gluon through the time
derivative coupling of the bottom quarks to the gluon field.
$V_{\mbox{\tiny BF}}$ represents the Breit-Fermi potential
which is known from higher order positronium calculations. It
describes the Darwin 
and spin-orbit interactions which are mediated by the longitudinal
gluons and also the so called hyperfine or tensor interactions
which are mediated by the transverse gluons in the instantaneous
approximation. Due to the $1/M_b^2$ suppression $V_{\mbox{\tiny BF}}$
already leads to NNLO effects in the cross section and will be taken
into account as a perturbation. For the same reason only the radiative
corrections to the Coulomb exchange of longitudinal gluons have to
be taken into account. We want to emphasize that these radiative
corrections are caused by the massless degrees of freedom in the NRQCD
Lagrangian. Because in the corresponding loops transverse gluon lines
can end at other massless lines the considerations given in the
preceding paragraph cannot be applied in this case. Thus, in general,
transverse gluons (or massless quarks) in all energy-momentum
configurations have to taken into account to calculate the radiative
corrections properly.
The calculation of these radiative corrections 
can be found in existing literature and we therefore just present the
results. 

\begin{figure}[t] % fignonanalytic
\begin{center}
\leavevmode
\epsfxsize=3cm
\epsffile[220 390 420 520]{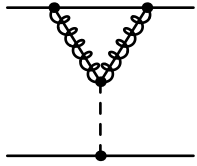}
\vskip  1.cm
 \caption{\label{fignonanalytic} Vertex diagram in Coulomb gauge
responsible for the potential non-Abelian potential $V_{\mbox{\tiny
NA}}$.} 
%\label{fignonanalytic}
 \end{center}
\end{figure}

At the one-loop level (and using the $\overline{\mbox{MS}}$
scheme for the strong coupling) the corrections read 
($\gamma_{\mbox{\tiny E}} = 0.57721566\ldots$ being the
Euler-Mascheroni constant) 
\begin{eqnarray}
V_c^{(1)}(\vec r) & = &
V_c^{(0)}(\vec r)\, \bigg(\frac{a_s}{4\,\pi}\bigg)\,\bigg[\,
2\,\beta_0\,\ln(\tilde\mu\,r) + a_1
\,\bigg]
\,,
\qquad
\tilde\mu \, \equiv \, e^{\gamma_{\mbox{\tiny E}}}\,\mu_{\rm soft}
\,,
\label{Vrunning1loop}
\end{eqnarray}
where
\begin{eqnarray}
\beta_0 & = & \frac{11}{3}\,C_A - \frac{4}{3}\,T\,n_l
\,,
\nonumber\\
a_1 & = &  \frac{31}{9}\,C_A - \frac{20}{9}\,T\,n_l
\,,
\nonumber\\[2mm]
n_l & = & 4
\,,
\label{b0a1def}
\end{eqnarray}
and
\begin{eqnarray}
V_{\mbox{\tiny NA}}(\vec r) & = & -\,
\frac{C_A\,C_F\,a_s^2}{2 \, M_b \, r^2}
\,.
\label{Vnonabelian}
\end{eqnarray}
$V_c^{(1)}$ represents the one-loop corrections to the Coulomb
potential $\propto 1/r$ and leads to NLO contributions in the cross
section. $V_c^{(1)}$ has been calculated by Fischler~\cite{Fischler1}
and Billoire~\cite{Billoire1}. $V_{\mbox{\tiny NA}}$, called
non-Abelian potential for the rest of this work, arises
from a non-analytic behavior of the vertex diagram
depicted in Fig.~\ref{fignonanalytic} $\propto (\vec k^2/M_b^2)^{1/2}$
where $\vec k$ is the three momentum exchanged between the bottom and
the antibottom quark.
Because the non-analytic term causes a behavior $\propto
1/|\vec k|$ for the non-Abelian potential
in momentum space representation, $V_{\mbox{\tiny NA}}$ is
proportional to $1/r^2$. We would like to point out that in Coulomb
gauge such a non-analytic behavior does not exist for Abelian
diagrams.  We refer the reader to~\cite{nonabelianpotential} for older
publications, where the non-Abelian potential has been determined. 
Due to the $a_s/M_b$ factor $V_{\mbox{\tiny NA}}$ is a NNLO
interaction and no further corrections to it have to taken into
account. 

\begin{figure}[t] % figinteractions
\begin{center}
\leavevmode
\epsfxsize=2cm
\epsffile[220 410 420 540]{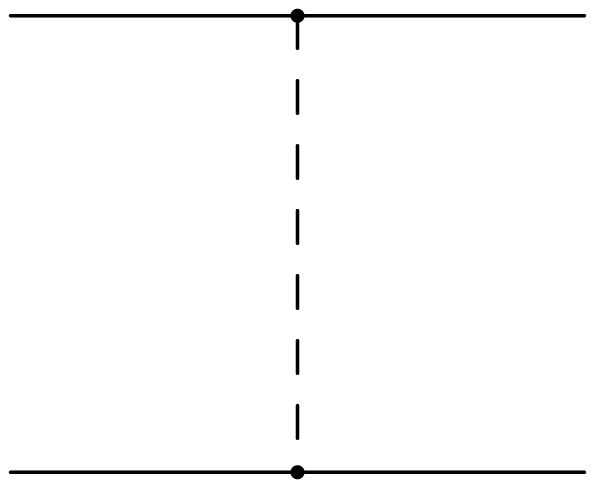}
\hspace{1cm}
\leavevmode
\epsfxsize=2cm
\epsffile[220 410 420 540]{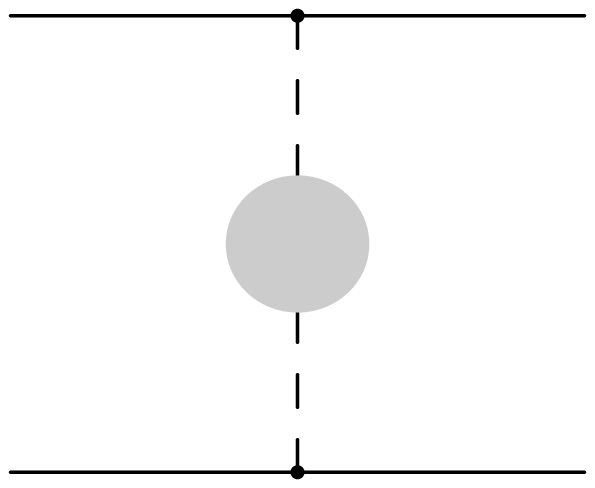}
\hspace{1cm}
\leavevmode
\epsfxsize=2cm
\epsffile[220 410 420 540]{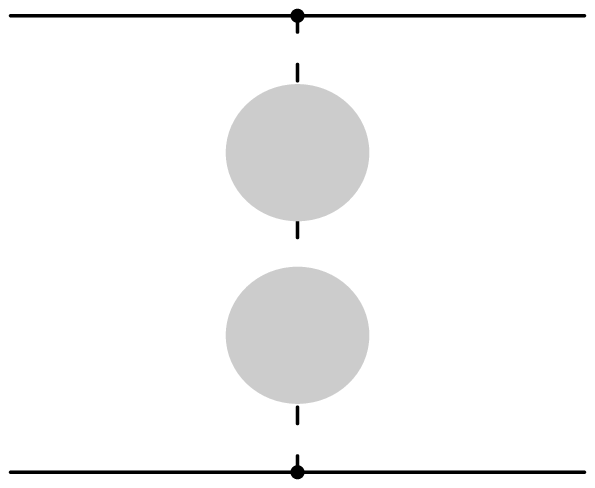}
\hspace{1cm}
\leavevmode
\epsfxsize=2cm
\epsffile[220 410 420 540]{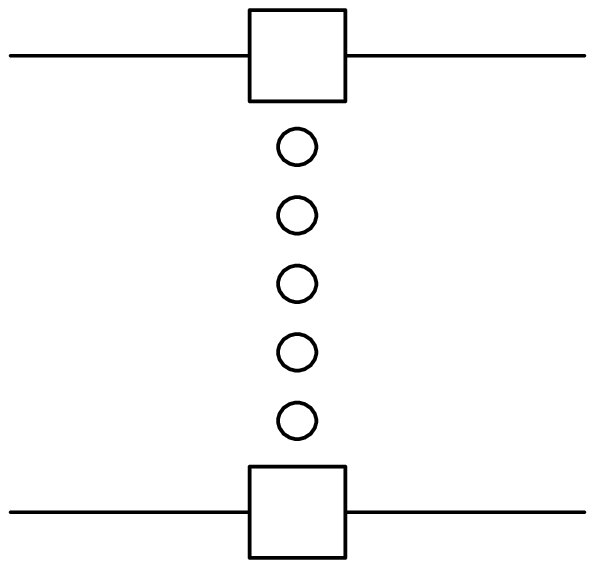}
\hspace{1cm}
\leavevmode
\epsfxsize=2cm
\epsffile[220 410 420 540]{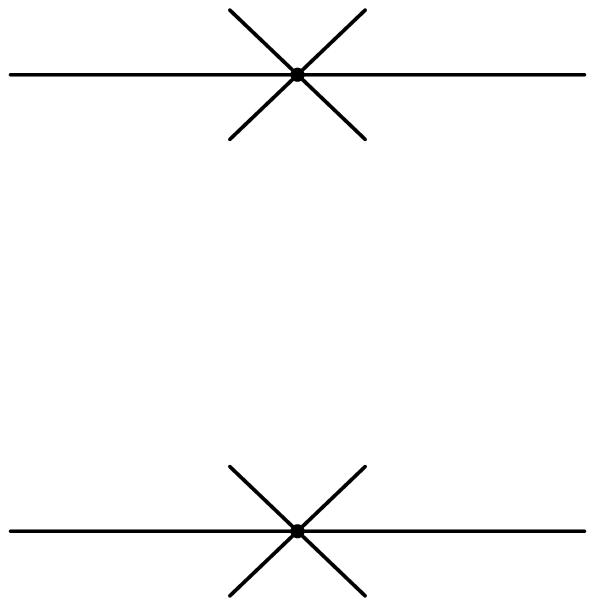}\\[1.2cm]
$V_c^{(0)}$
\mbox{\hspace{2.2cm}}
$V_c^{(1)}$
\mbox{\hspace{2.2cm}}
$V_c^{(2)}$ 
\mbox{\hspace{1.6cm}}
$V_{\mbox{\tiny BF}}+V_{\mbox{\tiny NA}}$
\mbox{\hspace{1.7cm}}
$\delta H_{\mbox{\tiny kin}}$
\vskip  .2cm
 \caption{\label{figinteractions} 
Symbols describing the interactions potentials $V_c^{(0)}$,
$V_c^{(1)}$, $V_c^{(2)}$, $V_{\mbox{\tiny BF}}$ and $V_{\mbox{\tiny
NA}}$ and the kinetic energy correction 
$\delta H_{\mbox{\tiny kin}} = -\vec\nabla^4/4 M_b^3$.}
%\label{figinteractions}
 \end{center}
\end{figure}

\begin{figure}[t] % figannihilation
\begin{center}
\leavevmode
\epsfxsize=3cm
\epsffile[220 380 420 510]{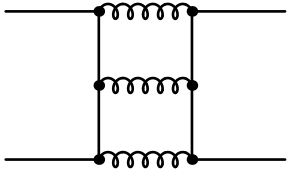}\\
\vskip  1.cm
 \caption{\label{figannihilation} 
The dominant annihilation diagram relevant for $b\bar b\to b\bar b$
scattering for a bottom-antibottom quark pair in a color singlet 
$J^{PC} = 1^{--}$, ${}^3\!S_1$ configuration. Its dominant
contribution leads to a potential 
$V_{\mbox{\tiny ann}}(\vec r) \propto a_s^3/M_b^2\,\delta(\vec r)$ and
to contributions in the cross section and the moments beyond the NNLO
level.}
%\label{figannihilation}
 \end{center}
\end{figure}

At the two-loop level only the corrections to the Coulomb
potential have to be considered. They have been calculated recently by
Peter~\cite{Peter1} and read (in the $\overline{\mbox{MS}}$ scheme)
\begin{eqnarray}
V_c^{(2)}(\vec r) & = &
V_c^{(0)}(\vec r)\, \bigg(\frac{a_s}{4\,\pi}\bigg)^2\,\bigg[\,
\beta_0^2\,\bigg(\,4\,\ln^2(\tilde\mu\,r) 
      + \frac{\pi^2}{3}\,\bigg) 
+ 2\,\Big(2\,\beta_0\,a_1 + \beta_1\Big)\,\ln(\tilde\mu\,r) 
+ a_2
\,\bigg]
\,,
\label{Vrunning2loop}
\end{eqnarray}
where
\begin{eqnarray}
\beta_1 & = & \frac{34}{3}\,C_A^2 
-\frac{20}{3}C_A\,T\,n_l
- 4\,C_F\,T\,n_l
\,,
\nonumber\\[2mm]
a_2 & = & 
\bigg(\,\frac{4343}{162}+6\,\pi^2-\frac{\pi^4}{4}
 +\frac{22}{3}\,\zeta_3\,\bigg)\,C_A^2 
-\bigg(\,\frac{1798}{81}+\frac{56}{3}\,\zeta_3\,\bigg)\,C_A\,T\,n_l
\nonumber\\[2mm] & &
-\bigg(\,\frac{55}{3}-16\,\zeta_3\,\bigg)\,C_F\,T\,n_l 
+\bigg(\,\frac{20}{9}\,T\,n_l\,\bigg)^2
\,.
\label{b1a2def}
\end{eqnarray}
For later reference we assign the symbols in
Fig.~\ref{figinteractions} to 
the potentials given above. We also would like to note that we do not
have to consider any annihilation effects. The leading annihilation
diagram is depicted in Fig~\ref{figannihilation}. Because the
annihilation process takes place at short distances, it produces local
four-fermion operators in the NRQCD Lagrangian, which can be written
as instantaneous potentials. The dominant annihilation potential which
comes from the three gluon annihilation diagram has the form
$V_{\mbox{\tiny ann}}(\vec r) \propto (a_s^3/M_b^2)
\delta^{(3)}(\vec r)$ and would lead to effects suppressed by $v^4$ in
the cross section.\\

\noindent
{\bf\underline{Recipe for the Calculation of Large $n$ Moments at
NNLO}}\\[1mm]
Based on the issues discussed above the calculation of the NNLO
nonrelativistic cross section $R_{\mbox{\tiny NNLO}}^{\mbox{\tiny
thr}}$ and the theoretical moments $P_n^{th}$ in terms of the
correlators ${\cal{A}}_1$ and ${\cal{A}}_2$ and
the short-distance coefficients $C_{1/2}$ proceeds in the following
three basic steps:

\vspace{2mm}\noindent\hspace{2mm}
Step 1: \begin{minipage}[t]{15cm} 
{\it Solution of the Schr\"odinger equation.} -- The Green function 
of the NNLO Schr\"odinger equation is calculated incorporating the
potentials displayed above and including the NNLO corrections to
the kinetic energy. The correlators ${\cal{A}}_1$ and ${\cal{A}}_2$
are directly related to the zero-distance Green function of the
Schr\"odinger equation.
\end{minipage}

\vspace{2mm}\noindent\hspace{2mm}
Step 2: \begin{minipage}[t]{15cm} 
{\it Matching calculation.} -- The short-distance constant $C_1$ is
determined at ${\cal{O}}(\alpha_s^2)$ by matching
expression~(\ref{crosssectionexpansion}) directly to the cross section
calculated in full QCD at the two-loop level and including terms up to
NNLO in an expansion in $v$ in the (formal) limit $\alpha_s\ll v\ll 1$.
\end{minipage}

\vspace{2mm}\noindent\hspace{2mm}
Step 3:  \begin{minipage}[t]{15cm}
{\it Dispersion Integration.} -- The
integration~(\ref{momentcrosssectionrelation}) is carried out.
\end{minipage}

\vspace{2mm}\noindent
For the rest of this section we briefly explain the strategies
and basic procedures for the steps 1 and 2. The explicit calculations
for steps~1 -- 3 are presented in detail in
Sections~\ref{subsectioncorrelators}, \ref{subsectionmatching} and
\ref{subsectionintegration}, respectively.

\vspace{2mm}\noindent\hspace{2mm}
{\it Solution of the Schr\"odinger equation:} 
The nonrelativistic correlators ${\cal{A}}_1$ and ${\cal{A}}_2$ are
calculated by determining the Green function of the Schr\"odinger
equation ($E\equiv \sqrt{q^2}-2 M_b$)
\begin{eqnarray}
& &
\bigg(\,
-\frac{\vec\nabla^2}{M_b} 
- \frac{\vec\nabla^4}{4\,M_b^3} 
+ \bigg[\,
  V_{c}^{(0)}(\vec r) + V_{c}^{(1)}(\vec r) +V_{c}^{(2)}(\vec r) 
  + V_{\mbox{\tiny BF}}(\vec r) + V_{\mbox{\tiny NA}}(\vec r)
\,\bigg]  
- E
\,\bigg)\,G(\vec r,\vec r^\prime, E)
\nonumber\\[2mm] & &
\mbox{\hspace{9cm}}
 = \, \delta^{(3)}(\vec r-\vec r^\prime) 
\,,
\label{Schroedingerfull}
\end{eqnarray}
where $V_{\tiny BF}$ is evaluated for the ${}^3\!S_1$ configuration
only. The relation between the correlator ${\cal{A}}_1$ and Green
function reads
\begin{eqnarray}
{\cal{A}}_1 & = & 6\,N_c\,\Big[\,
\lim_{|\vec r|,|\vec r^\prime|\to 0}\,G(\vec r,\vec r^\prime, E)
\,\Big]
\,.
\label{A1Greenfunctionrelation}
\end{eqnarray}
Eq.~(\ref{A1Greenfunctionrelation}) can be quickly derived from the
facts that $G(\vec r,\vec r^\prime,\tilde E)$ describes the propagation
of a bottom-antibottom pair which is produced and annihilated at
relative distances $|\vec r|$ and $|\vec r^\prime|$, respectively, and
that the bottom-antibottom quark pair is produced and annihilated
through the electromagnetic current at zero
distances. Therefore ${\cal{A}}_1$ must be proportional to $\lim_{|\vec
r|,|\vec r^\prime|\to 0}\,G(\vec r,\vec r^\prime, E)$. The correct
proportionality constant can then be determined by considering
production of a free (i.e.\ $\alpha_s=0$) bottom-antibottom pair in the
nonrelativistic limit. (In this case the Born cross section in full QCD
can be easily compared to the imaginary part of
the Green function of the free nonrelativistic Schr\"odinger equation.)
The correlator ${\cal{A}}_2$ is determined from ${\cal{A}}_1$ via
relation~(\ref{A2toA1}).
We would like to emphasize that the zero-distance Green function on
the RHS of Eqs.~(\ref{A1Greenfunctionrelation}) contains UV
divergences which have to regularized. In the actual calculations
carried out in Section~\ref{subsectioncorrelators} we impose the
explicit short-distance cutoff $\mu_{\rm fac}$. As mentioned
before, this is the reason why the correlators and the short-distance
constants depend explicitly on the (factorization) scale $\mu_{\rm
fac}$. In this work we solve  equation~(\ref{Schroedingerfull})
perturbatively by starting from well known Green function $G_c^{(0)}$
of the nonrelativistic Coulomb
problem~\cite{Wichmann1,Hostler1,Schwinger1}
\begin{equation}
\bigg(\,-\frac{\nabla^2}{M_b} - V_c^{(0)}(\vec r)
- E\,\bigg)\,G_c(\vec r,\vec r^\prime, E)
\, = \, \delta^{(3)}(\vec r-\vec r^\prime)
\label{SchroedingerNR}
\end{equation} 
and by incorporating all the higher order terms using TIPT. 

\vspace{2mm}\noindent\hspace{2mm}
{\it  Matching calculation:}  
After the nonrelativistic correlators ${\cal{A}}_1$ and ${\cal{A}}_2$
are calculated the determination of $C_1$ is achieved by considering
the (formal) limit $\alpha_s\ll v\ll 1$. In this limit fixed
multi-loop perturbation theory (i.e.\ an expansion in $\alpha_s$) as
well as the nonrelativistic approximation (i.e.\ a subsequent expansion
in $v$) are feasible. This means that multi-loop QCD (with an
expansion in $v$ {\it after} the loop integrations have been carried out)
and multi-loop NRQCD must give the same results. In our case we use
this fact to determine the constant $C_1$ up to terms of order
$\alpha_s^2$. For that we expand the NNLO NRQCD expression for the
cross section~(\ref{crosssectionexpansion}) for small $\alpha_s$ up to 
terms of order $\alpha_s^2$ and demand equality (i.e.\ match) 
to the total cross section obtained at the two-loop level in full QCD  
keeping terms up to NNLO in an expansion in $v$. 
Because NRQCD is an effective field theory of QCD
(i.e.\ it has the same infrared behavior as full QCD) for
the limit $v\ll 1$, $C_1$ contains only constant coefficients (modulo
logarithms of the ratios $M_b/\mu_{\rm fac}$ and $M_b/\mu_{\rm
hard}$). All the 
singular terms $\propto 1/v, \ln v$ are incorporated in the correlators
${\cal{A}}_1$ and ${\cal{A}}_2$ \\

\noindent
{\bf\underline{Comment on Nonperturbative Effects}}\\[1mm]
To conclude this section we would like to mention that nowhere in this
work nonperturbative effects in terms of phenomenological constants
like the gluon condensate $\langle\,0\,|
G_{\mu\nu}\,G^{\mu\nu}|\,0\,\rangle$~\cite{Shifman1} are taken into
account. 
In~\cite{Voloshin1,Voloshin2} it has been shown that the contribution
of the most important condensate $\langle\,0\,|
G_{\mu\nu}\,G^{\mu\nu}|\,0\,\rangle$ is at the per-mill level in the
moments $P_n^{th}$ for $4\le n\le 10$. As we show in
Section~\ref{sectionproperties}, this effect is completely negligible
compared to the theoretical uncertainties coming from the large
renormalization scale dependences of the NNLO moments $P_n^{th}$.
The condensates are therefore irrelevant from the purely
practical point of view. 

Nevertheless, we even think that the inclusion of
the condensates for the moments at the NNLO level would be 
conceptually unjustified. For the gluon condensate this can be seen
from the fact that it provides a phenomenological parameterization of
the average long-wavelength vacuum fluctuations of the gluon
field involving scales smaller than the relative three momentum of the
$b\bar b$ system~\cite{Voloshin3}. Thus, for the theoretical moments
$P_n^{th}$ ($4\le n\le 10$) (and also for heavy enough quarkonia in
general)
%, where the dynamics up to NNLO is determined perturbatively
%by instantaneous potentials,
%which are governed by scales of order of the relative
%momentum of the quark pair $\sim M_Q v\sim M_Q \alpha_s$, 
the condensates describe retardation-like effects~\cite{Voloshin2}. As
explained before, we neglect retardation effects because they formally
contribute beyond the NNLO level. We conclude that taking into
account the condensates would only be sensible in a complete
NNNLO analysis. In this respect the condensate contributions
might provide some estimates for the size of some NNNLO
effects. However, if the small size of the condensate effects in the
moments $P_n^{th}$ is compared to the large perturbative
uncertainties contained of the NNLO theoretical moments, it is seems
rather doubtful whether the condensates represent the dominant
contributions at the NNNLO level. 
\par
\vspace{0.5cm}
\section{Calculation of the Moments}
\label{sectioncalculatemoments}
In this section the determination of the theoretical moments $P_n^{th}$
is presented in detail. Because all conceptual issues have been
discussed is Section~\ref{sectionintroduction} we concentrate only on
the technical aspects. The task is split into three parts which are
described in the following three subsections. In
Section~\ref{subsectioncorrelators} the nonrelativistic correlators
${\cal{A}}_1$ and ${\cal{A}}_2$ are calculated and
Section~\ref{subsectionmatching} describes
the calculation of the short-distance constant $C_1$. In
Section~\ref{subsectionintegration} the dispersion
integration~(\ref{momentcrosssectionrelation}) is carried and the
final formulae for the theoretical moments are presented.
\subsection{Calculation of the Nonrelativistic Correlators}
\label{subsectioncorrelators}
To calculate the nonrelativistic correlators ${\cal{A}}_1$ and
${\cal{A}}_2$ the Green function $G$ of the Schr\"odinger
equation~(\ref{Schroedingerfull}) has to be determined. As explained
before, we start with the Green function $G_c^{(0)}$ of the
nonrelativistic Schr\"odinger equation~(\ref{SchroedingerNR}), called
``Coulomb Green function'' from now on, and determine the effects all
the higher order contributions through TIPT. The most general form of
the Coulomb Green function reads
($r\equiv |\vec r|$, $r^\prime\equiv |\vec r^\prime|$)
\begin{eqnarray}
\lefteqn{
G_c^{(0)}(\vec r,\vec r^\prime, E) \, = \,
-\,\frac{M_b}{4\,\pi\,\Gamma(1+i\,\rho)\,\Gamma(1-i\,\rho)}\,
\int\limits_0^1 \! dt \int\limits_1^\infty \! ds \,
\Big[\, s\,(1-t) \,\Big]^{i\,\rho}\,
\Big[\, t\,(s-1) \,\Big]^{- i\,\rho}\,\times
}
\nonumber\\[2mm] & &
\times\,\frac{\partial^2}{\partial t\,\partial s}\,
\bigg[\,
\frac{t\,s}{|\,s\,\vec r - t\,\vec r^\prime\,|}\,
\exp\bigg\{\,
i\,p\,\Big(\,
|\,\vec r^\prime\,|\,(1-t) + |\,\vec r\,|\,(s-1) +
|\,s\,\vec r - t\,\vec r^\prime\,|
\,\Big)
\,\bigg\}
\,\bigg]
\,,\quad r^\prime \, < \, r
\,,
\label{CoulombGreenfunctioncomplete}
\end{eqnarray}
where
\begin{equation}
p \, \equiv \, M_b\,v \, = \, \sqrt{M_b \, (E+i\,\epsilon)}
\,,\qquad
\rho \, \equiv \, \frac{C_F\,a_s}{2\,v}
\end{equation}
and $\Gamma$ is the gamma function. 
The case $r < r^\prime$ is obtained by interchanging $r$ and
$r^\prime$. $G_c^{(0)}(\vec r,\vec r^\prime,
E)$ represents the analytical expression for the sum of ladder
diagrams depicted in Fig.~\ref{figladder}. We refer the reader
interested in the derivation of $G_c^{(0)}$ to the classical
papers~\cite{Wichmann1,Hostler1,Schwinger1}. The analytic form of the
Coulomb Green function shown in
Eq.~(\ref{CoulombGreenfunctioncomplete}) has been taken from
Ref.~\cite{Wichmann1}. Fortunately we do not
need the Coulomb Green function in its most general form but only its
S-wave component
\begin{eqnarray}
\lefteqn{
G_c^{(0),S}(r,r^\prime, E) \, = \,
\frac{1}{4\,\pi}\,\int\! d\Omega \, G_c^{(0)}(\vec r,\vec r^\prime, E)
}
\nonumber\\[2mm] & = &
-\,\frac{2\,i\,M_b\,p}{4\,\pi\,\Gamma(1+i\,\rho)\,\Gamma(1-i\,\rho)}\,
\int\limits_0^1\! dt\,
\int\limits_1^\infty\! ds\,
\Big[\, s\,(1-t) \,\Big]^{i\,\rho}\,
\Big[\, t\,(s-1) \,\Big]^{- i\,\rho}\,\times
\nonumber\\[2mm] & & \qquad\qquad
\times\,\exp\Big\{\, i\,p\,\Big[\, 
r^\prime\,(1-2\,t) + r\,(2\,s-1)
\,\Big]
\,\Big\}
\,,\qquad r^\prime \, < \, r
\,.
\label{CoulombGreenfunctionS}
\end{eqnarray}
The case $r < r^\prime$ is again obtained by interchanging $r$ and
$r^\prime$. For $r^\prime=0$ the form of the Coulomb Green function is
particularly simple 
\begin{eqnarray}
G_c^{(0)}(0,r, E) & = & G_c^{(0)}(0,\vec r, E)
 \, = \, 
-\,i\,\frac{M_b\,p}{2\,\pi}\,e^{i\,p\,r}\,
\int\limits_1^\infty\! dt \,
e^{2\,i\,p\,r\,t}\,
\bigg(\,\frac{1+t}{t}
\,\bigg)^{i\,\rho}
\nonumber\\[2mm] & = & 
-\,i\,\frac{M_b\,p}{2\,\pi}\,e^{i\,p\,r}\,
\Gamma(1-i\,\rho)\,U(1-i\,\rho,2,-2\,i\,p\,r)
\nonumber\\[2mm] & = & 
\frac{M_b}{4\,\pi\,r}\,\Gamma(1-i\,\rho)\,
   W_{i\,\rho\,,\frac{1}{2}}(-2\,i\,p\,r)
\,.
\label{CoulombGreenfunctionzero}
\end{eqnarray} 
where $U(a,b,z)$ is a confluent hypergeometric function and
$W_{\kappa,\mu}(z)$ one of the Whittaker`s
functions~\cite{Abramowitz1,Gradshteyn1}.
It is an important fact that $ G_c^{(0)}(0,\vec r, E)$ diverges for
the limit $r\to 0$ because it contains power ($\propto 1/r$) and
logarithmic ($\propto \ln r$) divergences~\cite{Hoang4}. As explained in
Section~\ref{sectionintroduction} these ultraviolet
(UV) divergences are regularized by imposing the small distance
cutoff $\mu_{\rm fac}$. The regularized form of $\lim_{r\to
0}\,G_c(0,\vec r,E)$ reads
\begin{equation}
G_c^{(0),\,reg}(0,0,E) \, = \,
\frac{M_b^2}{4\,\pi}\,
\bigg\{\,
i\, v - C_F\,a_s\,\bigg[\,
\ln(-i \frac{M_b\,v}{\mu_{\rm fac}}) + \gamma_{\mbox{\tiny E}} 
  + \Psi\bigg( 1-i\,\frac{C_F\,a_s}{2 v} \bigg)
\,\bigg]
\,\bigg\}
\,,
\label{CoulombGreenfunctionregularized}
\end{equation}
where the superscript ``reg'' indicates the cutoff regularization and 
$\Psi(z)=d \ln\Gamma(z)/dz$ is the digamma function. For the
regularization we use the convention where all power divergences
$\propto \mu_{\rm fac}$ are freely dropped and only logarithmic
divergences
$\propto \ln(\mu_{\rm fac}/M_b)$ are kept. Further, we define
$\mu_{\rm fac}$ such that in the expression between the brackets 
all constants except the Euler-Mascheroni constant
$\gamma_{\mbox{\tiny E}}$ are
absorbed. The same convention is also employed for the calculation of
the higher order corrections to the Coulomb Green function which are
discussed below. The results for any other regularization scheme which
suppressed power divergences (like the $\overline{\mbox{MS}}$ scheme)
can be obtained by redefinition of the factorization scale.
\begin{figure}[t] % figcorrelators
\begin{center}
\leavevmode
\epsfxsize=2cm
\epsffile[220 410 420 540]{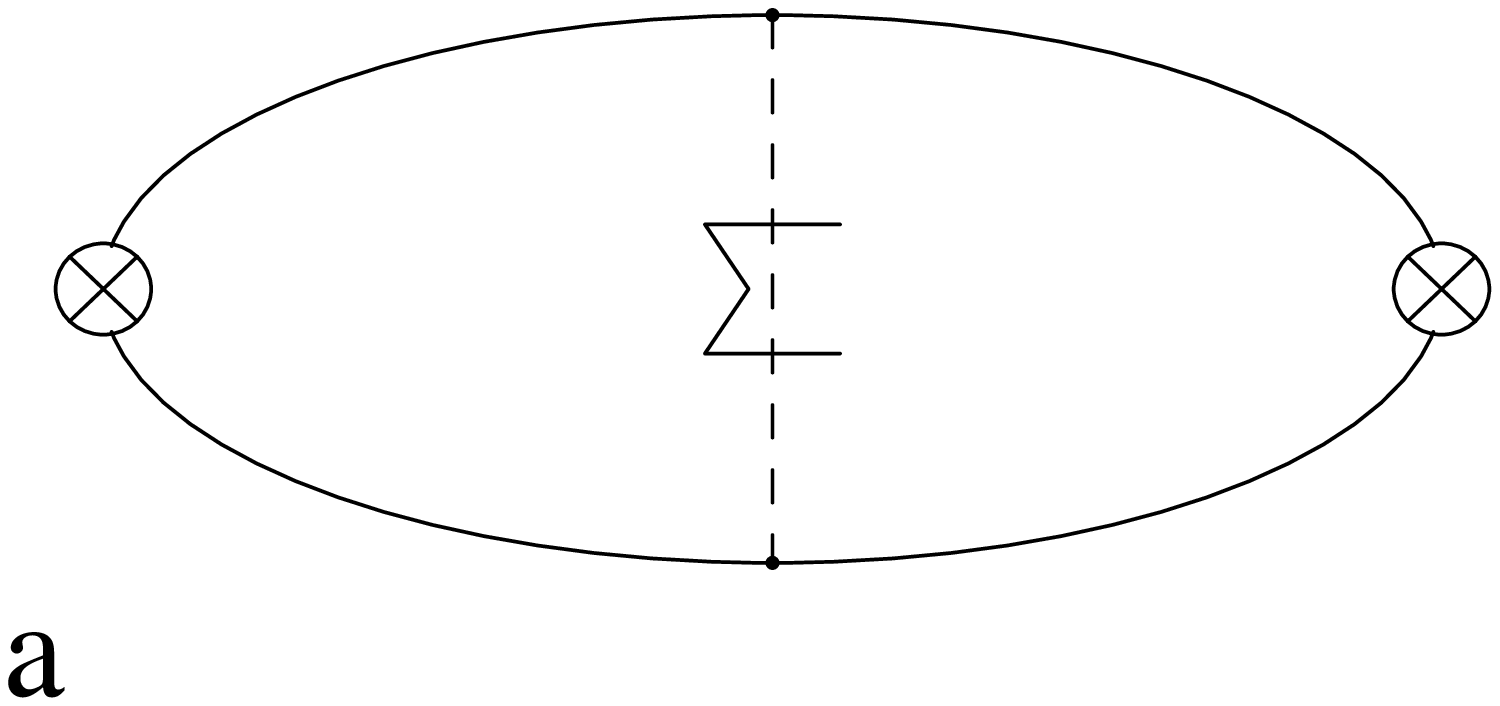}
\hspace{3cm}
\leavevmode
\epsfxsize=2cm
\epsffile[220 410 420 540]{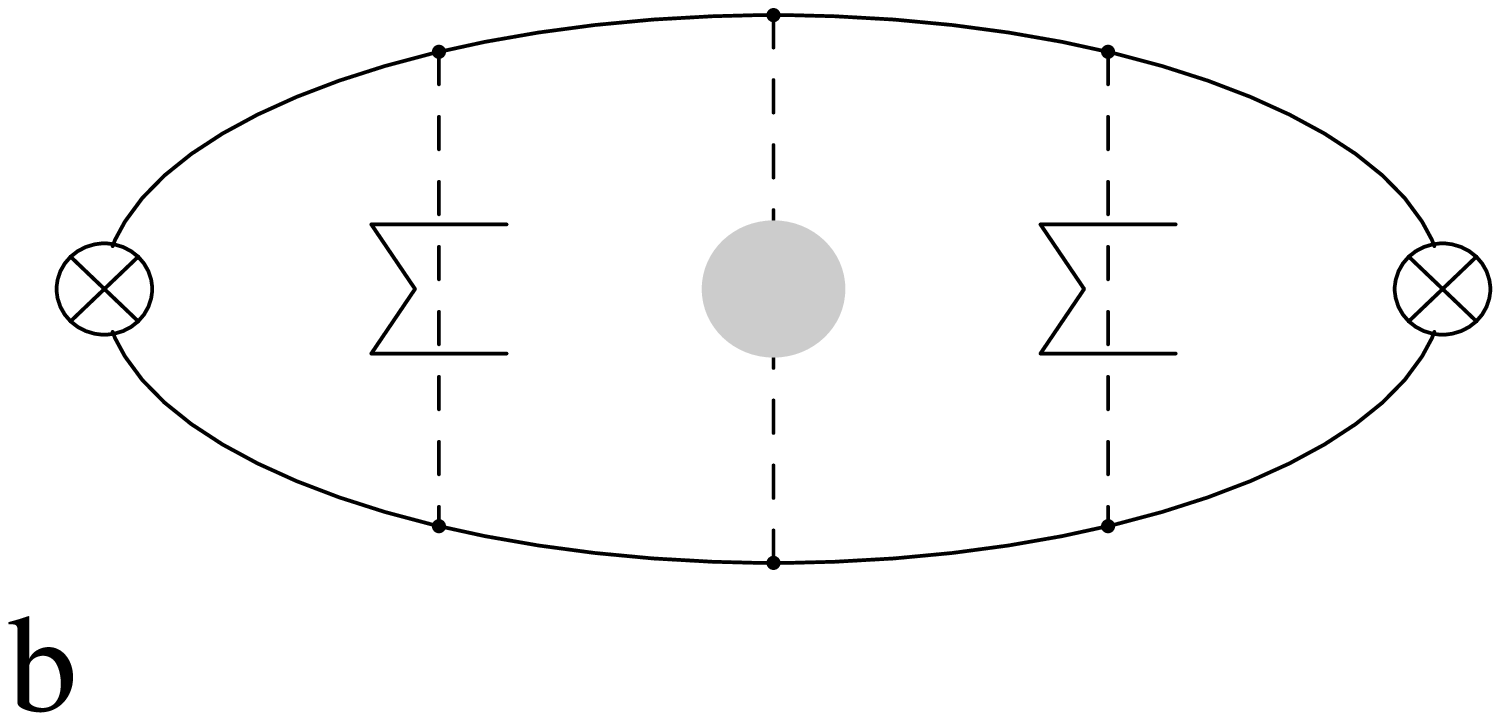}
\hspace{3cm}
\leavevmode
\epsfxsize=2cm
\epsffile[220 410 420 540]{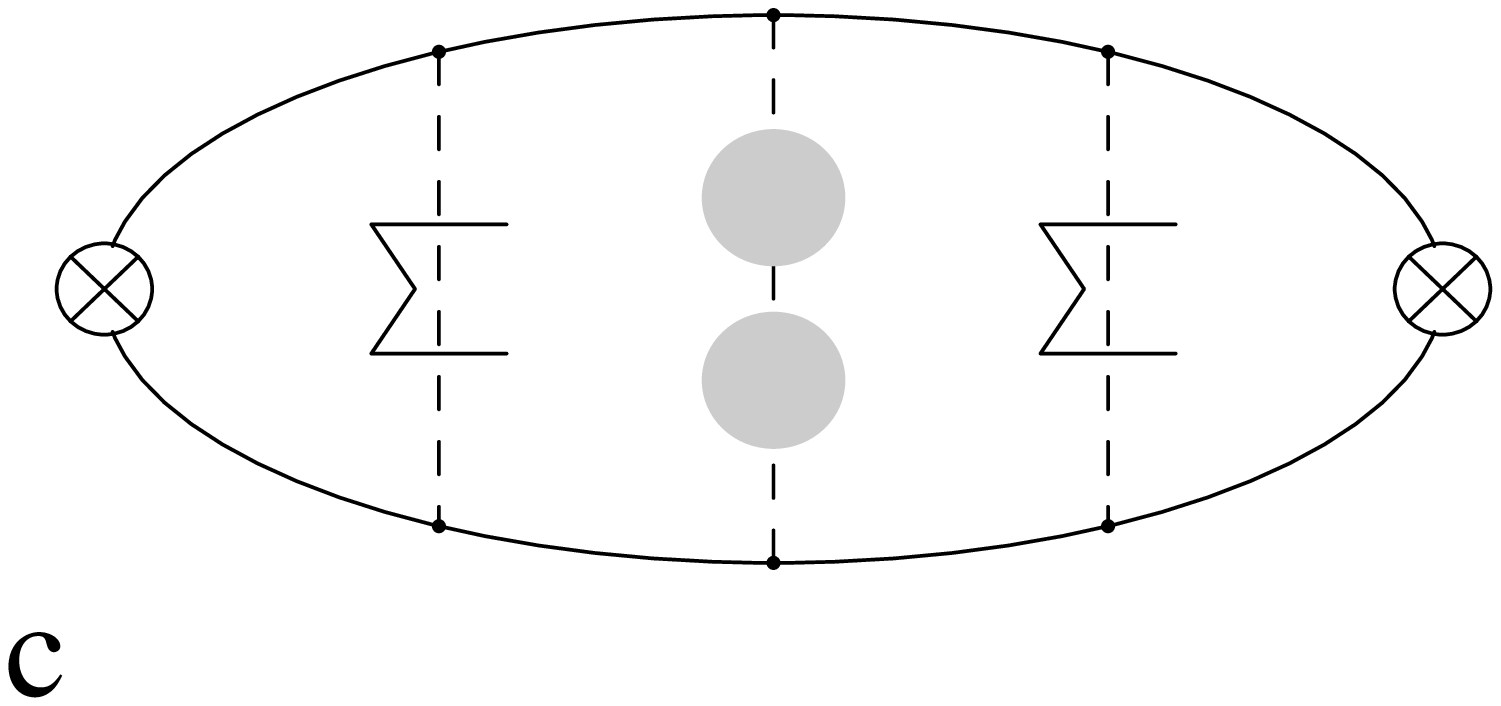}\\[1.5cm]
\leavevmode
\epsfxsize=2cm
\epsffile[220 410 420 540]{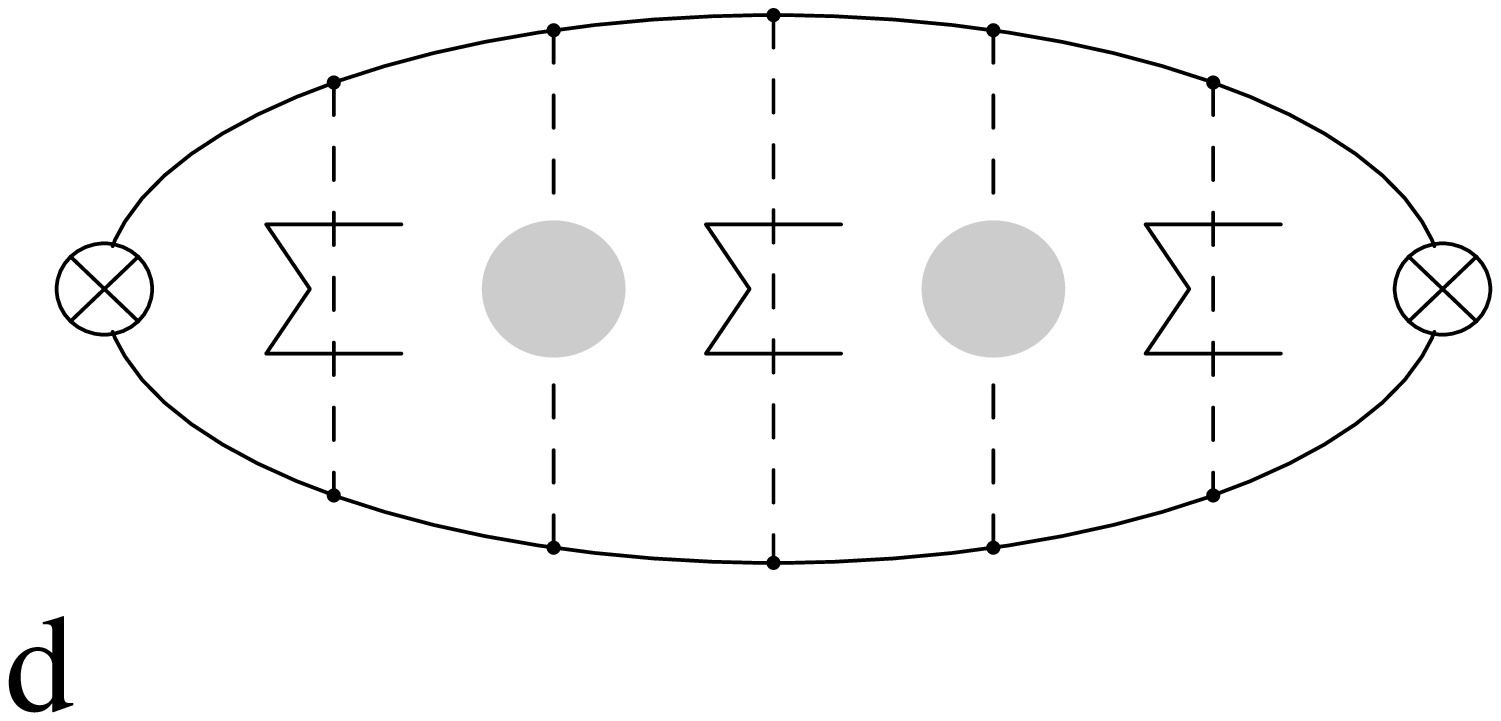}
\hspace{3cm}
\leavevmode
\epsfxsize=2cm
\epsffile[220 410 420 540]{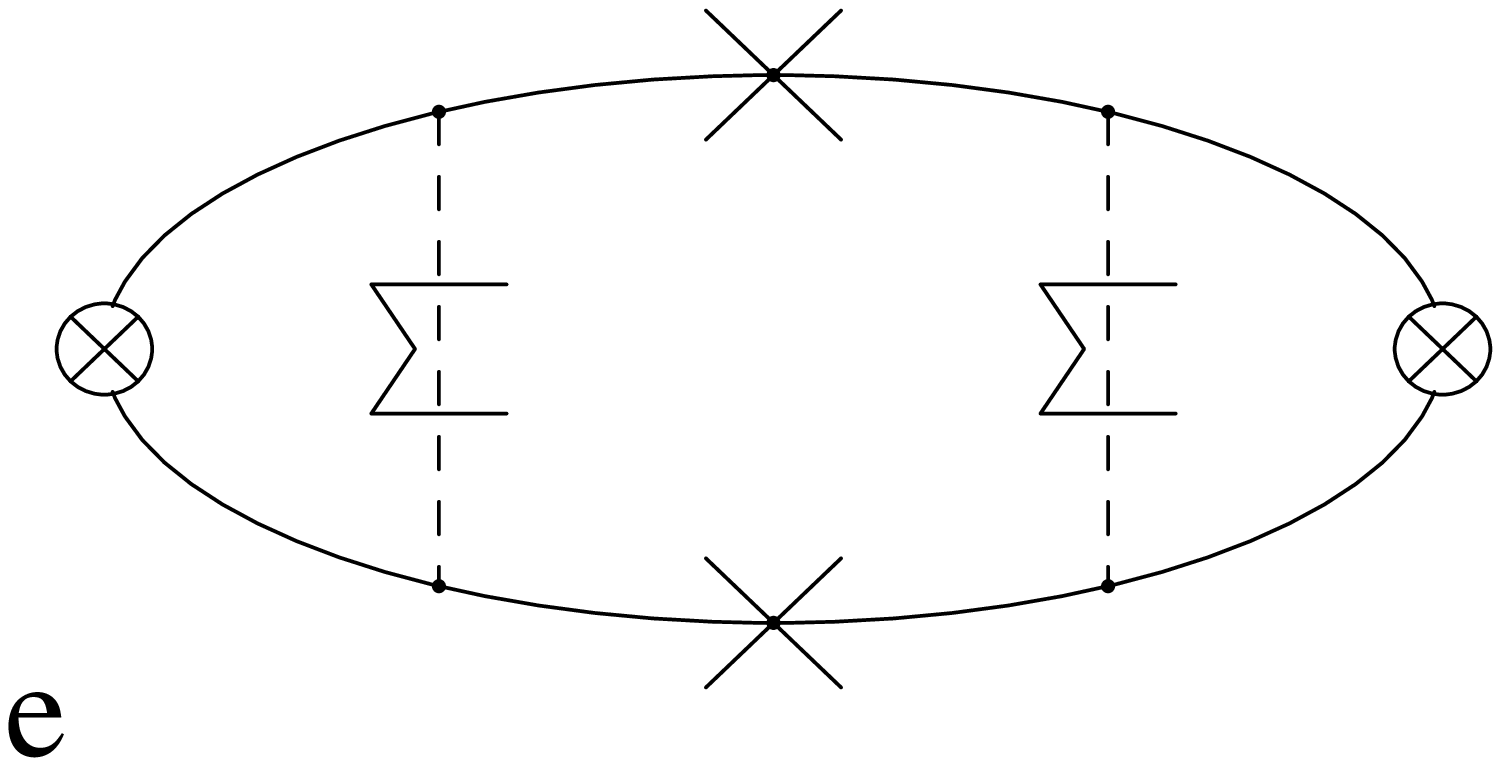}
\hspace{3cm}
\leavevmode
\epsfxsize=2cm
\epsffile[220 410 420 540]{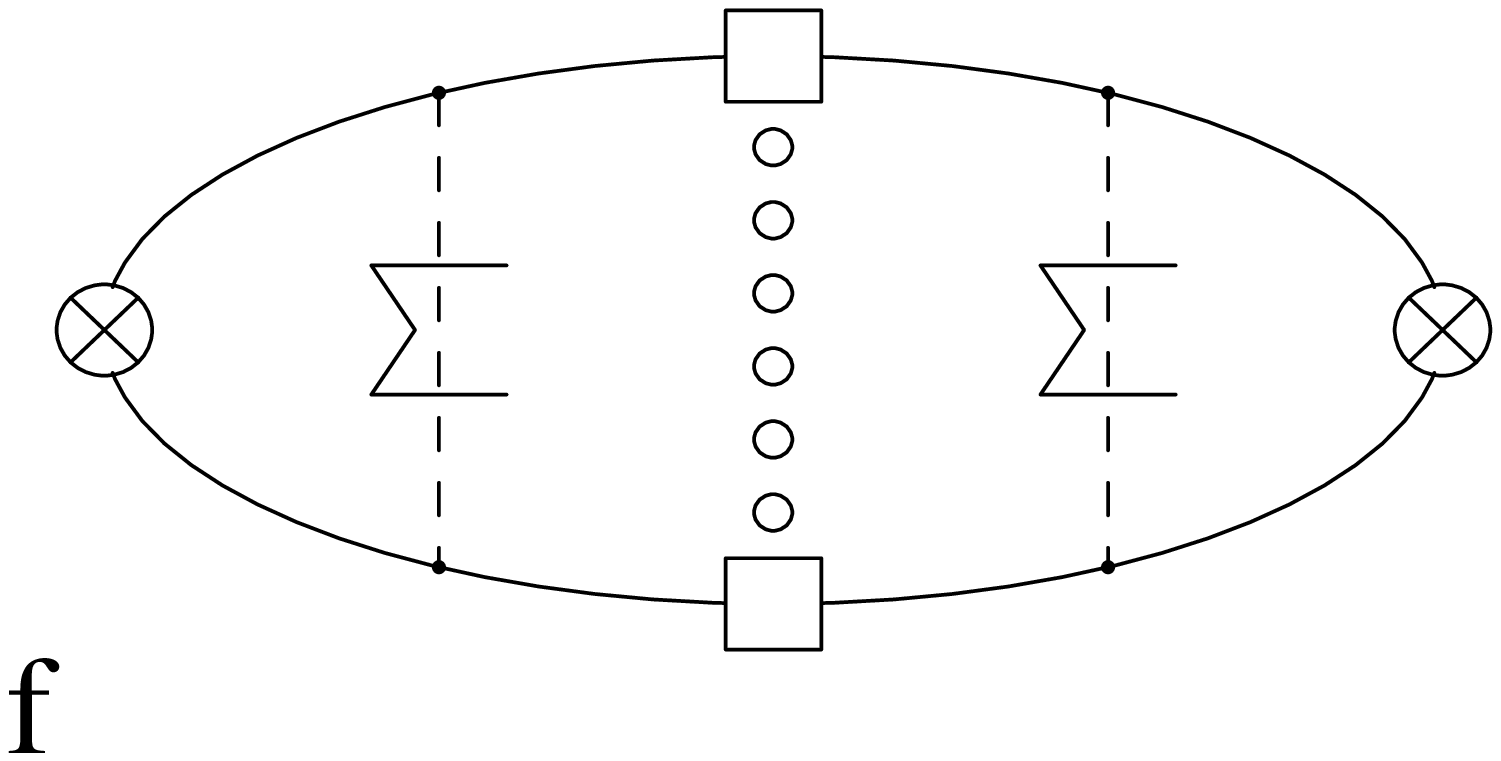}
\vskip  1.5cm
 \caption{\label{figcorrelators} 
Graphical representation of the vacuum polarization ladder diagrams
needed to determine the nonrelativistic cross section and the large
$n$ moments at NNLO.
}
%\label{figcorrelators}
 \end{center}
\end{figure}
Our apparently sloppy realization of the regularization procedure is
possible because in Section~\ref{subsectionmatching} we will match the
expression for the NNLO cross section in NRQCD directly to the
corresponding two-loop expression in full QCD. As a consequence
additional constant terms in the brackets on the
RHS of Eq.~(\ref{CoulombGreenfunctionzero}) do not affect the final
result for the cross section at NNLO in NRQCD because they merely
represent contributions which can be anyway freely shifted 
between the nonrelativistic correlators and the short-distance
coefficients. The resulting (small) ambiguity will be accounted for
during the fitting procedure by
varying the factorization scale $\mu_{\rm fac}$. However, the reader
should note that even with a more stringent regularization scheme like
the $\overline{\mbox{MS}}$ scheme this ambiguity cannot be avoided
because the factorization $\mu_{\rm fac}$ is by construction not fixed
to any value regardless which regularization scheme is used.\footnote{
In fact, using the $\overline{\mbox{MS}}$ scheme is a quite tricky
(but not impossible) task if one wants to avoid solving the
Schr\"odinger equation in $D$ dimensions. 
}
For later reference we call $G_c^{(0),\,reg}(0,0,E)$ ``zero-distance
Coulomb Green function''. A graphical representation of
$G_c^{(0),\,reg}(0,0,E)$ in terms of NRQCD Feynman diagrams is
displayed in Fig.~\ref{figcorrelators}a.
For convenience we suppress the superscript ``reg'' from now in
this work.

The Coulomb Green function contains $b\bar b$ bound state poles at the
energies $\sqrt{s}_n = 2 M_b - C_F^2 a_s^2 M_b/4 n^2$
($n=1,2,\ldots\infty$). These poles come from the digamma function in
Eq.~(\ref{CoulombGreenfunctionregularized}) and correspond to the
nonrelativistic positronium state poles known from
QED~\cite{Braun1}. They are located entirely {\it below}
the threshold point $\sqrt{s}_{\mbox{\tiny thr}} = 2 M_b$. 
This can be seen explicitly from the cross section in the
nonrelativistic limit,
\begin{eqnarray}
R_{\mbox{\tiny LO}}^{\mbox{\tiny thr}} & = & 
\frac{\pi\,Q_b^2}{M_b^2}\,
\mbox{Im}\Big[\,{\cal{A}}_1\,\Big]^{\mbox{\tiny LO}}
\, = \, \frac{6\,\pi\,N_c\,Q_b^2}{M_b^2}\,\mbox{Im}\Big[\,
G_c^{(0)}(0,0,E)
\,\Big]
\nonumber\\[2mm] & = &
\frac{24\,\pi^2\,N_c\,Q_b^2}{M_b}\,
\sum\limits_{n=1}^{\infty}|\Psi_n(0)|^2\,
       \delta(s-s_n) \, + \,
\Theta(E)\,\frac{3}{2}\,N_c\,Q_b^2\,\frac{C_F\,a_s\,\pi}
 {1-\exp(-\frac{C_F\,a_s\,\pi}{v})}
\,,
\label{Rthreshnonrelativistic}
\end{eqnarray}
where $|\Psi_n(0)|^2 = (M_b C_F a_s)^3/8 \pi n^3$ is the modulus
squared of the LO nonrelativistic bound state wave functions for the
radial quantum number $n$. The continuum contribution on the RHS of
Eq.~(\ref{Rthreshnonrelativistic}) is sometimes called ``Sommerfeld
factor'' of ``Fermi factor'' in the literature.
The resonance contributions are described
by the first term in the second line of
Eq.~(\ref{Rthreshnonrelativistic}). The corrections to the
zero-distance Coulomb Green function calculated below lead to higher
order contributions to the bound state energy levels, the residues at
the bound state poles and the continuum.  We would like to stress that
all these contributions must be included in the dispersion
integration~(\ref{momentcrosssectionrelation}) to arrive at reliable
results for the theoretical moments $P_n^{th}$.
Nevertheless, it is worth to note that the resonances are not
necessarily equivalent to the actual $\Upsilon$
resonances~\cite{Novikov1}.  In particular for large radial
excitations a direct comparison would be more than suspicious. In the
context of the calculation of the moments they have to be included for
mathematical rather than physical reasons.
(See also the discussion in Section~\ref{sectionproperties}.)
%\footnote{
%In Ref.~\cite{X} the bound state contributions have been neglected
%completely. As a consequence, the extracted values for $M_b$ and
%$\alpha_s$ in~\cite{X} contain large systematic errors in their
%central values. 
%}

Let us now come to the determination of the corrections to the
zero-distance Coulomb Green function coming from the remaining terms
in the Schr\"odinger equation~(\ref{Schroedingerfull}). At NLO only
the one-loop contributions to the Coulomb potential, $V_c^{(1)}$ (see
Eq.~(\ref{Vrunning1loop})), have to considered. Using first order
TIPT in configuration space representation the NLO corrections to 
$G_c^{(0)}(0,0,E)$ read
\begin{eqnarray}
G_c^{(1)}(0,0,E) & = &
-\,\int d^3\vec r \, 
G_c^{(0)}(0,r,E)\,V_c^{(1)}(\vec r)\,G_c^{(0)}(r,0,E)
\,.
\label{GreenfunctionNLO}
\end{eqnarray}
Expression~(\ref{GreenfunctionNLO}) is displayed graphically in
Fig.~\ref{figcorrelators}b. 
Further evaluation of the integration on the RHS of
Eq~(\ref{GreenfunctionNLO}) is possible but not presented here,
because it is already in a form suitable for the dispersion
integration~(\ref{momentcrosssectionrelation}) (see
Section~\ref{subsectionintegration}). At NNLO several contributions
have to be considered. The corrections from the two-loop contributions 
to the Coulomb potential,  $V_c^{(2)}$ (see
Eq.~(\ref{Vrunning2loop})), are calculated in analogy to the NLO
contributions using first order TIPT (Fig.~\ref{figcorrelators}c)
\begin{eqnarray}
\Big[\,G_c^{(2)}(0,0,E)\,\Big]_c^{2\,\mbox{\tiny loop}} & = &
-\,\int d^3\vec r \, 
G_c^{(0)}(0,r,E)\,V_c^{(2)}(\vec r)\,G_c^{(0)}( r,0,E)
\,.
\label{GreenfunctionNNLO2loop}
\end{eqnarray}
We also have to take into account the one-loop Coulomb potential
(Fig.~\ref{figcorrelators}(d)) in second order TIPT, 
\begin{eqnarray}
\lefteqn{
\Big[\,G_c^{(2)}(0,0,E)\,\Big]_c^{1\,\mbox{\tiny loop}} \, = \,
}
\nonumber\\[2mm] & = &
\int d^3\vec r_1 \,\int d^3\vec r_2 \,
G_c^{(0)}(0,r_1,E)\,V_c^{(1)}(\vec r_1)\,
G_c^{(0), S}(\vec r_1,\vec r_2,E)\,V_c^{(1)}(\vec r_2)\,
G_c^{(0)}(r_2,0,E)
\,.
\label{GreenfunctionNNLO1loop}
\end{eqnarray}
Because the Coulomb potential is angular independent, only the S-wave
components of the Coulomb Green function in the center of
expression~(\ref{GreenfunctionNNLO1loop}) are needed. Finally, we have
to determine the NNLO contributions to the zero-distance Green function
coming from the kinetic energy, $\delta H_{\mbox{\tiny kin}}=
-\vec\nabla^4/4 M_b^3$, the
Breit-Fermi potential $V_{\mbox{\tiny BF}}$ and the non-Abelian
potential $V_{\mbox{\tiny NA}}$ (see Figs.~\ref{figcorrelators}e and
f). These corrections are symbolized by 
$[G_c^{(2)}(0,0,E)]^{\mbox{\tiny kin+BF+NA}}$ in the following. A
method to determine them has been presented in an earlier
publication~\cite{Hoang1,Melnikov1}. Some details about this method
are presented in
Appendix~\ref{appendixcrosssection}. The final result for
$[G_c^{(2)}(0,0,E)]^{\mbox{\tiny kin+BF+NA}}$ reads
\begin{eqnarray}
\lefteqn{
G_c^{(0)}(0,0,E) + 
\Big[\,G_c^{(2)}(0,0,E)\,\Big]^{\mbox{\tiny kin+BF+NA}}
\, = \,
}
\nonumber\\[2mm] & = &
\frac{M_b^2}{4\,\pi}\,
\bigg\{
i\,v\,\Big(1+\frac{5}{8}\,v^2\Big)
 - C_F\,a_s\, \Big(1+2 \,v^2\Big) \bigg[
\ln(-i \frac{M_b\,v}{\mu_{\rm fac}}) + \gamma_{\mbox{\tiny E}}
   + \Psi\bigg( 1-i\,\frac{C_F\,a_s\,(1+\frac{11}{8} v^2)}{2\,v} \bigg)
\bigg]
\bigg\}
\nonumber\\[2mm] & & +
\frac{C_F\,a_s\,M_b^2}{12\,\pi}\,
\bigg(\, 1 + \frac{3}{2}\,\frac{C_A}{C_F}
\,\bigg)\,
\bigg\{\,
i \,v - C_F\,a_s \,\bigg[\,
\ln(-i \frac{M_b\,v}{\mu_{\rm fac}}) + \gamma_{\mbox{\tiny E}} 
  + \Psi\bigg( 1-i\,\frac{C_F\,a_s}{2\, v} \bigg)
\,\bigg]
\,\bigg\}^2
\,.
\label{GreenfunctionNNLOBF}
\end{eqnarray}
Because $[G_c^{(2)}(0,0,E)]^{\mbox{\tiny kin+BF+NA}}$ contains also
kinematic corrections to the zero-distance Coulomb Green function,
we found it convenient to add the zero-distance Coulomb Green
function~(\ref{CoulombGreenfunctionregularized}). The first term on
the RHS of Eq.~(\ref{GreenfunctionNNLOBF}) represents the
zero-distance Coulomb Green function including the NNLO kinematic
corrections and the second term
the remaining corrections. It is an interesting fact that these
remaining corrections can be written as the squared of the
zero-distance Coulomb Green function. This is a consequence of the
(renormalization group) invariance of the total cross
section~(\ref{crosssectionexpansion}) under variations of the
factorization scale $\mu_{\rm soft}$. (See also the comment after
Eq.~(\ref{rho1NNLOBF}).) 
Collecting all contributions the complete expression for the
nonrelativistic correlator ${\cal{A}}_1$ at NNLO reads
\begin{eqnarray}
{\cal{A}}_1 & = &
6\,N_c\,\bigg\{\,
G_c^{(0)}(0,0,E) + G_c^{(1)}(0,0,E)
\nonumber\\[2mm] & & \qquad
+\Big[\,G_c^{(2)}(0,0,E)\,\Big]_c^{1\,\mbox{\tiny loop}}
+\Big[\,G_c^{(2)}(0,0,E)\,\Big]_c^{2\,\mbox{\tiny loop}}
+\Big[\,G_c^{(2)}(0,0,E)\,\Big]^{\mbox{\tiny kin+BF+NA}}
\,\bigg\}
\,.
\label{A1final}
\end{eqnarray} 
The calculation of the correlator ${\cal{A}}_2$, on the other hand is
trivial using the equation of motion for the Green function, see
Eq.~(\ref{A2toA1}). Because ${\cal{A}}_2$ is multiplied by an explicit
factor $v^2$, Eq.~(\ref{A2toA1}), its form is particularly simple,
\begin{equation}
{\cal{A}}_2 \, = \, 
v^2\,\frac{3 \,M_b^4}{2\,\pi}\,
\bigg\{\,
i\, v - C_F\,a_s\,\bigg[\,
\ln\Big(-i \frac{M_b\,v}{\mu_{\rm fac}}\Big) + \gamma_{\mbox{\tiny E}} 
  + \Psi\Big( 1-i\,\frac{C_F\,a_s}{2\,v} \Big)
\,\bigg]
\,\bigg\}
\,.
\label{A2final}
\end{equation}
\subsection{Determination of the Short-distance Coefficients}
\label{subsectionmatching}
\begin{figure}[t] % figQCDdiagrams
\begin{center}
\leavevmode
\epsfxsize=2cm
\epsffile[220 410 420 540]{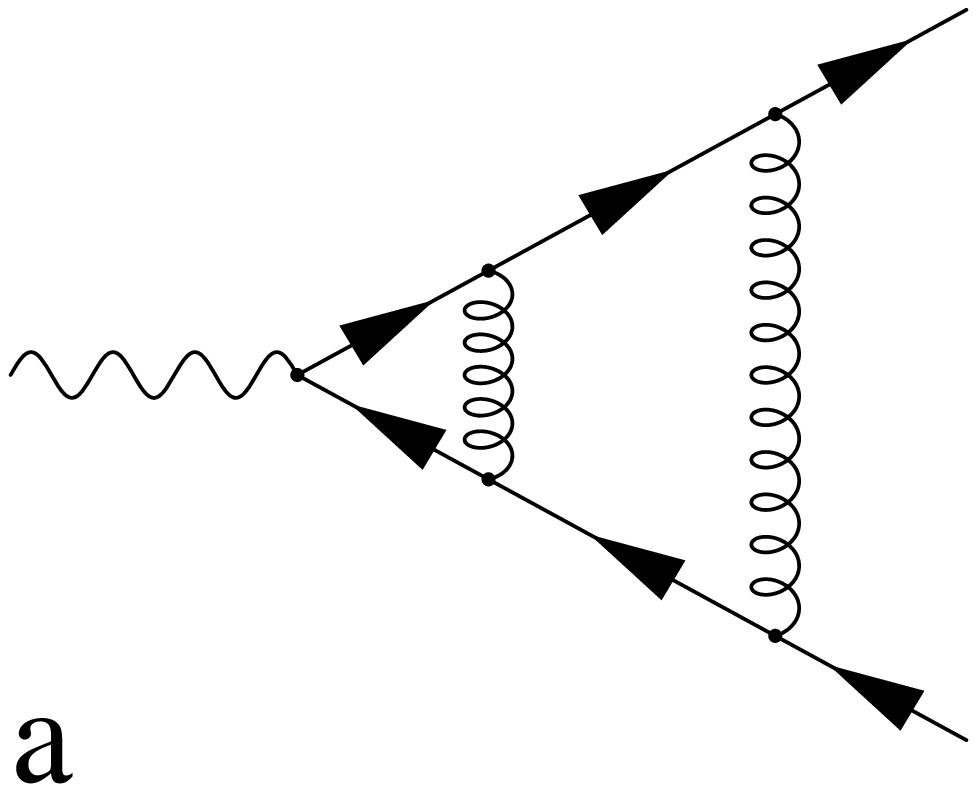}
\hspace{2cm}
\leavevmode
\epsfxsize=2cm
\epsffile[220 410 420 540]{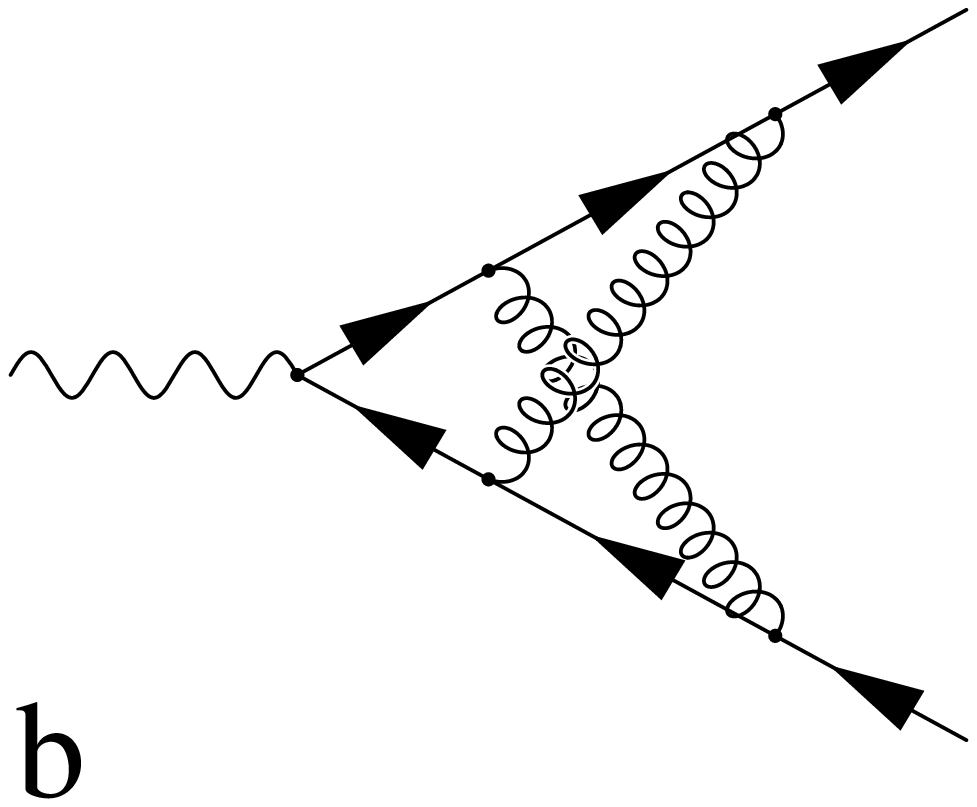}
\hspace{2cm}
\leavevmode
\epsfxsize=2cm
\epsffile[220 410 420 540]{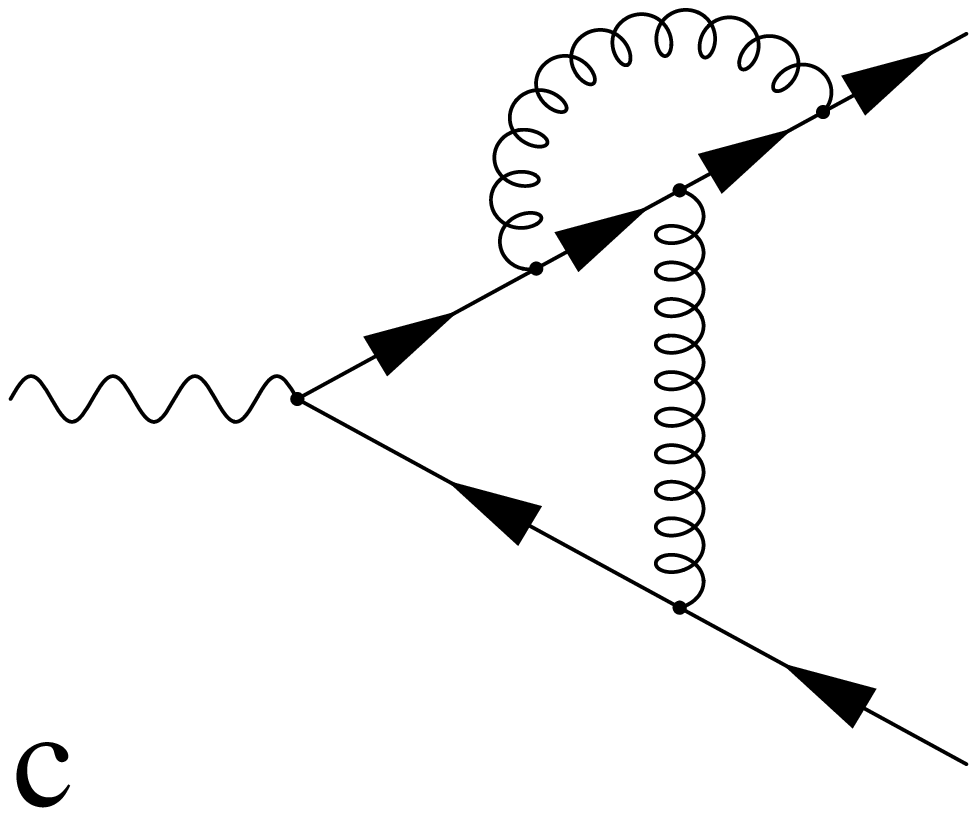}
\hspace{2cm}
\leavevmode
\epsfxsize=2cm
\epsffile[220 410 420 540]{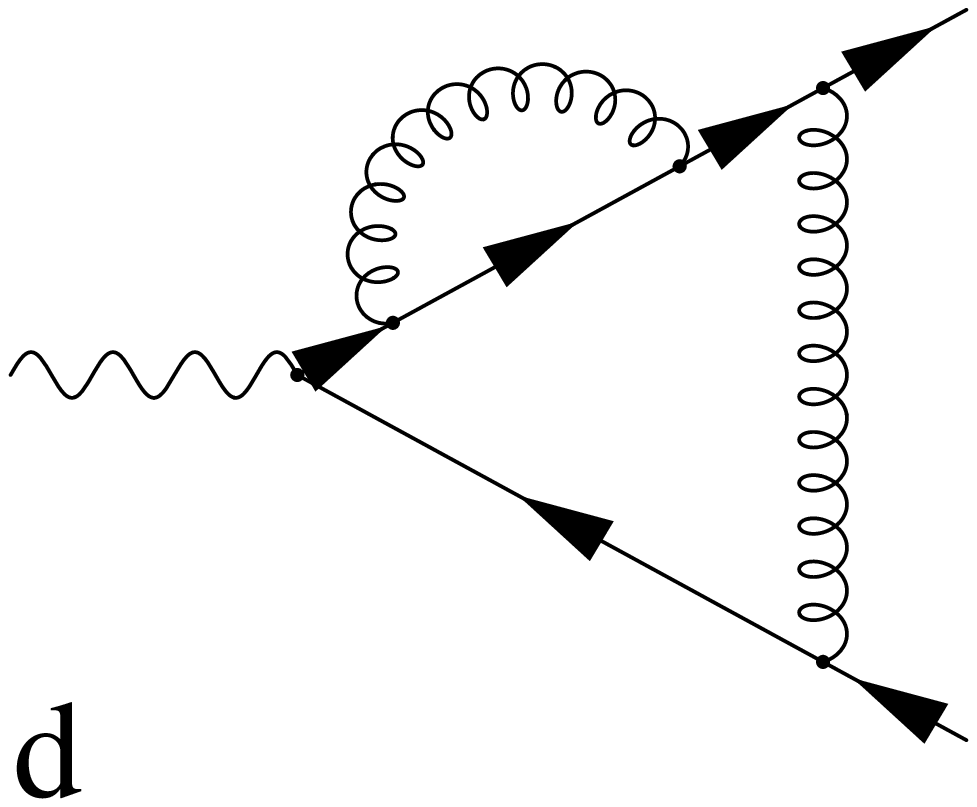}\\[1.5cm]
\leavevmode
\epsfxsize=2cm
\epsffile[220 410 420 540]{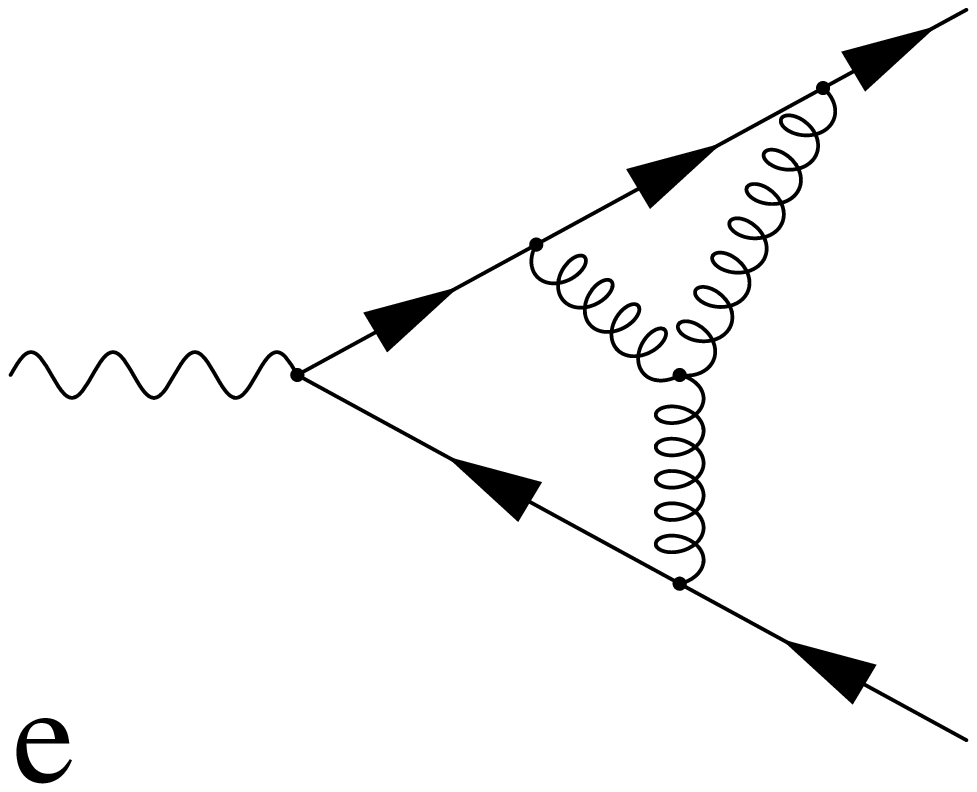}
\hspace{2cm}
\leavevmode
\epsfxsize=2cm
\epsffile[220 410 420 540]{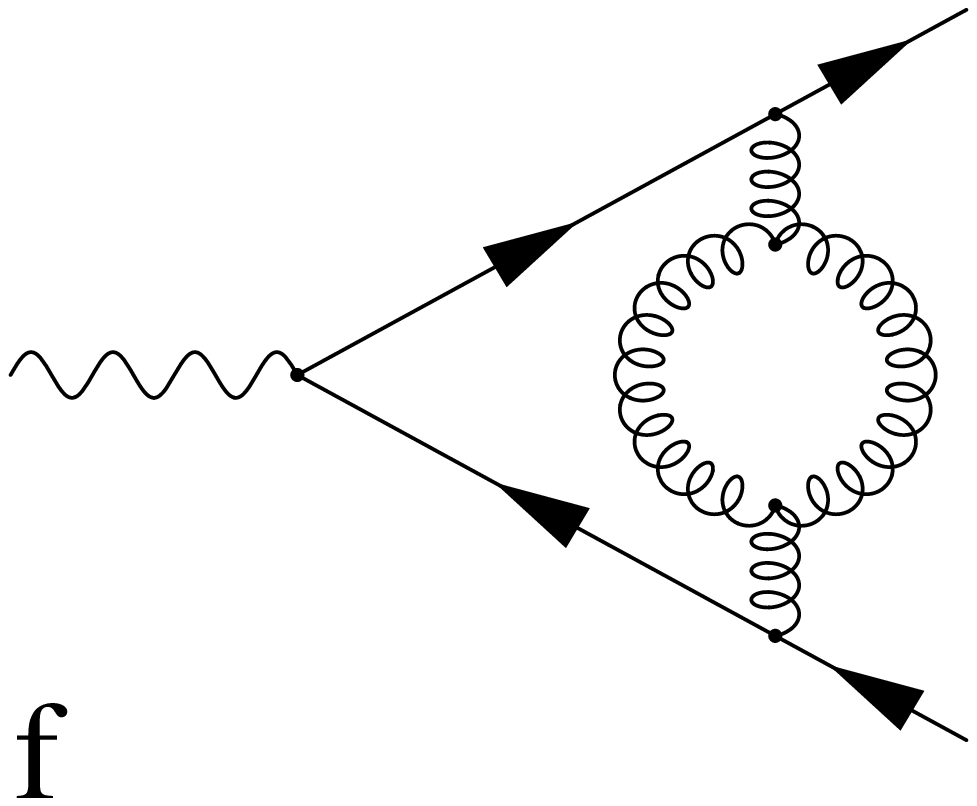}
\hspace{2cm}
\leavevmode
\epsfxsize=2cm
\epsffile[220 410 420 540]{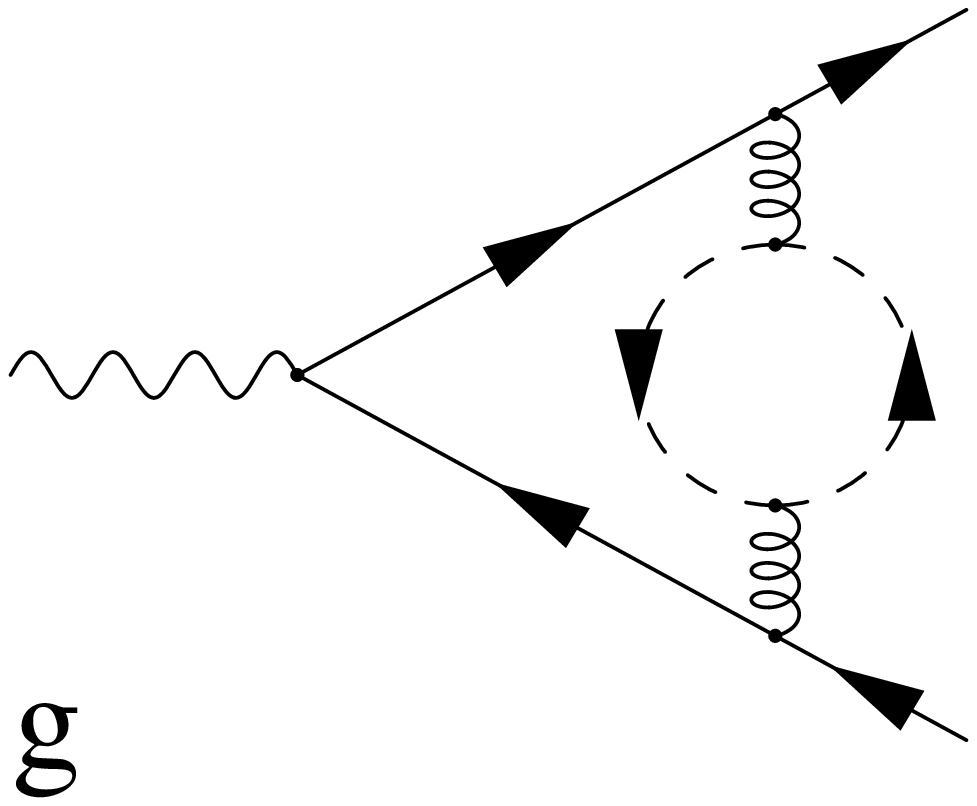}
\hspace{2cm}
\leavevmode
\epsfxsize=2cm
\epsffile[220 410 420 540]{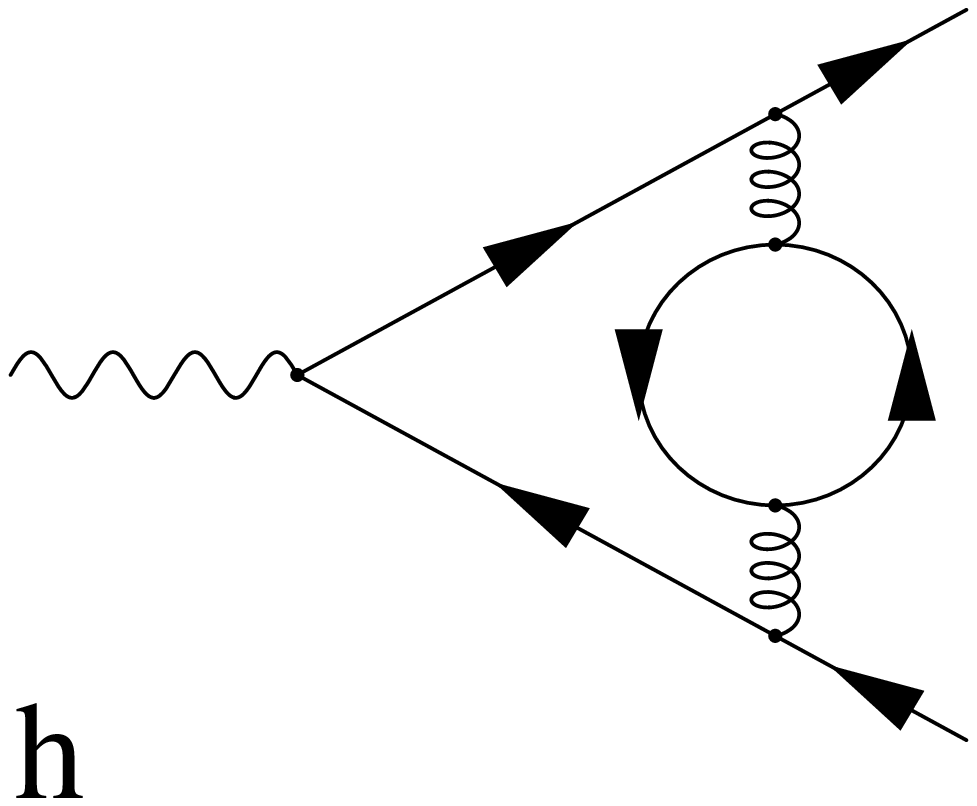}
\vskip  1.5cm
 \caption{\label{figQCDdiagrams} 
QCD Feynman diagrams relevant for the calculation of the cross section
at the two-loop level. The calculation of these diagrams is needed for
the matching calculation which leads to the determination of the
short-distance coefficient $C_1$. Feynman diagrams needed for the wave
function renormalization are not displayed.
}
%\label{figQCDdiagrams}
 \end{center}
\end{figure}

The short-distance coefficient $C_1$ and $C_2$ are determined by
matching the NNLO cross 
section~(\ref{crosssectionexpansion}) in NRQCD to the same cross
section calculated in full QCD (in the limit $\alpha_s\ll
v\ll 1$) at the two-loop level and including
terms in the velocity expansion up to NNLO. 
It is convenient to parameterize the higher order
contributions to $C_1$ in the form ($a_h\equiv\alpha_s(\mu_{\rm hard})$)
\begin{equation}
C_1(M_b,\mu_{\rm hard},\mu_{\rm fac}) \, = \,
1 + \Big(\frac{a_h}{\pi}\Big)\,c_1^{(1)}
+ \Big(\frac{a_h}{\pi}\Big)^2\,c_1^{(2)}(\mu_{\rm hard},\mu_{\rm fac})
+ \ldots
\,.
\label{C1series}
\end{equation}
Due to renormalization group invariance only the
${\cal{O}}(\alpha_s^2)$  coefficient of $C_1$ depends on the hard
scale $\mu_{\rm hard}$. We have already anticipated that the
${\cal{O}}(\alpha_s)$ coefficient does not depend on the factorization
scale $\mu_{\rm fac}$. For $C_2$, on the other hand, no higher order
contributions are needed because the correlator ${\cal{A}}_2$ is
already of NNLO, 
\begin{equation}
C_2 \, = \, 1
\,.
\label{C2series}
\end{equation}
The expansion of the NNLO cross section in NRQCD, $R_{\mbox{\tiny
NNLO}}^{\mbox{\tiny thr}}$, Eq.~(\ref{crosssectionexpansion}),
keeping terms up to order $\alpha_s^2$, reads
\begin{eqnarray}
\lefteqn{
R_{\mbox{\tiny NNLO}}^{\mbox{\tiny thr}} 
\, \stackrel{\alpha_s\ll 1}{=} \,
N_c\,Q_b^2\,\bigg\{\,\bigg[\,
\frac{3}{2}\,v-\frac{17}{16}\,v^3 
\,\bigg] +
\frac{C_F\,a_h}{\pi}\,\bigg[\,
\frac{3\,\pi^2}{4}-6\,v+\frac{\pi^2}{2}\,v^2 
\,\bigg]
}
\nonumber\\[2mm] & & 
+\, a_h^2\,\bigg[\,
\frac{C_F^2\,\pi^2}{8\, v} 
+ \frac{3}{2}C_F\,\bigg(\,
- 2 \,C_F 
+ C_A \Big(\,
    -\frac{11}{24}\ln\frac{4 \,v^2 M_t^2}{\mu_{\rm hard}^2}+\frac{31}{72}  
\,\Big)
+ T\,n_l \,\Big(\, 
    \frac{1}{6}\ln\frac{4 \,v^2 M_t^2}{\mu_{\rm hard}^2}-\frac{5}{18}  
\,\Big)
\,\bigg) 
\nonumber\\[2mm] & & \qquad
+ \bigg(\, 
\frac{49\,C_F^2\,\pi^2}{192}  
  + \frac{3}{2}\,\frac{c_1^{(2)}}{\pi^2}
  - C_F\Big( C_F+\frac{3}{2}\, C_A \Big)\,\ln\frac{M_b\,v}{\mu_{\rm fac}}
\,\bigg)\,v
\,\bigg]
 + {\cal{O}}(\alpha_s^3)
\,\bigg\}
\,.
\label{Rthreshexpansion}
\end{eqnarray}
where we have set $\mu_{\rm soft}=\mu_{\rm hard}$ because in the limit
$\alpha_s\ll v\ll 1$ a distinction between soft and hard scale is
irrelevant. (We want to emphasize that the choice $\mu_{\rm
soft}=\mu_{\rm hard}$ implicitly means that the hard scale is, like
the soft scale, defined in the $\overline{\mbox{MS}}$ scheme.) 
The corresponding expression for the two-loop cross
section calculated in full QCD reads\footnote{
The two-loop contributions from secondary radiation
of a $b\bar b$ pair off a light quark-antiquark pair are kinematically
suppressed and do not contribute at NNLO in the velocity expansion.
}
\begin{eqnarray}
\lefteqn{
R_{\mbox{\tiny 2loop QCD}}^{\mbox{\tiny NNLO}} 
\, \stackrel{v\ll 1}{=} \,
N_c\,Q_b^2\,\bigg\{\,\bigg[\,
\frac{3}{2}\,v-\frac{17}{16}\,v^3 + {\cal{O}}(v^4)
\,\bigg] +
\frac{C_F\,a_h}{\pi}\,\bigg[\,
\frac{3\,\pi^2}{4}-6\,v+\frac{\pi^2}{2}\,v^2 + {\cal{O}}(v^3)
\,\bigg]
}
\label{RthreshfullQCD}
\\[2mm] & & 
+\, a_h^2\,\bigg[\,
\frac{C_F^2\,\pi^2}{8\, v} 
+ \frac{3}{2}C_F\,\bigg(\,
- 2 \,C_F 
+ C_A \Big(\,
    -\frac{11}{24}\ln\frac{4\, v^2 M_t^2}{\mu_{\rm hard}^2}+\frac{31}{72}  
\,\Big)
+ T\,n_l \,\Big(\, 
    \frac{1}{6}\ln\frac{4\, v^2 M_t^2}{\mu_{\rm hard}^2}-\frac{5}{18}  
\,\Big)
\,\bigg) 
\nonumber\\[2mm] & & \qquad
+ \bigg(\, 
\frac{49\,C_F^2\,\pi^2}{192}  
  + \frac{3}{2}\,\kappa 
  + \frac{C_F}{\pi^2}\,
       \Big( \frac{11}{2}\, C_A - 2\, T\,n_l \Big)\,
      \ln\frac{M_t^2}{\mu_{\rm hard}^2} 
  - C_F\Big( C_F+\frac{3}{2}\, C_A \Big)\,\ln v
\,\bigg)\,v
\,\bigg] + {\cal{O}}(v^2)
\,\bigg\}
\,,
\nonumber
\end{eqnarray}
where 
\begin{eqnarray}
\kappa & = &
C_F^2\,\bigg[\, \frac{1}{\pi^2}\,\bigg(\,
\frac{39}{4}-\zeta_3 \,\bigg) +
\frac{4}{3} \ln 2 - \frac{35}{18}
\,\bigg] 
\nonumber\\[2mm] & & -
C_A\,C_F\,\bigg[\,  \frac{1}{\pi^2} \,\bigg(
\frac{151}{36} + \frac{13}{2} \zeta_3 \,\bigg) +
\frac{8}{3} \ln 2 - \frac{179}{72} \,\bigg] 
\nonumber\\[2mm] & & +
C_F\,T\,\bigg[\,
\frac{4}{9}\,\bigg(\, \frac{11}{\pi^2} - 1\,\bigg)
\,\bigg] +
C_F\,T\,n_l\,\bigg[\, \frac{11}{9\,\pi^2} \,\bigg] 
\,.
\label{kappadef}
\end{eqnarray}
The Born and one-loop contributions in Eq.~(\ref{RthreshfullQCD}) are
standard~\cite{Kallensabry1,Schwinger2}. The two-loop contributions
are presented with
the various combinations of the SU(3) group theoretical factors
$C_F=4/3$, $C_A=3$ and $T=1/2$. The terms proportional to $C_F^2$ come
from the QED-like, Abelian exchange of two gluons and have been
calculated analytically in~\cite{Hoang6}. The result has been confirmed
numerically in~\cite{Adkins1} and analytically
in~\cite{Czarnecki1,Beneke1}. The corresponding
Feynman diagrams (in the covariant gauge) are displayed in
Fig.~\ref{figQCDdiagrams}(a), (b),
(c) and (d). 
The $C_A C_F$ terms correspond to the non-Abelian exchange of two
gluons, i.e.\ involving the triple gluon vertex, ghost fields and
topologies with crossed gluon lines (Figs.~\ref{figQCDdiagrams}(b),
(c), (d), (e), (f), (g)). These contributions have been determined
in~\cite{Czarnecki1,Beneke1}. The $C_F T n_l$ contributions are from
diagrams with a vacuum 
polarization of massless quarks (Fig.~\ref{figQCDdiagrams}(h)) and
have been calculated in~\cite{Hoang7}.
The contributions proportional to $C_F T$, finally, correspond to the
diagram where the vacuum polarization is from the bottom quarks
(Fig.~\ref{figQCDdiagrams}(g)) and have been calculated
in~\cite{Karshenboim1,Hoang7}. The virtual top quark contributions are
suppressed by a factor $(M_b/M_t)^2\sim 0.001$) and are neglected. 

The constants $c_1^{(1)}$ and $c_1^{(2)}$ defined in
Eq.~(\ref{C1series}) can now be easily determined by demanding
equality of expressions~(\ref{Rthreshexpansion}) and
(\ref{RthreshfullQCD}). This constitutes the ``direct matching''
procedure~\cite{Hoang1,Hoang4} and leads to
\begin{eqnarray}
c_1^{(1)} & = & -\,4\,C_F
\,,
\label{c11short}
\\[2mm]
c_1^{(2)} & = & 
\pi^2\,\bigg[\,
\kappa 
+ \frac{C_F}{\pi^2}\,\bigg(\,
\frac{11}{3}\,C_A - \frac{4}{3}\,T\,n_l
\,\bigg)\,\ln\frac{M_b^2}{\mu_{\rm hard}^2}
+ C_F\,\bigg(\,
\frac{1}{3}\,C_F + \frac{1}{2}\,C_A
\,\bigg)\,
\ln \frac{M_b^2}{\mu_{\rm fac}^2}
\,\bigg]
\,.
\label{c12short}
\end{eqnarray}
The constant $c_1^{(1)}$ is the ${\cal{O}}(\alpha_s)$ short-distance
contributions which is well known from the single photon annihilation
contributions to the positronium hyperfine splitting~\cite{Karplus1}
and from corrections to electromagnetic quarkonium
decays~\cite{Barbieri1}. We want to mention again that $\mu_{\rm
hard}$ and $\mu_{\rm fac}$ are independent and defined in different
regularization schemes. 

To conclude this subsection we would like to point out that the
short-distance coefficients $C_1$ and $C_2$ determined above are
not sufficient to determine the vacuum polarization function
(Eq.~(\ref{vacpoldef})) in the threshold regime at NNLO, because
they have been determined via matching at the level of the cross
section only, i.e.\ at the level of the imaginary part of the
vacuum polarization function.
The expressions for the correlators still contain overall UV
divergences $\propto \ln(M_b/\mu_{\rm fac})$ in their real
parts~\cite{Voloshin3,Braun1}, see
e.g.\ Eq.~(\ref{CoulombGreenfunctionregularized}). For the large $n$
moments calculated in this work these ambiguities are irrelevant
because the
divergent contributions in the real parts do not contribute to the
large $n$ moments.
The relation between the nonrelativistic correlators and the vacuum
polarization function at NNLO in the threshold regime, including the
proper short-distance contributions for the real part, has the form
\begin{eqnarray}
\lefteqn{
\frac{1}{3\,q^2}\,\Pi_\mu^{\,\,\mu}(q) 
\, \stackrel{q^2\to 4M_b^2}{\longrightarrow} \,
}
\nonumber\\[1mm] & &
\frac{1}{12\,M_b^2}\,C_1(\mu_{\rm hard},\mu_{\rm fac})\,
{\cal{A}}_1(E,\mu_{\rm soft},\mu_{\rm fac})
 - \,\frac{1}{9 M_b^4}\,
C_2(\mu_{\rm hard},\mu_{\rm fac})\,
{\cal{A}}_2(E,\mu_{\rm soft},\mu_{\rm fac})
+ \ldots
\nonumber\\[1mm] & & +
h_1 +
\frac{C_F\,a_h}{4\,\pi}\,\bigg[\,\frac{1}{2}\,\ln\Big(\frac{M_b}{\mu_{\rm fac}}\Big)
+ h_2  \,\bigg] + \ldots
\,.
\label{vacpolthresh}
\end{eqnarray}
The constants $h_1$ and $h_2$ can be determined via (direct) matching
to the one and two-loop vacuum polarization function in full
QCD at threshold, i.e.\ for $q^2\to 4M_b^2$. This work has
been carried out in a previous publication~\cite{Hoang4} and leads to 
$h_1 = \frac{2}{9\pi^2}$ and
$h_2 = \frac{1}{4\pi^2}(3-\frac{21}{2}\zeta_3)+\frac{11}{32}-
\frac{3}{4}\ln 2$.
For the complete expression of the vacuum polarization function in the
threshold regime at NNLO in the nonrelativistic expansion also the
${\cal{O}}(\alpha_s^2)$ and ${\cal{O}}(\alpha_s^3)$ short-distance
contributions would have to be calculated. This would require the
calculation of the three- and four loop the vacuum polarization
functions in full QCD in the threshold regime. This task has not been
accomplished yet and remains to be done.\footnote{
In Refs.~\cite{Steinhauser1} numerical approximations for the three
loop vacuum
polarization valid for all energies has been obtained based on the
Pad\'e method. Unfortunately numerical approximations are of little
use for a precise extraction the ${\cal{O}}(\alpha_s^2)$
short-distance constants due to the presence of singular terms
$\propto \ln v$ and $\ln^2 v$ in the real part of the three loop
vacuum polarization function close to the threshold.
}
\subsection{The Dispersion Integration}
\label{subsectionintegration}
After the nonrelativistic correlators ${\cal{A}}_1$
and ${\cal{A}}_2$ and the short-distance constants $C_1$ and $C_2$
are calculated we are now ready to carry out the dispersion
integration~(\ref{momentcrosssectionrelation}). This task is
quite cumbersome if the complete covariant form of the integration
measure $d s/s^{n+1}$ is used. Fortunately the integration can be
simplified because we are only interested in NNLO accuracy in the
nonrelativistic expansion in $v=(E/M_b)^{1/2}$. Changing the
integration variable to the energy $E=\sqrt{s}-2 M_b$ and expanding up
to NNLO in $v$, where combination $(E/M_b) n$ is considered of order
one, the resulting integration measure reads
\begin{eqnarray}
\frac{d s}{s^{n+1}} & = & 
\frac{1}{(4\,M_b^2)^n}\, \frac{d E}{M_b}\,\exp\bigg\{\,
-(2\,n+1)\,\ln\Big(1+\frac{E}{2\,M_b}\Big)
\,\bigg\}
\nonumber\\[2mm] 
& \stackrel{E \ll M_b}{\longrightarrow} &
\frac{1}{(4\,M_b^2)^n}\, \frac{d E}{M_b}\,\exp\bigg\{\,
-\frac{E}{M_b}\,n
\,\bigg\}\,\bigg(\,
1 - \frac{E}{2\,M_b} + \frac{E^2}{4\,M_b^2}\,n
+ {\cal{O}}\bigg(\frac{E^2}{M_b^2},\frac{E^3}{M_b^3}\,n,
     \frac{E^4}{M_b^4}\,n^2\bigg)
\,\Bigg)
\,.
\nonumber\\&&
\label{integrationmeasure}
\end{eqnarray}
The dispersion integration for the theoretical moments $P_n^{th}$ at
NNLO then takes the form
\begin{equation}
P_n^{th} \, = \,
\frac{1}{(4\,M_b^2)^n}\,\int\limits_{E_{\rm bind}}^\infty 
\frac{d E}{M_b} \,\exp\bigg\{\,
-\frac{E}{M_b}\,n
\,\bigg\}\,\bigg(\,
1 - \frac{E}{2\,M_b} + \frac{E^2}{4\,M_b^2}\,n
\,\bigg)\,R_{\mbox{\tiny NNLO}}^{\mbox{\tiny thr}}(E)
\,,
\label{Pnexpression1}
\end{equation}
where $E_{\rm bind}$ is the (negative) binding energy of the lowest
lying resonance. We would like to point out that 
expansion~(\ref{integrationmeasure}) leads to an asymptotic series,
which means that including more an more terms in the expansion can
improve the approximation only up to a certain point beyond which the
series starts diverging. We have checked that for all values of $n$
employed in this work the expansion is still well inside the
converging regime. It should also be noted that for increasing values
of $n$ the expansion provides better and better approximations only as
long as the condition $(E_{\rm bind}/M_b) n < 1$ is satisfied. In our
case, where the $b\bar b$ system is treated as Coulombic, i.e.\
$E_{\rm bind}= M_b C_F^2 \alpha_s^2/4 +\ldots$. this condition is
always satisfied. (See also discussion at the end of
Section~\ref{sectionproperties}.)
\begin{figure}[t] % figcomplexintegration
\begin{center}
\leavevmode
\epsfxsize=5.5cm
\epsffile[220 420 420 550]{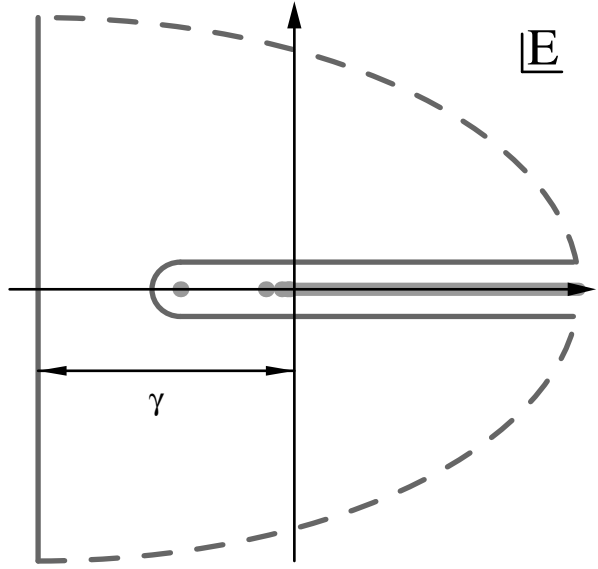}\\
\vskip  5cm
 \caption{\label{figcomplexintegration} 
Path of integration to calculate expression~(\ref{Pnexpression1}) for
the theoretical moments $P_n^{th}$. The dashed line closes the contour
at infinity and does not contribute to the integration. The constant
$\gamma$ is chosen large enough to be safely away from the bound state
poles which are indicated by the gray dots on the negative energy
axis. The thick gray line on the positive energy axis represents the
continuum. 
}
%\label{figcomplexintegration}
 \end{center}
\end{figure}
Integration~(\ref{Pnexpression1}) is carried out most efficiently by
deforming the path of integration into the negative complex energy
plane as shown in Fig.~\ref{figcomplexintegration}. 
Because the (dashed) line which closes the contour at infinity does
not contribute and because we take $\gamma$ large enough to be safely
away from the bound state poles ($\gamma\gg E_{\rm bind}$), we can
rewrite expression~(\ref{Pnexpression1}) as
\begin{eqnarray}
\lefteqn{
P_n^{th} \, = \,
\frac{-2\, i\,Q_b^2\,\pi}{(4M_b^2)^{n+1}}
\int\limits_{-\gamma-i\infty}^{-\gamma+i\infty} \frac{d E}{M_b} 
\exp\bigg\{-\frac{E}{M_b}n\bigg\}\,\bigg(
1 - \frac{E}{2M_b} + \frac{E^2}{4M_b^2} n
\bigg)\,\bigg[\,
C_1{\cal{A}}_1(E) -\frac{4}{3M_b^2} C_2 {\cal{A}}_2(E)
\,\bigg]
}
\nonumber
\\[2mm] & = &
\frac{4Q_b^2\pi^2}{(4M_b^2)^{n+1}}\,\frac{1}{2\pi i}\,
\int\limits_{\gamma-i\infty}^{\gamma+i\infty} \frac{d \tilde E}{M_b} \,
\exp\bigg\{\frac{\tilde E}{M_b}n\bigg\}\,\bigg(\,
1 + \frac{\tilde E}{2M_b} + \frac{\tilde E^2}{4M_b^2}\,n
\,\bigg)\bigg[\,
C_1 \,{\cal{A}}_1(-\tilde E) -\frac{4}{3M_b^2}\, 
C_2 \,{\cal{A}}_2(-\tilde E)
\,\bigg]
\,,
\nonumber\\&&
\label{Pnexpression2}
\end{eqnarray}
where in second line the change of variables $E\to-\tilde E$ has been
performed. The reader should note that for the integration in the
negative complex energy plane also the real part of the correlators
${\cal{A}}_1$ and ${\cal{A}}_2$ is needed.
The expression in the second line of
Eq.~(\ref{Pnexpression2}) offers three advantages which make it much
easier to calculate than expression~(\ref{Pnexpression1}):
\begin{enumerate}
\item Because the integration path is far away from bound state
energies, the integrand can be expanded in $\alpha_s$. This avoids
that we have to integrate over complicated special function like the
digamma function $\Psi$. 
\item We do not have to integrate separately over the resonances and
the continuum. Both contributions are in a convenient way calculated
at the same time.
\item The expression in the second line of Eq.~(\ref{Pnexpression2})
is nothing else than an inverse Laplace transform for which a vast
number of tables exist in literature (see
e.g.~\cite{Gradshteyn1}). 
\end{enumerate}
We want to stress that the advantages described above are merely
technical in nature and just simplify the calculation. The results of
the integration are not affected.
 
The final result for the theoretical moments including all
contributions up to NNLO in the nonrelativistic expansion can be cast
into the form
\begin{equation}
P_n^{th} \, = \,
\frac{3\,N_c\,Q_b^2\,\sqrt{\pi}}{4\,(4\,M_b^2)^n \, n^{3/2}}\,
\bigg\{\,
C_1(\mu_{\rm hard},\mu_{\rm fac})\,\varrho_{n,1}(\mu_{\rm
soft},\mu_{\rm fac}) +
C_2\,\varrho_{n,2}
\,\bigg\}
\label{Pntheoryfinal}
\end{equation}
where $\varrho_{n,1}$ comes from the integration of the correlator
${\cal{A}}_1$ (including LO, NLO and NNLO contributions in the
nonrelativistic expansion) and $\varrho_{n,2}$ originate from the
integration of ${\cal{A}}_2$ which is of NNLO only. To illustrate 
the technical aspects of the integration~(\ref{Pnexpression2}) let
first present some of the details of the calculation of the LO
contribution to $\varrho_{n,1}$. The LO contributions to
$\varrho_{n,1}$ originates from the
zero-distance Coulomb Green function in
Eq.~(\ref{CoulombGreenfunctionregularized}). The corresponding
integration takes the form
\begin{eqnarray}
\Big[\,\varrho_{n,1}\,\Big]^{\mbox{\tiny LO}} & = &
\frac{8\,\pi^{3/2}\,n^{3/2}}{M_b^2}\,\frac{1}{2\,\pi\,i}\,
\int\limits_{\gamma-i\infty}^{\gamma+i\infty} \frac{d \tilde E}{M_b}\,
\exp\bigg\{\frac{\tilde E}{M_b}\,n\bigg\}\,
G_c^{(0)}(0,0,-\tilde E)
\nonumber\\[2mm] & = &
2\,\sqrt{\pi}\,n^{3/2}\,\frac{1}{2\,\pi\,i}\,
\int\limits_{\gamma-i\infty}^{\gamma+i\infty} \frac{d \tilde E}{M_b}\,
\exp\bigg\{\frac{\tilde E}{M_b}\,n\bigg\}\,
\bigg[\,
-\tilde v - C_F\,a_s\,\ln\tilde v 
+ C_F\,a_s\,\sum\limits_{p=2}^{\infty}\zeta_p\,
 \bigg(\frac{C_F\,a_s}{2\,\tilde v}\bigg)
\,\bigg]
\,,
\nonumber\\&&
\label{rho1LOexpression1}
\end{eqnarray}
where 
\begin{equation}
\tilde v \, \equiv \, \sqrt{\frac{\tilde E}{M_b}}
\end{equation}
and $\zeta_p$ is the Riemann zeta function for the argument $p$.
Because $|C_F a_s/2 \tilde v|\ll 1$ along the integration path we have
expanded the digamma function in $G_c^{(0)}(0,0,-\tilde E)$ for small
$\alpha_s$. The resulting expression is now immediately ready for
the application of inverse Laplace transforms. Here, we only
need the relations
\begin{eqnarray}
\frac{1}{2\pi i}\,\int\limits_{\gamma-i\infty}^{\gamma+i\infty}
\frac{1}{x^\nu}\,e^{x\,t}\,dx & = & \frac{t^{\nu-1}}{\Gamma(\nu)}
\,,
\nonumber
\\[2mm]
\frac{1}{2\pi i}\,\int\limits_{\gamma-i\infty}^{\gamma+i\infty}
\frac{\ln x}{x^\nu}\,e^{x\,t}\,dx & = & 
\frac{t^{\nu-1}}{\Gamma(\nu)}\,\Big[\,
\Psi(\nu) - \ln t
\,\Big]
\,.
\label{Laplacetext}
\end{eqnarray}
The result for $[\varrho_{n,1}]^{\mbox{\tiny LO}}$ reads
\begin{equation}
\Big[\,\varrho_{n,1}\,\Big]^{\mbox{\tiny LO}} \, = \,
1 + 2\,\sqrt{\pi}\,\phi + 
4\,\sqrt{\pi}\,\sum\limits_{p=2}^{\infty}\,\phi^p\,
\frac{\zeta_p}{\Gamma(\frac{p-1}{2})}
\label{rho1LOexpression2}
\end{equation}
where
\begin{equation}
\phi \, \equiv \, \frac{C_F\,a_s\,\sqrt{n}}{2}
\,.
\end{equation}
Expression~(\ref{rho1LOexpression2}) can be rewritten in the form
\begin{equation}
\Big[\,\varrho_{n,1}\,\Big]^{\mbox{\tiny LO}} \, = \,
1 + 2\,\sqrt{\pi}\,\phi + \frac{2\,\pi^2}{3}\,\phi^2 +
4\,\sqrt{\pi}\,\sum\limits_{p=1}^{\infty}\,
\bigg(\frac{\phi}{p}\bigg)^3\,
\exp\bigg\{\bigg(\frac{\phi}{p}\bigg)^2\bigg\}\,
\bigg[\,
1+{\rm erf}\bigg(\frac{\phi}{p}\bigg)
\,\bigg]
\,
\label{rho1LOexpression3}
\end{equation}
where ${\rm erf}$ is the error function defined as
${\rm erf}(z)=\frac{2}{\sqrt{\pi}}\int_0^z\exp(-t^2)
dt$. Expression~(\ref{rho1LOexpression3}) agrees with the result
obtained by Voloshin~\cite{Voloshin2}. The infinite series
defined in Eq.~(\ref{rho1LOexpression2}) is absolute convergent with
an infinite
radius of convergence. For the values of $n$ employed in this work
($4\le n\le 10$), however, convergence is somewhat slow and a
large number of terms have to be taken into account. 
\begin{table}[thb]  % tab1
\vskip 7mm
\begin{center}
\begin{tabular}{|c||c|c|c|c|c|c|} \hline
$\phi$ & $0.5$ &  $0.6$ & $0.7$ & $0.8$ & $0.9$ & $1.0$ \\ \hline\hline 
$[\varrho_{n,1}]^{\mbox{\tiny LO}}$  
 & $6.38$ & $9.44$ & $14.07$  & $21.16$ & $32.10$ & $49.12$ \\ \hline
first three terms in Eq.~(\ref{rho1LOexpression2})  
 & $4.42$ & $5.50$ & $6.71$ & $8.05$ & $9.52$ & $11.12$ \\ \hline
\end{tabular}
\caption{\label{tab1} 
Comparison of the series for $[\varrho_{n,1}]^{\mbox{\tiny LO}}$ with
the sum of Born, one- and two-loop contributions in the series on the
RHS of Eq.~(\ref{rho1LOexpression2}) for the values of $\phi$ employed
in this work.
}
\end{center}
\vskip 3mm
\end{table}
This fact is illustrated in Tab.~\ref{tab1} where the sum of the first
three terms (corresponding to Born, one- and two-loop contributions) in
the series~(\ref{rho1LOexpression2}) is compared to the total sum for
values of $\phi$ between $0.5$ and $1.0$, which represent the range of
$\phi$ values used in this work. Tab.~\ref{tab1} shows that the
resummation of  
higher orders in $\alpha_s$ is essential to arrive at sensible results
in particular for larger values of $n$. This feature remains true for
all contributions to $\varrho_{n,1}$ and $\varrho_{n,2}$ and shows
that a naive fixed order (multi-loop) calculation for the moments is
unreliable for large values of $n$. 

Along the lines of the calculation of $[\varrho_{n,1}]^{\mbox{\tiny
LO}}$ it is now straightforward to determine $\varrho_{n,2}$ and the
NLO and NNLO contributions to $\varrho_{n,1}$. The contributions to
$\varrho_{n,1}$ coming from the one- and two-loop corrections to the
Coulomb potential, $V_c^{(1)}$ and $V_c^{(2)}$, have the form
\begin{eqnarray}
\lefteqn{
\Big[\,\varrho_{n,1}\,\Big]_c^{\mbox{\tiny NLO+NNLO}} \, = \,
\frac{8\,\pi^{3/2}\,n^{3/2}}{M_b^2}\,\frac{1}{2\,\pi\,i}\,
\int\limits_{\gamma-i\infty}^{\gamma+i\infty} \frac{d \tilde E}{M_b}\,
\exp\bigg\{\frac{\tilde E}{M_b}\,n\bigg\}\,
\bigg\{\,
G_c^{(1)}(0,0,-\tilde E) 
}
\nonumber\\[1mm]
& & \qquad\qquad\qquad\qquad
+ \Big[\,G_c^{(2)}(0,0,-\tilde E)\,\Big]_c^{\mbox{\tiny 1 loop}}
+ \Big[\,G_c^{(2)}(0,0,-\tilde E)\,\Big]_c^{\mbox{\tiny 2 loop}}
\,\bigg\}
\nonumber\\[2mm] & = &
4\,\sqrt{\pi}\,\delta_1\,\phi\,\bigg\{\,
\frac{1}{2}\,\ln\Big(\frac{\mu_1\,e^{\gamma_{\mbox{\tiny E}}/2}\,
  \sqrt{n}}{2\,M_b}\Big)
+ \sum\limits_{p=1}^\infty \phi^p\,\bigg[\,w^1_p + w^0_p \, {\rm
cln}\Big(M_b,n,\frac{2}{\mu_1},p\Big)
\,\bigg]
\,\bigg\}
\nonumber\\[1mm] & &
+ 4\,\sqrt{\pi}\,\delta_2\,\phi\,\bigg\{\,  
\frac{1}{2}\,\ln^2\Big(\frac{\mu_2\,e^{\gamma_{\mbox{\tiny E}}/2}\,
 \sqrt{n}}{2\,M_b}\Big)
+ \frac{\pi^2}{16}  
\nonumber\\[1mm] & & \qquad\qquad
+ \sum\limits_{p=1}^\infty \phi^p\,\bigg[\,
w^2_p - 
2\,w^1_p\, {\rm cln}\Big(M_b,n,\frac{2}{\mu_2},p\Big) -
w^0_p \, {\rm cln2}\Big(M_b,n,\frac{2}{\mu_2},p\Big)
\,\bigg]
\,\bigg\}
\nonumber\\[1mm] & &
8\,\sqrt{\pi}\,\delta_3^2\,\phi^2\,
\sum\limits_{p=0}^\infty\,\phi^p\,
 \bigg[\, \tilde w^2_p \,
 {\rm
csin}\Big(M_b,n,\frac{2}{\mu_3},\frac{\sqrt{n}}{M_b\,\pi\,\phi},p\Big)
+\tilde w^1_p \,
 {\rm
csinln}\Big(M_b,n,\frac{2}{\mu_3},\frac{\sqrt{n}}{M_b\,\pi\,\phi},p\Big)
\nonumber\\[1mm] & & \qquad\qquad
+\tilde w^0_p \,
{\rm
csinln2}\Big(M_b,n,\frac{2}{\mu_3},\frac{\sqrt{n}}{M_b\,\pi\,\phi},p\Big)
\,\bigg]
\,,
\label{rho1NNLOC}
\end{eqnarray}
where
\begin{eqnarray}
\delta_1 & = &
\bigg(\frac{a_s}{4\,\pi}\bigg)\, 2\,\beta_0
+ 2\,\bigg(\frac{a_s}{4\,\pi}\bigg)^2\,\Big(\,
2\,\beta_0\,a_1+\beta_1
\,\Big)
\,,
\nonumber\\[2mm]
\delta_2 & = &
\bigg(\frac{a_s}{4\,\pi}\bigg)^2\, 4\,\beta_0^2
\,,
\nonumber\\[2mm]
\delta_3 & = & 
\bigg(\frac{a_s}{4\,\pi}\bigg)\, 2\,\beta_0
\,,
\nonumber\\[2mm]
\mu_1 & = &
\mu_{\rm soft}\,\exp\bigg\{\,
\frac{1}{\delta_1}\,\bigg[\,
\bigg(\frac{a_s}{4\,\pi}\bigg)\,a_1 +
\bigg(\frac{a_s}{4\,\pi}\bigg)^2\,\Big(\,
\frac{\pi^2}{3}\,\beta_0^2+a_2
\,\Big)
\,\bigg]
\,\bigg\}
\,,
\nonumber\\[2mm]
\mu_2 & = & 
\mu_{\rm soft}
\,,
\nonumber\\[2mm]
\mu_3 & = & 
\mu_{\rm soft}\,\exp\bigg(\,\frac{a_1}{2\,\beta_0}
\,\bigg)
\,,
\label{deltamuNNLO}
\end{eqnarray}
and
\begin{eqnarray}
{\rm cln}(m,n,a,p) & \equiv & \ln\Big(\frac{a\,m}{\sqrt{n}}\Big)+
 \frac{1}{2}\,\Psi\bigg(\frac{p}{2}\bigg)
\,,
\\[2mm]
{\rm cln2}(m,n,a,p) & \equiv & 
\bigg[\,\ln\Big(\frac{a\,m}{\sqrt{n}}\Big)+
 \frac{1}{2}\,\Psi\Big(\frac{p}{2}\Big)
\,\bigg]^2 - \frac{1}{4}\,\Psi^\prime\bigg(\frac{p}{2}\bigg)
\,,
\\[2mm]
{\rm csin}(m,n,a,b,p) & \equiv & {}_0\rm F_2\Big(
\frac{3}{2},\frac{p+1}{2}, -\frac{n}{(2\,b\,m)^2}
\Big) 
\,,
\\[2mm]
{\rm csinln}(m,n,a,b,p) & \equiv & 
\bigg[\,\ln\Big(\frac{a\,m}{\sqrt{n}}\Big)+
 \frac{1}{2}\,\Psi\Big(\frac{p+1}{2}\Big)
\,\bigg]\,{}_0\rm F_2\Big(
  \frac{3}{2},\frac{p+1}{2}, -\frac{n}{(2\,b\,m)^2}
\Big) 
\nonumber\\[1mm] & & \qquad
-\frac{d}{dp}\,{}_0\rm F_2\Big(
  \frac{3}{2},\frac{p+1}{2}, -\frac{n}{(2\,b\,m)^2}
\Big) 
\,,
\\[2mm]
{\rm csinln2}(m,n,a,b,p) & \equiv & 
\bigg\{\,\bigg[\,\ln\Big(\frac{a\,m}{\sqrt{n}}\Big)+
 \frac{1}{2}\,\Psi\bigg(\frac{p+1}{2}\bigg)
\,\bigg]^2 - \frac{1}{4}\,\Psi^\prime\Big(\frac{p+1}{2}\Big)\,
\bigg\}\,
{}_0\rm F_2\Big(
  \frac{3}{2},\frac{p+1}{2}, -\frac{n}{(2\,b\,m)^2}
\Big) 
\nonumber\\[1mm] & & \qquad
-2\,\bigg[\,\ln\Big(\frac{a\,m}{\sqrt{n}}\Big)+
 \frac{1}{2}\,\Psi\bigg(\frac{p+1}{2}\bigg)
\,\bigg]\,\frac{d}{dp}\,{}_0\rm F_2\Big(
  \frac{3}{2},\frac{p+1}{2}, -\frac{n}{(2\,b\,m)^2}
\Big) 
\nonumber\\[1mm] & & \qquad
+\frac{d^2}{dp^2}\,{}_0\rm F_2\Big(
  \frac{3}{2},\frac{p+1}{2}, -\frac{n}{(2\,b\,m)^2}
\Big) 
\,.
\end{eqnarray}
The coefficients of the beta function, $\beta_{0,1}$ and the constants
$a_{1,2}$ are given in Eqs.~(\ref{b0a1def}) and (\ref{b1a2def}). 
The function $\Psi^\prime$ is the derivative of the digamma function
and ${}_0\rm F_2$ is a generalized hypergeometric
function~\cite{Gradshteyn1}. 
The constants $w^{0,1,2}_p$ and $\tilde w^{0,1,2}_p$ are given in
Appendix~\ref{appendixconstants}. 
For the calculation of expression~(\ref{rho1NNLOC}) the table of
inverse Laplace transforms given in Appendix~\ref{appendixLaplace} has
been used
extensively. The term proportional to $\delta_1$ in
Eq.~(\ref{rho1NNLOC}) contains the NLO
contributions coming from $V_c^{(1)}$ and the NNLO
contributions coming from the terms $\propto 1/r$ and $\propto
\ln(\mu_{\rm soft}\,e^\gamma_{\mbox{\tiny E}}\,r)/r$ in
$V_c^{(2)}$ in first order TIPT. The term
proportional to $\delta_2$ contains the remaining NNLO corrections
coming from the term $\propto \ln^2(\mu_{\rm
soft}\,e^\gamma_{\mbox{\tiny E}}\,r)/r$ in
$V_c^{(2)}$. The expression proportional to $\delta_3$, finally,
arises from the second order interaction in TIPT of $V_c^{(1)}$. 
The NNLO
contributions to $\varrho_{n,1}$ originating from the kinetic energy
corrections, the Breit-Fermi potential, the non-Abelian potential (see
Eq.~(\ref{GreenfunctionNNLOBF}) for the corresponding corrections to
the zero-distance Green function) and the kinematic correction factor
$(1+\tilde E/2M_b + (\tilde E^2/4 M_b^2)n)$ from
Eq.~(\ref{Pnexpression1}) read
\begin{eqnarray}
\lefteqn{
\Big[\,\varrho_{n,1}\,\Big]^{\mbox{\tiny LO}} + 
\Big[\,\varrho_{n,1}\,\Big]^{\mbox{\tiny NNLO}}_{\mbox{\tiny
kin+BF+NA}} \, = \,
}
\nonumber\\[1mm] & = & 
1 + \frac{9}{8\,n} + 2\,\sqrt{\pi}\,\phi\,\bigg[\,1+\frac{2}{n}\,\bigg] + 
4\,\sqrt{\pi}\,\sum\limits_{p=2}^{\infty}\,\phi^p\,
  \frac{\zeta_p}{\Gamma(\frac{p-1}{2})}\,\bigg[
  \,1+\frac{(3-p)\,(3+5\,p)}{8\,n}\,\bigg]
\nonumber\\[1mm] & & 
+\frac{8}{3\,n}\,\phi^2\,\bigg\{
 -\bigg[\,1-\frac{\gamma_{\mbox{\tiny E}}}{2}-\ln(2\,\sqrt{n})\,\bigg]
 +2\,\sqrt{\pi}\,\phi\,\bigg[\,\frac{\gamma_{\mbox{\tiny E}}}{2}
  +\ln\sqrt{n}\,\bigg]
\nonumber\\[1mm] & & \qquad\qquad
 -2\,\sqrt{\pi}\,\sum\limits_{p=2}^\infty\,
\frac{\phi^p}{\Gamma(\frac{p-1}{2})}\,\bigg[\,
\zeta_p\,\bigg(\Psi\Big(\frac{p-1}{2}\Big)-2\,\ln\sqrt{n}\,\bigg)
+\zeta_{p+1}
\,\bigg]
\nonumber\\[1mm] & & \qquad\qquad
+2\,\sqrt{\pi}\,\sum\limits_{p,q=2}^\infty\,
\phi^{p+q-1}\,\frac{\zeta_p\,\zeta_q}{\Gamma(\frac{p+q-2}{2})}
\,\bigg\}
\nonumber\\[1mm] & &   
-\Big[\,\varrho_{n,1}\,\Big]^{\mbox{\tiny LO}}\,\bigg[\,
a_s^2\,\bigg(\,\frac{1}{3}\,C_F^2+\frac{1}{2}\,C_A\,C_F\,\bigg)\,
  \ln\frac{M_b^2}{\mu_{\rm fac}^2}
\,\bigg]
\,,
\label{rho1NNLOBF}
\end{eqnarray} 
where, for convenience, also the LO result from
Eq.~(\ref{rho1LOexpression2}) has been added. From the last line of
expression~(\ref{rho1NNLOBF}) one can easily determine a
renormalization group equation (with respect to the factorization
scale $\mu_{\rm fac}$) for $\varrho_{n,1}$ and $\varrho_{n,2}$, which
would allow for a resummation of the corrections coming from the
kinetic energy corrections, the Breit-Fermi potential and the
non-Abelian potential to all orders in TIPT. Although it is quite
tempting to carry out this resummation, we refrain from doing so
because a resummation of those corrections would not account for the
retardation effects mentioned in Section~\ref{sectionbasicidea}. 
The complete expression for $\varrho_{n,1}$ has the form
\begin{equation}
\varrho_{n,1} \, = \, 
\Big[\,\varrho_{n,1}\,\Big]^{\mbox{\tiny LO}}
+ \Big[\,\varrho_{n,1}\,\Big]^{\mbox{\tiny NLO+NNLO}}_c
+ \Big[\,\varrho_{n,1}\,\Big]^{\mbox{\tiny NNLO}}_{\mbox{\tiny
kin+BF+NA}}
\,.
\label{rho1final}
\end{equation}
Finally, the result for $\varrho_{n,2}$ coming from the integration of
${\cal{A}}_2$, Eq.~(\ref{A2final}), reads
\begin{equation}
\varrho_{n,2} \, = \,
\frac{1}{n}\,\bigg[\,
-2 -\frac{8}{3}\,\sqrt{\pi}\,\phi + 
4\,\sqrt{\pi}\,\sum\limits_{p=2}^{\infty}\,\phi^p\,
\frac{2\,(p-3)}{3}\,
\frac{\zeta_p}{\Gamma(\frac{p-1}{2})}
\,\bigg]
\label{rho2final}
\end{equation}

From from expression~(\ref{Pntheoryfinal}) for the theoretical moments
at NLO one can easily recover the moments at NNLO by setting
\begin{eqnarray}
C_1 & = & 1 + \Big(\frac{a_h}{\pi}\Big)\,c_1^{(1)}
\,,
\nonumber\\[2mm]
C_2 & = & 0
\,,
\nonumber\\[2mm]
\delta_1 & = &
\bigg(\frac{a_s}{4\,\pi}\bigg)\, 2\,\beta_0
\,,
\nonumber\\[2mm]
\delta_2 & = & \delta_3 \, = \, 0
\,,
\nonumber\\[2mm]
\mu_1 & = &
\mu_{\rm soft}\,\exp\bigg(\,\frac{a_1}{2\,\beta_0}
\,\bigg)
\,,
\label{deltamuNLO}
\end{eqnarray}
and by ignoring the corrections $[\varrho_{n,1}]^{\mbox{\tiny
NNLO}}_{\mbox{\tiny kin+BF+NA}}$.
The resulting expression for the NLO moments is
identical to the one obtained by Voloshin~\cite{Voloshin1}.
\par
\vspace{0.5cm}
\section{Some Comments to the Moments}
\label{sectionproperties}
In this section we will spend some time to discuss some interesting 
properties of the theoretical moments $P_n^{th}$ which have been
calculated in Section~\ref{sectioncalculatemoments}. We will address
three issues: (i) the relation between the strong dependence of the
moments on $M_b$ and $\alpha_s$ and the dependences of the moments on
the scales $\mu_{\rm soft}$, $\mu_{\rm hard}$ and $\mu_{\rm fac}$,
(ii) the properties of the resonance and continuum contributions and
(iii) the quality of the nonrelativistic expansion.

It is a characteristic feature of the moments that they depend very
strongly on the bottom quark mass $M_b$ and the strong coupling
$\alpha_s$. This is illustrated in Tab.~\ref{tab2a} where the moments
$P_n^{th}$ are displayed for $n=4,6,8,10,20$ and for various
values of $M_b$ and $\alpha_s(M_z)$ while the renormalization scales
are fixed to $\mu_{\rm soft}=2.5$~GeV and $\mu_{\rm hard}=\mu_{\rm
fac}=5$~GeV.
\begin{table}[t] % tab2a
\vskip 7mm
\begin{center}
\begin{tabular}{|r@{$/$}l||c|c|c|c||c|c|c|c|} \hline
\multicolumn{2}{|c||}{Moment} 
   & \multicolumn{4}{|c||}{$M_b/[GeV]$}
   & \multicolumn{4}{|c|}{$\alpha_s(M_z)$} \\ \hline
\multicolumn{2}{|c||}{}
 & $4.6$ &  $4.8$ & $5.0$ & $5.2$ & $0.10$ & $0.11$ 
                          & $0.12$ & $0.13$ \\ \hline\hline 
$P_4^{th}$&$[10^{-xx}\,\mbox{GeV}^{-8}]$
 & $0.51$ & $0.37$ & $0.27$ & $0.20$ 
 & $0.19$ & $0.27$ & $0.41$ & $0.74$  \\ \hline
$P_6^{th}$&$[10^{-xx}\,\mbox{GeV}^{-12}]$
 & $0.67$ & $0.41$ & $0.25$ & $0.16$ 
 & $0.17$ & $0.26$ & $0.46$ & $0.97$  \\ \hline
$P_8^{th}$&$[10^{-xx}\,\mbox{GeV}^{-16}]$
 & $0.95$ & $0.49$ & $0.26$ & $0.14$ 
 & $0.18$ & $0.29$ & $0.57$ & $1.37$  \\ \hline
$P_{10}^{th}$&$[10^{-xx}\,\mbox{GeV}^{-20}]$
 & $1.42$ & $0.61$ & $0.27$ & $0.13$ 
 & $0.19$ & $0.34$ & $0.73$ & $1.99$  \\ \hline
$P_{20}^{th}$&$[10^{-xx}\,\mbox{GeV}^{-40}]$
 & $12.96$ & $2.37$ & $0.47$ & $0.10$ 
 & $0.42$ & $1.00$ & $3.07$ & $13.93$  \\ \hline\hline
\multicolumn{2}{|c||}{} 
   & \multicolumn{4}{c||}{$\alpha_s(M_z) = 0.118$}
   & \multicolumn{4}{c|}{$M_b=4.8$~GeV} \\ \hline
\multicolumn{2}{|c||}{} 
   &      \multicolumn{8}{c|}{$\mu_{\rm soft}=2.5\,\mbox{GeV}\,,\quad
          \mu_{\rm hard}=\mu_{\rm fac}=5$~GeV} \\\hline
\end{tabular}
\caption{\label{tab2a} 
The theoretical moments $P_n^{th}$ for $n=4,6,8,10,20$ and fixed
$\mu_{\rm soft}=2.5$~GeV and $\mu_{\rm hard}=\mu_{\rm fac}=5$~GeV 
for various values of $M_b$ and $\alpha_s(M_z)$.
The two-loop running for the strong coupling has been employed.
}
\end{center}
\vskip 3mm
\end{table}
The dependence on $M_b$ is powerlike ($P_n^{th}\sim
M_b^{-2n}$) for dimensional reasons (see
definition~(\ref{momentdef})). The dependence on $\alpha_s$ is
exponentially (see e.g.\ Eq.~(\ref{rho1LOexpression3})) and comes from
the resummations of the ladder diagrams containing the exchange of
longitudinal Coulomb gluons. At this point one might conclude that
fitting the theoretical moments to the experimental ones would allow
for an extremely precise extraction of $M_b$ and
$\alpha_s$, in particular if $n$ is chosen very large. Unfortunately
this conclusion is wrong. It is wrong from the conceptual point of
view because for increasing $n$ the effective smearing range $\Delta E$ in the
integral~(\ref{momentcrosssectionrelation}) becomes smaller and
smaller, which makes the perturbative calculations for the
moments become less trustworthy~\cite{Poggio1}. In
Section~\ref{sectionbasicidea} we
have used this argument to determine an upper bound on the allowed
values on $n$. However, besides the conceptual arguments, the
perturbative series for the moments itself contains a mechanism which
prevents an arbitrarily precise determination of $M_b$ and $\alpha_s$
for large values of $n$.
\begin{table}[t] % tab2
\vskip 7mm
\begin{center}
\begin{tabular}{|r@{$/$}l||c|c|c||c|c|c||c|c|c|} \hline
\multicolumn{2}{|c|}{Moment} 
   & \multicolumn{3}{c||}{$\mu_{\rm soft}/[\mbox{GeV}]$}
   & \multicolumn{3}{c||}{$\mu_{\rm hard}/[\mbox{GeV}]$}
   & \multicolumn{3}{c|}{$\mu_{\rm fac}/[\mbox{GeV}]$} \\ \hline
\multicolumn{2}{|c|}{}
 & $1.5$ &  $2.5$ & $3.5$ & $2.5$ & $5.0$ & $10.0$ 
                          & $2.5$ & $5.0$ & $10.0$ \\ \hline\hline 
$P_4^{th}$&$[10^{-8}\,\mbox{GeV}^{-8}]$
 & $0.94$ & $0.37$ & $0.27$ 
 & $0.31$ & $0.37$ & $0.43$ 
 & $0.45$ & $0.37$ & $0.25$ \\ \hline
$P_6^{th}$&$[10^{-12}\,\mbox{GeV}^{-12}]$
 & $1.16$ & $0.41$ & $0.28$ 
 & $0.34$ & $0.41$ & $0.47$ 
 & $0.51$ & $0.41$ & $0.27$ \\ \hline
$P_8^{th}$&$[10^{-16}\,\mbox{GeV}^{-16}]$
 & $1.53$ & $0.49$ & $0.33$ 
 & $0.41$ & $0.49$ & $0.56$ 
 & $0.62$ & $0.49$ & $0.32$ \\ \hline
$P_{10}^{th}$&$[10^{-20}\,\mbox{GeV}^{-20}]$
 & $2.10$ & $0.61$ & $0.39$ 
 & $0.51$ & $0.61$ & $0.70$ 
 & $0.79$ & $0.61$ & $0.39$ \\ \hline
$P_{20}^{th}$&$[10^{-40}\,\mbox{GeV}^{-40}]$
 & $11.89$ & $2.37$ & $1.28$ 
 & $1.98$ & $2.37$ & $2.72$ 
 & $3.17$ & $2.37$ & $1.47$ \\ \hline\hline
\multicolumn{2}{|c|}{} 
   & \multicolumn{3}{c||}{$\mu_{\rm hard}=5$~GeV}
   & \multicolumn{3}{c||}{$\mu_{\rm soft}=2.5$~GeV}
   & \multicolumn{3}{c|}{$\mu_{\rm soft}=2.5$~GeV} \\ 
\multicolumn{2}{|c|}{} 
   & \multicolumn{3}{c||}{$\mu_{\rm fac}=5$~GeV}
   & \multicolumn{3}{c||}{$\mu_{\rm fac}=5$~GeV}
   & \multicolumn{3}{c|}{$\mu_{\rm hard}=5$~GeV} \\\hline
\end{tabular}
\caption{\label{tab2} 
The theoretical moments $P_n^{th}$ for $n=4,6,8,10,20$ and fixed
$\alpha_s(M_z)=0.118$ and $M_b=4.8$~GeV for various
choices of the renormalization scales $\mu_{\rm soft}$, $\mu_{\rm hard}$ and
$\mu_{\rm fac}$. The two-loop running for the strong coupling has been
employed.
}
\end{center}
\vskip 3mm
\end{table}
In Tab.~\ref{tab2} the theoretical moments $P_n^{th}$,
$n=4,6,8,10,20$, are displayed for different choices
for $\mu_{\rm soft}$, $\mu_{\rm hard}$ and 
$\mu_{\rm fac}$ and for $\alpha_s(M_z)=0.118$ and
$M_b=4.8$~GeV. It is obvious that
the dependence of the moments on the renormalization scales, and in
particular on the soft scale, is becoming increasingly strong for
larger values of $n$. As an example, the moment $P_{20}^{th}$
($P_{10}^{th}$) can change by a factor of ten (five) if the soft scale
is varied between $1.5$ and $3.5$~GeV. These huge scale dependences
are mainly caused by the large NNLO contributions to the large $n$
moments coming from the two-loop corrections to the Coulomb potential,
$V_c^{(2)}$, the second iteration of one-loop corrections to the
Coulomb potential, $V_c^{(1)}$, and the non-Abelian potential,
$V_{\mbox{\tiny NA}}$. During the fitting procedure, when all
renormalization scales are scanned of the
ranges~(\ref{choiceofscales}), the large scale dependencies
effectively compensate the strong
dependence of the moments on $M_b$ and $\alpha_s$. In
Section~\ref{subsectionunconstraint} it is shown that this 
affects mostly the extraction of $\alpha_s$ rendering the sum rule, at
least at the present stage, a rather powerless tool as far as
precision determinations of the strong coupling are concerned. We want
to stress that this compensation represents a very delicate balance
which, if at all, can only be trusted if $n$ is not chosen too large.
We believe that this balance is still under controll for the values of
$n$ used in this work ($4\le n\le 10$), although no proof for this
assumption can be given. However, it is certain that for even larger
values of $n$ the extracted values of $M_b$ and $\alpha_s$ might
contain sizable systematic errors.

We also would like to make one comment on the fact that the
theoretical moments contain contributions from below ($E<0$) and above
($E>0$) the threshold point. As shown in
Eq.~(\ref{Rthreshnonrelativistic}), the former contributions come
from the resonance poles whereas the latter arise from the
continuum. To demonstrate the size of the resonance and the continuum
contributions let us examine the LO contribution to $\varrho_{n,1}$
which respect to this aspect. The contributions to
$[\varrho_{n,1}]^{\mbox{\tiny LO}}$ from $E<0$ and $E>0$ can be
calculated separately from Eq.~(\ref{Pnexpression1}) using the
LO nonrelativistic expression for the cross
section from Eq.~(\ref{Rthreshnonrelativistic}), ($\phi=
C_F\,a_s\,\sqrt{n}/2$)
\begin{eqnarray}
\Big[\,\varrho_{n,1}\,\Big]^{\mbox{\tiny LO}}_{E<0} & = &
8\,\sqrt{\pi}\,\sum\limits_{p=1}^{\infty}\,
\bigg(\frac{\phi}{p}\bigg)^3\,
\exp\bigg\{\bigg(\frac{\phi}{p}\bigg)^2\bigg\}
\,,
\label{rho1LOresonances}
\\[2mm]
\Big[\,\varrho_{n,1}\,\Big]^{\mbox{\tiny LO}}_{E>0} & = &
1 + 2\,\sqrt{\pi}\,\phi + \frac{2\,\pi^2}{3}\,\phi^2 +
4\,\sqrt{\pi}\,\sum\limits_{p=1}^{\infty}\,
\bigg(\frac{\phi}{p}\bigg)^3\,
\exp\bigg\{\bigg(\frac{\phi}{p}\bigg)^2\bigg\}\,
\bigg[\,
-1+{\rm erf}\bigg(\frac{\phi}{p}\bigg)
\,\bigg]
\,.
\label{rho1LOcontinuum}
\end{eqnarray}
\begin{table}[t]  % tab4
\vskip 7mm
\begin{center}
\begin{tabular}{|l||c|c|c|c|c|c|c|c|c|c|c|} \hline
\multicolumn{1}{|c||}{$\phi$} 
  & $0.0$ & $0.1$ & $0.2$ & $0.3$ & $0.4$ & $0.5$ 
  & $0.6$ & $0.7$ & $0.8$ & $0.9$ & $1.0$  \\ \hline\hline 
$[\varrho_{n,1}]^{\mbox{\tiny LO}}_{E<0}$  
 & $0.00$ & $0.02$ & $0.14$ & $0.50$ & $1.25$ & $2.65$ 
 & $5.05$ & $9.01$ & $15.42$ & $25.66$ & $42.00$  \\ \hline
$[\varrho_{n,1}]^{\mbox{\tiny LO}}_{E>0}$  
 & $1.00$ & $1.41$ & $1.92$ & $2.48$ & $3.09$ & $3.73$ 
 & $4.39$ & $5.06$ & $5.74$ & $6.43$ & $7.13$   \\ \hline
\end{tabular}
\caption{\label{tab4} 
The resonance ($E<0$) and continuum ($E>0$) contributions to 
the function $[\varrho_{n,1}]_{\mbox{\tiny LO}}$ for $0.0\le\phi\le
1.0$.
}
\end{center}
\vskip 3mm
\end{table}
In Tab.~\ref{tab4} expressions~(\ref{rho1LOresonances}) and
(\ref{rho1LOcontinuum}) are evaluated for $0.0\le\phi\le 1.0$. For
$\phi\approx 0.5$ resonance and continuum contributions are
approximately equal in size, whereas for larger values of $\phi$ the
resonance contributions dominate. This shows explicitly that for
large values of $n$ (where $n>4$ can already considered as large) the
resonance effects cannot be neglected. In particular, any sum rule
analysis which is based on the large $n$ moments and
ignores the resonance contributions will lead to
a bottom quark mass which is too low. 
%We will come back to this point in Section~\ref{sectioncomments}. 
From Eq.~(\ref{rho1LOresonances}) it
is also conspicuous that there are no
resonance contributions proportional to $\alpha_s^n$ with $n=0,1,2$. This
shows that in conventional multi-loop perturbation theory bound state
contributions to the heavy-quark-antiquark production cross section
(in lepton pair collisions) are produced by Feynman diagrams
containing three and more loops. Because the two-loop level represents
the current
state of the art in covariant multi-loop calculations where the full quark
mass and energy dependence is taken into account (see Ref.~\cite{Kuhn2}
for a review and \cite{Chetyrkin1} far a recent publications on this
subject),
these peculiar contributions to the cross section sitting below the
threshold point have not been observed so far. At the three-loop level,
however, the cross section {\it will} have singular contributions
$\propto \alpha_s/v^2$ below the
threshold. In fact, these contributions are required by the analyticity
of the vacuum polarization function in the nonrelativistic regime. [Of
course, for a proper description of the bound state regime fixed order
multi-loop perturbation theory is insufficient and a resummation of the singular
terms to all orders in $\alpha_s$ has to be carried out.]

Finally, we also would like to address the question how well the
nonrelativistic (and asymptotic)
expansion at NNLO for the cross section $R$
(Eq.~(\ref{Rdefinitioncovariant})) and the integration
measure $ds/s^{n+1}$ in the dispersion
integral~(\ref{momentcrosssectionrelation}) can 
approximate a complete covariant calculation of the large $n$ moments,
where all mass and energy dependences would be accounted for
exactly. Strictly speaking,
this question cannot be answered entirely because a complete
covariant calculation of the moments, Eq.~(\ref{momentdef}), for large
values of $n$ is certainly an impossible task. (If it were possible, we
would not use the nonrelativistic expansion and NRQCD in the first
place.) However, a partial answer can be given by comparing the
terms proportional to $\alpha_s^n$ with $n=0,1,2$ in
$P_n^{th}$, Eq.~(\ref{Pntheoryfinal}), to the corresponding
contributions calculated in full QCD. For simplicity we only present a
comparison of the Born and one-loop contributions in the
following. The two-loop
contributions lead to the same conclusions. The Born and the one-loop
contributions from $P_n^{th}$ read
\begin{eqnarray}
\Delta_{n,\mbox{\tiny NRQCD}}^{\mbox{\tiny Born}} & \equiv &
\bigg\{\,
\bigg[\,
\frac{3\,N_c\,Q_b^2\,\sqrt{\pi}}{4\,(4\,M_b^2)^n \, n^{3/2}}
\,\bigg]^{-1}\,P_n^{th}
\,\bigg\}^{{\cal{O}}(1)}
\, = \, 1 - \frac{7}{8\,n}
\,,
\label{deltaBORNNNLO}
\\[2mm]
\Delta_{n,\mbox{\tiny NRQCD}}^{\mbox{\tiny 1 loop}} & \equiv &
\bigg\{\,
\bigg[\,
\frac{3\,N_c\,Q_b^2\,\sqrt{\pi}}{4\,(4\,M_b^2)^n \, n^{3/2}}
\,\Big(\frac{C_F\,\alpha_s}{\pi}\Big)
\,\bigg]^{-1}\,P_n^{th}
\,\bigg\}^{{\cal{O}}(\alpha_s)}
\nonumber\\[1mm] & = &  
\pi^{3/2}\,\sqrt{n}\,\bigg(1+\frac{2}{3\,n}\bigg)  
- 4\,\bigg(1+ \frac{9}{8\,n}\bigg) 
\,.
\label{deltaalphasNNLO}
\end{eqnarray}
The complete covariant versions of expressions~(\ref{deltaBORNNNLO})
and (\ref{deltaalphasNNLO}) in full QCD can be
determined from the well known Born and one-loop formulae for the
cross section~\cite{Kallensabry1,Schwinger2},
\begin{eqnarray}
\lefteqn{
R^{\mbox{\tiny Born}}(q^2) \, = \,
\frac{N_c\,Q_b^2}{2}\,\beta\,(3-\beta^2)
\,,
}
\label{RBornexact}
\nonumber\\[2mm]
\lefteqn{
R^{\mbox{\tiny 1 loop}}(q^2) \, = \,
N_c\,Q_b^2\,\bigg(\frac{C_F\,\alpha_s}{\pi}\bigg)\,\bigg\{\, 
  \frac{3\,\beta\,\left( 5 - 3\,{\beta^2} \right) }{8}
}
\nonumber\\[2mm] & & \mbox{} - 
  \beta\,( 3 - {\beta^2} ) \,
   \Big( 2\,\ln(1 - p) + \ln(1 + p) \Big)  - 
  \frac{\left( 1 - \beta \right) \,
     \left( 33 - 39\,\beta - 17\,{\beta^2} + 7\,{\beta^3} \right) }{16}\,
   \ln p 
\nonumber\\[2mm] & & \mbox{}
+ \frac{\left( 3 - {\beta^2} \right) \,\left( 1 + {\beta^2} \right) }{2
    }\,\bigg[\, 2\,\mbox{Li}_2(p) + \mbox{Li}_2({p^2}) + 
     \ln p\,\Big( 2\,\ln(1 - p) + \ln(1 + p) \Big) 
      \,\bigg] 
\,\bigg\}
\,,
\label{Ralphasexact}
\end{eqnarray}
where $\beta=(1-4 M_b^2/q^2)^{1/2}$ and $p=(1-\beta)/(1+\beta)$ and
$\mbox{Li}_2$ is the dilogarithm, and the covariant
form of the dispersion relation for the moments,
Eq.~(\ref{momentdef}),
\begin{eqnarray}
\Delta_{n,\mbox{\tiny QCD}}^{\mbox{\tiny Born}} & \equiv &
\bigg[\,
\frac{3\,N_c\,Q_b^2\,\sqrt{\pi}}{4\,(4\,M_b^2)^n \, n^{3/2}}
\bigg]^{-1}\,\int\limits_{4M_b^2}^\infty \frac{ds}{s^{n+1}}\,
R^{\mbox{\tiny Born}}(s)
\,,
\label{deltaBORNQCD}
\\[2mm]
\Delta_{n,\mbox{\tiny QCD}}^{\mbox{\tiny 1 loop}} & \equiv &
\bigg[\, 
\frac{3\,N_c\,Q_b^2\,\sqrt{\pi}}{4\,(4\,M_b^2)^n \, n^{3/2}}
\,\Big(\frac{C_F\,\alpha_s}{\pi}\Big)
\bigg]^{-1}\,\int\limits_{4M_b^2}^\infty \frac{ds}{s^{n+1}}\,
R^{\mbox{\tiny 1 loop}}(s)
\,.
\label{deltaalphasQCD}
\end{eqnarray}
Expressions~(\ref{deltaBORNQCD}) and (\ref{deltaalphasQCD}) can be
easily calculated numerically. In Tab.~\ref{tab5} 
$\Delta_{n,\mbox{\tiny NRQCD}}^{\mbox{\tiny Born}}$, 
$\Delta_{n,\mbox{\tiny QCD}}^{\mbox{\tiny Born}}$,
$\Delta_{n,\mbox{\tiny NRQCD}}^{\mbox{\tiny 1 loop}}$ and
$\Delta_{n,\mbox{\tiny QCD}}^{\mbox{\tiny 1 loop}}$ are presented for
$n=1,\ldots,10$. 
\begin{table}[t]  % tab5
\vskip 7mm
\begin{center}
\begin{tabular}{|l||c|c|c|c|c|c|c|c|c|c|} \hline
\multicolumn{1}{|c||}{$n$} 
  & $1$ &  $2$ & $3$ & $4$ & $5$
  & $6$ &  $7$ & $8$ & $9$ & $10$  \\ \hline\hline
$\Delta_{n,\mbox{\tiny NRQCD}}^{\mbox{\tiny Born}}$
  & $0.13$ & $0.56$ & $0.71$ & $0.78$ & $0.83$ & $0.85$ 
  & $0.88$ & $0.89$ & $0.90$ & $0.91$ \\ \hline
$\Delta_{n,\mbox{\tiny QCD}}^{\mbox{\tiny Born}}$
  & $0.60$ & $0.73$ & $0.79$ & $0.83$ & $0.86$ & $0.88$ 
  & $0.89$ & $0.91$ & $0.91$ & $0.92$ \\ \hline\hline
$\Delta_{n,\mbox{\tiny NRQCD}}^{\mbox{\tiny 1 loop}}$
  & $0.78$ & $4.25$ & $6.29$ & $7.87$ & $9.21$ & $10.41$ 
  & $11.49$ & $12.50$ & $13.44$ & $14.33$ \\ \hline
$\Delta_{n,\mbox{\tiny QCD}}^{\mbox{\tiny 1 loop}}$
  & $2.28$ & $4.25$ & $5.91$ & $7.36$ & $8.65$ & $9.84$ 
  & $10.92$ & $11.94$ & $12.90$ & $13.81$ \\ \hline
\end{tabular}
\caption{\label{tab5} 
The Born and one-loop contributions to the theoretical moments
calculated in the nonrelativistic expansion (NRQCD) at NNLO and in
full QCD for $n=1,\ldots,10$. 
}
\end{center}
\vskip 3mm
\end{table}
The difference for the Born (one-loop) contributions amounts to 6\%
(7\%) for $n=4$ and quickly decreases for larger values of
$n$. Thus, for the values of $n$ employed in this work the
asymptotic expansion in the velocity and, in particular, the use of
NRQCD, lead to a sufficiently good approximation to the exact covariant
results for the cases where a comparison can be carried out. [At this
point one has to compare the quality of the approximation to the large
scale variations of the moments discussed at the beginning of this
section.] This strengthens our
confidence that our method to calculate the theoretical moments is
sufficient at the level of the remaining theoretical uncertainties. In
particular, we cannot confirm the claims in~\cite{Jamin1} that the
nonrelativistic expansion would behave badly and would represent a
good approximation only for $n\sim 100$.
\par
\vspace{0.5cm}
\section{Experimental Moments the Fitting Procedure} 
\label{sectionfitting}
In this section we will describe how the moments are calculated from
experimental data and present our method to fit the
experimental moments, $P_n^{ex}$, to the theoretical ones, $P_n^{th}$.

The experimental moments are determined using the available data on
the $\Upsilon$ masses $M_{\Upsilon(nS)}$ and electronic partial widths
$\Gamma_{\Upsilon(nS)}\equiv\Gamma(\Upsilon(nS)\to e^+e^-)$ for
$n=1,\ldots,6$. For a compilation of all experimental numbers
see Tab.~\ref{tab6}.
\begin{table}[t!]  % tab6
\vskip 7mm
\begin{center}
\begin{tabular}{|l||c|c|c|} \hline
\multicolumn{1}{|c||}{$nS$} 
  & $1S$ & $2S$ & $3S$ 
 \\ \hline\hline 
$M_{nS}/[\mbox{GeV}]$  
 & $9.460$ & $10.023$ & $10.355$    \\ \hline
$\Gamma_{nS}/[\mbox{keV}]$ 
 & $1.32\pm 0.04\pm 0.03$ & $0.52\pm 0.03\pm 0.01$ 
 & $0.48\pm 0.03\pm 0.03$ \\ \hline\hline\hline
\multicolumn{1}{|c||}{$nS$} 
  & $4S$ & $5S$ & $6S$ 
 \\ \hline\hline 
$M_{nS}/[\mbox{GeV}]$  
 & $10.58$ & $10.87$ & $11.02$   \\ \hline
$\Gamma_{nS}/[\mbox{keV}]$ 
 & $0.25\pm 0.03\pm 0.01$ 
 & $0.31\pm 0.05\pm 0.07$ & $0.13\pm 0.03\pm$ 0.03  \\ \hline\hline\hline
\multicolumn{4}{|c|}
{$\tilde \alpha_{em}^{-1} = \alpha_{em}^{-1}(10$~GeV$) = 131.8(1\pm0.005)\,,
\quad (\sqrt{s})_{B\bar B} = 2\times 5.279$~GeV} \\ \hline
\end{tabular}
\caption{\label{tab6} 
The Experimental number for the $\Upsilon$ masses and electronic decay
widths used for the calculation of the experimental moments
$P_n^{ex}$. For the widths the first error is statistical and the
second systematic. The errors for $\Upsilon_{1S}$ and $\Upsilon_{2S}$
are taken from~\cite{Albrecht1}. All the other errors are estimated
from the numbers presented in~\cite{PDG}. The errors in the $\Upsilon$
masses and the $B\bar B$ threshold $(\sqrt{s})_{B\bar B}$ are neglected.
}
\end{center}
\vskip 3mm
\end{table}
The formula for the experimental moments reads
\begin{eqnarray}
P_n^{ex} & = &  
\frac{9\,\pi}{\tilde\alpha^2_{em}}\,\sum\limits_{k=1}^6\,
\frac{\Gamma_{kS}}{M_{kS}^{2n+1}} 
\, + \,
\int\limits_{\sqrt{s}_{B\bar B}}^\infty
\frac{ds}{s^{n+1}}\,r_{cont}(s)
\,.
\label{Pnexperiment}
\end{eqnarray}
The first term on the RHS of Eq.~(\ref{Pnexperiment}) is obtained by
using the narrow width approximation for all the known resonances
\begin{equation}
R_{res}(s) \, = \,
\frac{9\,\pi}{\tilde \alpha_{em}^2}\,
\sum\limits_{n=1}^\infty\,\Gamma_{nS}\,
M_{nS}\,\delta(s-M_{nS}^2)
\,.
\label{Rresonances}
\end{equation}
$\tilde \alpha_{em}$ is the electromagnetic running coupling at the
scale $10$~GeV (see Tab.~\ref{tab6}) which divides out the effects of
the photonic vacuum polarization contained in the electromagnetic
decay width.\footnote{
To be more accurate, the electromagnetic coupling should be evaluated
for each resonance individually at the corresponding resonance
mass. The resulting differences, however, are smaller then the assumed
error in $\tilde \alpha_{em}$ itself and therefore neglected.
}
The second term describes the contribution from the continuum above
the $B\bar B$ threshold. We approximate the continuum cross section by
a constant with a $50\%$
error
\begin{equation}
r_{cont}(s) = r_c\,(1 \pm 0.5)
\,.
\label{Rexperimentcontinuum}
\end{equation}
This simplifies the treatment of experimental errors in the continuum
regime significantly but also represents an reasonably good
approximation because for $n\le 4$ the
continuum is already sufficiently suppressed that a more detailed
description of it is not necessary. During the fitting procedure we
vary the constant $r_c$ between $0.5$ and $1.5$ which certainly covers
all the experimental uncertainties. [In fact, this prescription
renders the resonances $4S$, $5S$ and $6S$, which lie above the $B\bar
B$ threshold practically irrelevant.]

For the fit we use the standard least squares method as described
in~\cite{PDG}. The $\chi^2$ function which has to be minimized reads
\begin{equation}
\chi^2(M_b,\alpha_s) \, = \,
\sum\limits_{\{n\},\{m\}}\,
\Big(\,P_n^{th}-P_n^{ex}\,\Big)\,(S^{-1})_{n m}\,
\Big(\,P_m^{th}-P_m^{ex}\,\Big)
\,.
\label{x2general}
\end{equation}
$\{n\}$ represents the set of $n$`s for which the fit shall be carried
out and $S^{-1}$ is the inverse covariance matrix describing the
experimental errors
and the correlation between the experimental moments. To construct the
covariance matrix we use the errors in the electronic decay widths
(where statistical and systematic errors are added quadratically), in
the electromagnetic coupling $\tilde \alpha_{em}$ (see
Tab.~\ref{tab6}) and the error in the continuum cross section,
Eq.~(\ref{Rexperimentcontinuum}), which we also treat as
experimental. The tiny errors in the $\Upsilon$ masses are
neglected. At this
point it is important to note that the errors in the electronic widths
are certainly not uncorrelated due to common sources of systematic
errors in the $e^+e^-$~collider experiments (mostly CLEO) where the
widths have been determined. Unfortunately an analysis of these
correlations cannot be found in the corresponding publications
(see~\cite{PDG} for references). We therefore assume that the
correlations between two widths can be written as
\begin{equation}
\overline{\delta\Gamma_{nS}\,\delta\Gamma_{mS}} 
\, = \, 
a_{cor} \, \delta\Gamma_{nS}^{sys}\,\delta\Gamma_{mS}^{sys}
\,,
\label{gammacorrelation}
\end{equation}
where $\delta\Gamma_{nS}$ is the systematic error in the electronic
width $\Gamma_{nS}$ as given in Tab.~\ref{tab6} and $a_{cor}$ is a
parameter which allow to switch the correlation on and off to check
its impact on the extraction of $M_b$ and $\alpha_s$. During the
fitting procedure $a_{cor}$ is varied between zero (no correlation)
and one (complete positive correlation of all systematic
errors). Collecting all the quantities for which we take experimental
errors into account into the vector 
\begin{equation}
y_i \, = \, \Big\{\,
\Gamma_{1S}, \Gamma_{2S}, \Gamma_{3S}, \Gamma_{4S}, 
\Gamma_{5S}, \Gamma_{6S}, \tilde \alpha_{em}, r_{cont}
\,\Big\}\,,
\qquad 
i=1,\ldots,8\,,
\end{equation}
and using the standard error propagation formulae (see e.g.~\cite{PDG})
the covariance matrix reads
\begin{equation}
S_{nm} \, = \,
\sum\limits_{i,j=1}^8 \,
\frac{\partial P_n^{ex}}{\partial y_i}\bigg|_{\hat y}
\frac{\partial P_m^{ex}}{\partial y_j}\bigg|_{\hat y}
\,V_{i j}
\,,
\label{correlationmatrix}
\end{equation}
where 
\begin{equation}
V_{i j} \, = \,
\left(
\begin{array}{cccccc}
(\delta\Gamma_{1S})^2 
  & \overline{\delta\Gamma_{1S}\,\delta\Gamma_{2S}} 
  & \cdots
  & \overline{\delta\Gamma_{1S}\,\delta\Gamma_{6S}}
  & 0 & 0 \\
\overline{\delta\Gamma_{2S}\,\delta\Gamma_{1S}}
  & (\delta\Gamma_{2S})^2 
  & \cdots
  & \overline{\delta\Gamma_{1S}\,\delta\Gamma_{6S}}
  & 0 & 0 \\
\vdots & \vdots & \ddots & \vdots & \vdots & \vdots \\
\overline{\delta\Gamma_{6S}\,\delta\Gamma_{1S}}
  & \overline{\delta\Gamma_{6S}\,\delta\Gamma_{2S}}
  & \cdots
  & (\delta\Gamma_{2S})^2 
  & 0 & 0 \\
0 & 0 & \cdots & 0 & (\delta\tilde\alpha_{em})^2 & 0 \\
0 & 0 & \cdots & 0 & 0 & (\delta r_{c})^2 
\end{array}
\right)
\,.
\label{covariancematrix}
\end{equation}
The symbol $|_{\hat y}$ indicates that the functions are evaluated
at the corresponding central values.

The fitting procedure if complicated by the fact that the theoretical
uncertainties (coming from the dependence of the theoretical moments
on the renormalization scales $\mu_{\rm soft}$, $\mu_{\rm hard}$ and
$\mu_{\rm fac}$) are much larger than the experimental errors, which are
dominated by the errors in $\Gamma_{1S}$, $\Gamma_{2S}$ and
$\Gamma_{3S}$. Further, while it is reasonable to assume that the
errors in the experimental data can be treated as Gaussian, this is
certainly not the case for the ``uncertainties'' (or better
``freedom'') in the choices of the renormalization scales for which
just a ``reasonable'' window can be given. It would therefore be
inconsistent to
include the theoretical uncertainties into the covariance matrix
$S$. Nevertheless, it is important to have some means to combine both
types of errors, the experimental and the theoretical ones. 
In this work this is realized by scanning all scales over the ranges
given in Eqs.~(\ref{choiceofscales}). We will carry out two kind of
fits. First, we fit for 
$M_b$ and $\alpha_s$ simultaneously  without taking into account any
constraints on $\alpha_s$, i.e.\ ignoring all existing determinations
of the strong coupling (Section~\ref{subsectionunconstraint}), and,
second, we fit for $M_b$ assuming that $\alpha_s$ is a known parameter
i.e.\ taking into account a constraint on $\alpha_s$
(Section~\ref{subsectionconstraint}).

To fit for $M_b$ and $\alpha_s$ simultaneously we
employ a strategy closely related to the one suggested by
Buras~\cite{Buras1} and adopted by the BaBar
collaboration~\cite{BaBar1} as a method to 
extract Cabibbo-Kobayashi-Maskawa matrix elements from various
B-decays. Our strategy consists of the following two steps:
\begin{itemize}
\item[(a)]
We first choose the range over which the renormalization scales
$\mu_{\rm soft}$, $\mu_{\rm hard}$ and $\mu_{\rm fac}$ have to be
scanned individually. For 
convenience we also count the constant $r_c$, the correlation
parameter $a_{cor}$ and the various sets of $n$`s for which the fits
shall be carried out as theoretical parameters. The individual ranges
employed in this work are as follows,
\begin{eqnarray}
1.5\,\mbox{GeV} \, \le & \mu_{\rm soft} & \le \, 3.5\,\mbox{GeV} 
\nonumber\\[1mm]
2.5\,\mbox{GeV} \, \le & \mu_{\rm hard} & \le \, 10\,\mbox{GeV} 
\nonumber\\[1mm]
2.5\,\mbox{GeV} \, \le & \mu_{\rm fac} & \le \, 10\,\mbox{GeV} 
\nonumber\\[1mm]
0.5 \, \le & r_c & \le \, 1.5 
\nonumber\\[1mm]
0 \, \le & a_{cor} & \le \, 1 
\,.
\label{parameterranges}
\end{eqnarray}
The sets of $n$`s for which we perform the fits are
\begin{equation}
\{n\} \, = \,
\{4,5,6,7\}\,,\{7,8,9,10\}\,,\{4,6,8,10\}
\,.
\label{nsets}
\end{equation}
The scanning over the ranges and sets given above is carried out
by using a Monte-Carlo generator.
\item[(b)]
Then, for each set of theoretical parameters 
\begin{equation}
{\cal{M}} \, = \, \{
\mu_{\rm soft}\,,\mu_{\rm hard}\,,\mu_{\rm fac}\,,
r_c\,,a_{cor}\,,\{n\}
\}
\,,
\end{equation}
called a ``model'', we construct the $\chi^2$ function as
describe before and determine the $95\%$ confidence level (CL) contour 
in the $M_b$-$\alpha_s$-plane by calculating the minimum
$\chi^2$, $\chi^2_{min}$, and drawing the contour 
$\chi^2(M_b,\alpha_s)=\chi^2_{min}+6$.
The external envelope of the contours obtained for all models
generated by scan represent the ``overall $95\%$ CL contour'' which we
will refer to as the ``allowed
range for $M_b$ and $\alpha_s$''. It should be mentioned that
we do not impose a $\chi^2$ cut which would
eliminate models for which the probability of $\chi_{min}$ would be
smaller than $5\%$. We will come back to this point in
Section~\ref{sectionresults}.
\end{itemize}
We would like to emphasize that the allowed region for $M_b$ and
$\alpha_s$ obtained by the procedure described above should not be
understood in any statistical sense. In fact, it is quite difficult to
ascribe any accurately defined meaning to the allowed region at all
without reference to the method how it has been
obtained. This is a consequence of the fact that the theoretical
uncertainties, which cannot be apprehended statistically, dominate
over the experimental ones.  
%Nevertheless, we consider the size of the uncertainties on
%$M_b$ and $\alpha_s$ resulting from our method much more reliable than
%the (much smaller) ones coming from the previous analyses~\cite{XXX}.
%A detailed discussion on the previous analyses~\cite{XXX} is
%presented in Section~\ref{sectioncomments}.
%

For the fit for $M_b$ where $\alpha_s$ is assumed to be a known
parameter we treat $\alpha_s$ like the theoretical parameters 
$\mu_{\rm soft}\,,\mu_{\rm hard}\,,\mu_{\rm fac}\,,
r_c\,,a_{cor}$ and $\{n\}$, i.e.\ we also scan over the given range of
$\alpha_s$. The fit for $M_b$ is then carried out in the same way as
for the unconstraint fit described before. The only difference is that
in this case the $95\%$ confidence level ``contour'' for each model is
determined by the equation $\chi^2(M_b)=\chi^2_{min}+4$ because this
method does represent only a one parameter fit. Some more remarks to
this method can be found in Section~\ref{subsectionconstraint}.
\par
\vspace{0.5cm}
\section{Numerical Results and Discussion}
\label{sectionresults}
In this section we present the numerical results for the bottom quark
pole mass $M_b$  gained from fitting
the theoretical moments at NNLO calculated in
Section~\ref{sectioncalculatemoments} to the experimental moments
obtained from experimental data. In
Section~\ref{subsectionunconstraint} we discuss the result if $M_b$
and $\alpha_s$ are fitted simultaneously (``unconstraint fit'') and in
Section~\ref{subsectionconstraint} we present the result for $M_b$ if
$\alpha_s$ is taken as an input (``constraint fit''). 
\subsection{Determination of $M_b$ and $\alpha_s$ without Constraints}
\label{subsectionunconstraint}
\begin{figure}[thb] % fignnlofull
\begin{center}
\leavevmode
\epsfxsize=5cm
\epsffile[220 420 420 550]{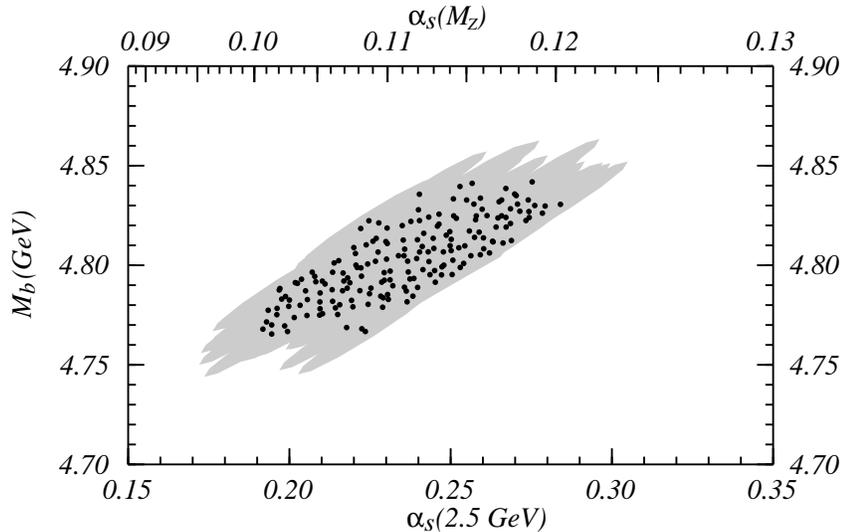}
%
%\centerline{\epsfig{file=got.ps,height=3.5in,width=3.5in}}
%\vspace{10pt}
%
\vskip  4.0cm
 \caption{\label{fignnlofull} 
Result for the allowed region in the $M_b$-$\alpha_s$ plane for the
unconstraint fit based on the theoretical moments at NNLO. The grey
shaded region represents the allowed region. Experimental errors are
included at the $95\%$ CL level. The dots represent point of minimal
$\chi^2$ for a large number of models.
}
%\label{fignnlofull}
 \end{center}
\end{figure}

The result for the allowed region for $M_b$ and $\alpha_s$ when both
parameters are fitted simultaneously and no previous determination of
$\alpha_s$ is taken into account is displayed in
Fig.~\ref{fignnlofull}.
The gray shaded region represents the allowed region in the
$M_b$-$\alpha_s$ plane. To illustrate that the allowed region does not
have any well defined statistical meaning we have also shown the dots
representing the best fits (i.e.\ the points in the $M_b$-$\alpha_s$
plane with the lowest $\chi^2$ value for a large number of models.
In fact, the
region covered by the dots for the best fits is a measure for the size
of the theoretical uncertainties inherent to our result. The latter
uncertainties, which cannot be apprehended statistically, clearly
dominate over the experimental ones, which are contained in the grey
shaded region not covered by any dots. For convenience of the reader
we
have shown the result for $\alpha_s$ at the scale $\mu=2.5$~GeV (lower
frame axis) and $\mu=M_z$ (upper frame axis) where we have used
two-loop running for the strong coupling. From the shape and
orientation of the gray shaded region in Fig.~\ref{fignnlofull} it is
evident that $M_b$ and $\alpha_s$ are positively correlated. This can
be easily understood from the fact that the theoretical moments are
monotonically increasing functions of $\alpha_s$ but monotonically
decreasing functions of $M_b$ (see Tab.~\ref{tab2a}). However, we
refrain from presenting a
numerical value for the correlation because, as already mentioned, the
allowed region for $M_b$ and $\alpha_s$ does not have any statistical
meaning. 

For the bottom quark pole mass and the strong coupling we obtain
\begin{center}
\begin{minipage}{17cm}
\begin{eqnarray}
4.74\,\, \mbox{GeV}\quad  \le & M_b & \le  \quad 4.87\,\,\mbox{GeV}
\,,
\label{nnloMbottom}\\
0.096\quad   \le & \alpha_s(M_z) & \le \quad 0.124
\,,
\label{nnloalphasMz}\\
0.175\quad   \le & \alpha_s(2.5\,\,\mbox{GeV}) & \le \quad 0.308
\,.
\label{nnloalphas25}
\end{eqnarray} 
\end{minipage}\\[0.3cm]
\mbox{(NNLO analysis, $M_b$ and $\alpha_s$ are fitted simultaneously)}
\end{center}
Because the uncertainties for $M_b$ and $\alpha_s$ are not Gaussian we
only present the allowed ranges obtained from Fig.~\ref{fignnlofull}.
We would like to emphasize that in this context the inequality sign
``$\le$'' does not have any mathematical meaning. It is only used to
describe the bounds on $M_b$ and $\alpha_s$ which are obtained from
our fitting procedure. The allowed range for $M_b$, which spans over
$120$~MeV,
can be definitely called a precise determination of the bottom quark
pole mass. The allowed range obtained for the strong coupling, on the
other hand, is consistent with the current world average, but much
wider than the uncertainties of the latter. In addition, most of the
allowed range for $\alpha_s$ is located below the current world
average. Taking the size of allowed ranges for $M_b$ and $\alpha_s$ as
their uncertainty we arrive at
\begin{eqnarray}
\frac{\Delta M_b}{M_b}\qquad & \sim & 2.5\,\,\%
\,,
\label{nnloMbottomdelta}\\[2mm]
\frac{\Delta \alpha_s(M_z)}{\alpha_s(M_z)}\quad & \sim & 25\,\,\%
\,,
\label{nnloalphaMzdelta}\\[2mm]
\frac{\Delta \alpha_s(2.5\,\,\mbox{GeV})}{\alpha_s(2.5\,\,\mbox{GeV})}
& \sim & 50\,\,\% 
\,,
\label{nnloalpha25delta}
\end{eqnarray}
for the relative uncertainties in our determination of $M_b$ and
$\alpha_s$. It is evident that the sum rule based on the large $n$
moments, Eq.~(\ref{momentdef}), is much more sensitive to the
bottom quark mass than to the strong coupling. At least at the present
stage one can certainly conclude that this sum rule does not belong to
most powerful methods to determine $\alpha_s$ as far as precision is
concerned. 

From Eqs.~(\ref{nnloMbottom}) and (\ref{nnloalphasMz}) we can
calculate the value for the running bottom quark mass. Using the
two-loop relation between the pole and running
mass~\cite{Broadhurst1} (see also Ref.~\cite{Kuhn2} and
references therein) and taking into account the correlation
between the pole mass and the strong coupling we get
\begin{eqnarray}
4.09\,\, \mbox{GeV}\quad  \le & m_b(M_{\Upsilon(1S)}/2) & 
  \le  \quad 4.32\,\,\mbox{GeV}
\,, \\
4.17\,\, \mbox{GeV}\quad  \le & m_b(m_b) & 
  \le  \quad 4.35\,\,\mbox{GeV}
\,.
\label{nnlombrunning}
\end{eqnarray} 
This result is in excellent agreement with a recent determination of
the running bottom quark mass obtained from the three-jet rate in
$b\bar b$ events at the CERN $e^+e^-$ LEP experiment
DELPHI~\cite{Rodrigo1,DELPHI1}, $m_b(M_{\Upsilon(1S)}/2)=4.16\pm
0.14$~GeV. The uncertainty in the result for the running quark mass,
Eq.~(\ref{nnlombrunning}), is larger than for our pole mass result,
Eq.~(\ref{nnloMbottom}), because of the correlation between $M_b$ and
$\alpha_s$, which has to be taken into account in the conversion
formula.

We have checked that the allowed region for $M_b$ and $\alpha_s$
presented in Fig.~\ref{fignnlofull}
is insensitive to the particular choices of the scanning ranges for
the renormalization scale $\mu_{\rm hard}$ and the constants $a_{cor}$
and $r_c$, which parameterize the correlation of the experimental data
for the electronic widths and the continuum cross section above the
$B\bar B$ threshold, respectively. However, the results depend on the
choice of the ranges for the soft scale $\mu_{\rm soft}$ and the
factorization scale $\mu_{\rm fac}$. 
\begin{figure}[t!] % fignnlodependence
\begin{center}
\leavevmode
\epsfxsize=4cm
\epsffile[220 420 420 550]{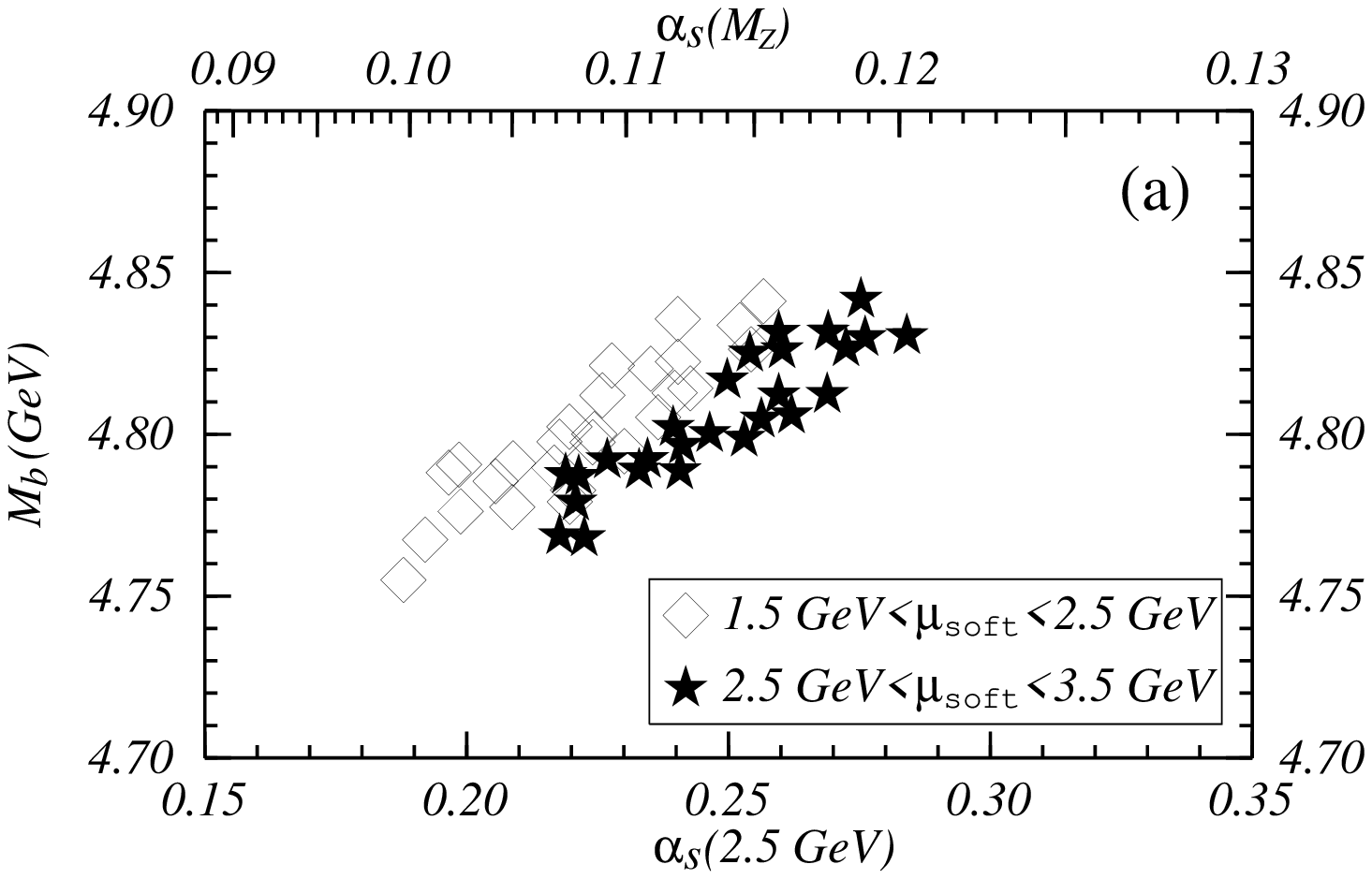}\\
\vskip 3.cm
\epsfxsize=4cm
\leavevmode
\epsffile[220 420 420 550]{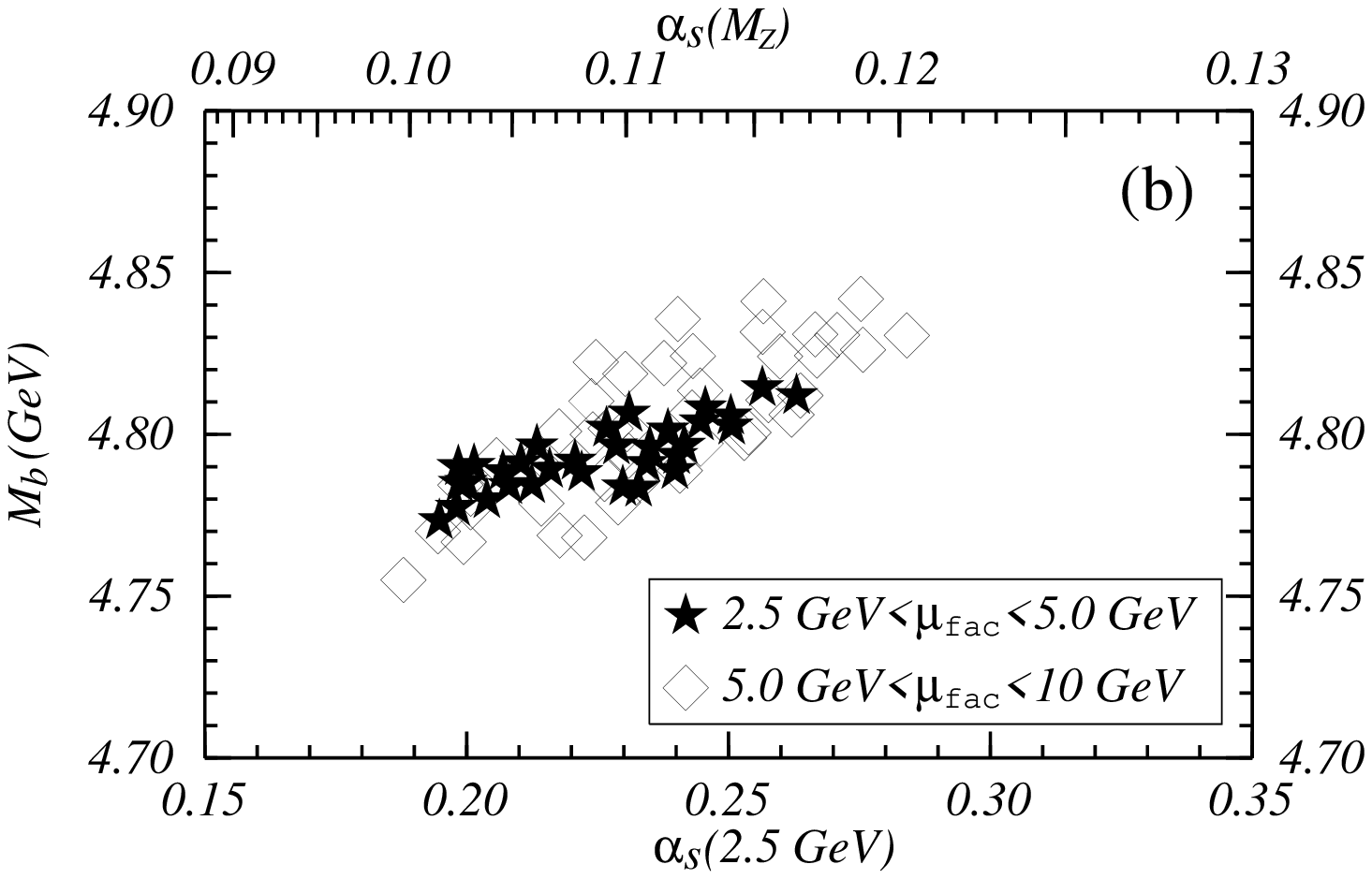}
%
%\centerline{\epsfig{file=got.ps,height=3.5in,width=3.5in}}
%\vspace{10pt}
%
\vskip  3.0cm
 \caption{\label{fignnlodependence} 
Typical distribution of points representing the best fits (a) for
models with $1.5\,\mbox{GeV} \le \mu_{\rm soft} \le 2.5\,\mbox{GeV}$
and $2.5\,\mbox{GeV} \le \mu_{\rm soft} \le 3.5\,\mbox{GeV}$ and (b)
for models with
$2.5\,\mbox{GeV} \le \mu_{\rm fac} \le 5.0\,\mbox{GeV}$ and
$5.0\,\mbox{GeV} \le \mu_{\rm soft} \le 10\,\mbox{GeV}$
based on the theoretical moments at NNLO. The other parameters are
scanned over the ranges given in Eqs.~(\ref{parameterranges}).
}
%\label{fignnlodependence}
 \end{center}
\end{figure}
This dependence is illustrated in Fig.~\ref{fignnlodependence} where
we have displayed points for the best fits (a) for models
with $1.5\,\mbox{GeV} \le \mu_{\rm soft} \le 2.5\,\mbox{GeV}$ and
$2.5\,\mbox{GeV} \le \mu_{\rm soft} \le 3.5\,\mbox{GeV}$ and (b) for
models with $2.5\,\mbox{GeV} \le \mu_{\rm fac} \le 5.0\,\mbox{GeV}$ and
$5.0\,\mbox{GeV} \le \mu_{\rm soft} \le 10\,\mbox{GeV}$ with different
symbols. In both figures the other parameters have been scanned over
the ranges given in Eqs.~(\ref{parameterranges}). From
Fig.~\ref{fignnlodependence}(a) we see that the allowed range for
$M_b$ does not depend significantly on the choice for the soft scale,
whereas the allowed range for $\alpha_s$ tends toward larger values if
the soft scale is larger. Fig.~\ref{fignnlodependence}(b), on the
other hand, shows that the size of the allowed range for $M_b$ could
be reduced if smaller factorization scales would be
chosen. In that case the allowed range for $\alpha_s$ would be only
mildly affected. From this observation it might be tempting to choose
the scanning range for $\mu_{\rm soft}$ at higher scales and for
$\mu_{\rm fac}$ at lower scales because this would lead to a seemingly
more precise determination of $M_b$ and higher values for
$\alpha_s$. However, we take the position that the choice of the
scanning ranges for the renormalization scales should not depend on
such considerations to represent a ``reasonable choice''. In fact, we
consider it inappropriate to tune or ``optimize'' renormalization
scales in some specific way if no good physical reason for that can be
given. In our case the choice for the scanning ranges for the soft scale
was motivated by the fact that it governs the
nonrelativistic correlators for which (at NNLO) the relative
momentum of the bottom quarks (which is of order $M_b\alpha_s$)
represents the only relevant physical scale. Our choice for
factorization scale $\mu_{\rm fac}$, on the other hand, is inspired by
the belief that is can take any value between the relative
momentum of the bottom quarks and the hard scale which is of order the
bottom quark mass (see Section~\ref{sectionbasicidea}). We will come
back to this issue in Section~\ref{sectioncomments}.

It is very interesting to compare the results of our NNLO analysis
presented above to an analogous analysis based on the NLO moments,
i.e.\ ignoring all the NNLO contributions. [See the end of
Section~\ref{sectioncalculatemoments} for a prescription how the NLO
moments can be recovered from the NNLO ones.]
\begin{figure}[thb] % fignlofull
\begin{center}
\leavevmode
\epsfxsize=5cm
\epsffile[220 420 420 550]{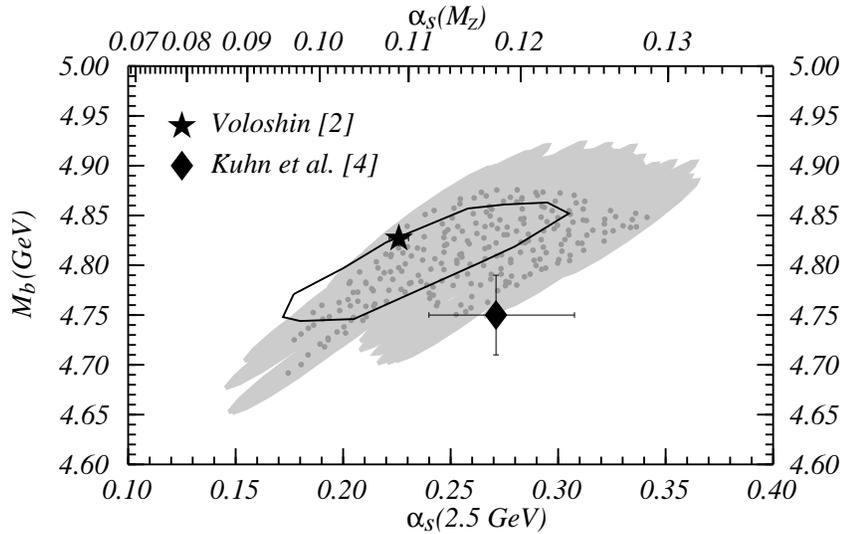}
%
%\centerline{\epsfig{file=got.ps,height=3.5in,width=3.5in}}
%\vspace{10pt}
%
\vskip  5.0cm
 \caption{\label{fignlofull} 
Result for the allowed region in the $M_b$-$\alpha_s$ plane for the
unconstraint fit based on the theoretical moments at NLO. The grey
shaded region represents the allowed region. Experimental errors are
included at the $95\%$ CL level. The dots represent point of minimal
$\chi^2$ for a large number of models. The star and the diamond
represent the results obtained by Voloshin~\cite{Voloshin1} and
K\"uhn {\it et al.}~\cite{Kuhn1}, respectively.
The error-bars quoted by Voloshin are smaller than the symbol used to
display his central value. 
The polygon represents the allowed region obtained from the NNLO
analysis.
}
%\label{fignlofull}
 \end{center}
\end{figure}
The result for the allowed range for $M_b$ and $\alpha_s$ based on the
NLO moments is displayed in Fig.~\ref{fignlofull}. The gray shaded
region and the dots have been obtained in the exactly the same way as
described for the NNLO analysis. For comparison we have also indicated
the allowed region obtained from the NNLO analysis by a
polygon. Evidently the allowed region for $M_b$ and $\alpha_s$ covers
a much larger area for the NLO analysis than for the NNLO one. At NLO
the result for bottom quark pole mass and the strong coupling read
\begin{center}
\begin{minipage}{17cm}
\begin{eqnarray}
4.64\,\, \mbox{GeV}\quad  \le & M_b & \le  \quad 4.92\,\,\mbox{GeV}
\,,
\label{nloMbottom}\\
0.086\quad   \le & \alpha_s(M_z) & \le \quad 0.132 
\,,
\label{nloalphasMz}\\
0.144\quad   \le & \alpha_s(2.5\,\,\mbox{GeV}) & \le \quad 0.368
\,.
\label{nloalphas25}
\end{eqnarray} 
\end{minipage}\\[0.3cm]
\mbox{(NLO analysis, $M_b$ and $\alpha_s$ are fitted simultaneously)}
\end{center}
From Fig,~\ref{fignlofull} and 
Eqs.~(\ref{nloMbottom}) -- (\ref{nloalphas25}) it is evident that the
inclusion of the NNLO contributions of the moments leads to a
considerable improvement upon a pure NLO analysis. We would like to
point out that the uncertainties in $M_b$ and $\alpha_s$ from our NLO
analysis are much larger then the uncertainties quoted by
Voloshin~\cite{Voloshin1} and K\"uhn {\it et al.}~\cite{Kuhn1}.
For comparison we have also displayed
the results from Refs.~\cite{Voloshin1,Kuhn1} in
Fig.~\ref{fignlofull}. Because the theoretical moments used in
Refs.~\cite{Voloshin1,Kuhn1} and the NLO moments used to
generate the allowed region for $M_b$ and $\alpha_s$ displayed in
Fig.~\ref{fignlofull} are equivalent, we consider the small
uncertainties quoted in Refs.~\cite{Voloshin1,Kuhn1} as a consequence
of an inappropriate treatments of the large theoretical uncertainties
inherent to the perturbative calculations of the moments. (See
Section~\ref{sectioncomments} for a more detailed discussion.)
Another way to see that that the NNLO contributions lead to a
considerable improvement is to compare the distributions of best
$\chi^2$ values which are achieved by the models based on NNLO and NLO
moments, respectively. 
\begin{table}[t!]  % tabchi2distribution
\vskip 7mm
\begin{center}
\begin{tabular}{|c||c|c|c|c|c|c|c|c|c|} \hline
$\chi^2_{min}$ & $0-3$ &  $3-6$ & $6-10$ & $10-15$ & $15-20$ & 
                 $20-30$ & $30-50$ & $50-100$ & $100-\infty$ \\ \hline 
NNLO   
 & $28\,\%$ & $17\,\%$  & $16\,\%$ & $22\,\%$ & $8\,\%$ &
   $4\,\%$ & $2\,\%$  & $3\,\%$ & $0\,\%$  \\ \hline
NLO   
 & $0\,\%$ & $0\,\%$  & $0\,\%$ & $0\,\%$ & $0\,\%$ &
   $1\,\%$ & $7\,\%$  & $35\,\%$ & $57\,\%$  \\ \hline
\end{tabular}
\caption{\label{tabchi2distribution} 
Distributions of best $\chi^2$ values for a NNLO and NLO analysis
based on, at each case, 1300 randomly generated models within the
ranges~(\ref{parameterranges}). 
}
\end{center}
\vskip 3mm
\end{table}
In Tab.~\ref{tabchi2distribution} the fraction (in percent) of best
$\chi^2$ values within certain intervals is displayed for the NNLO
and the NLO analysis based on, at each case, 1300 randomly generated
models within the scanning ranges in
Eqs.~\ref{parameterranges}. Whereas for
the NNLO analysis more than $60\,\%$ of the models have a best
$\chi^2$ value below $10$, the bulk of the best $\chi^2$ values for
the NLO analysis is larger than $50$. We would like to emphasize that,
because the uncertainties of the analysis are dominated by theory, the
distributions of  best $\chi^2$ values in
Tab.~\ref{tabchi2distribution} represent only a measure for the
quality of the theoretical expression for the moments, but do not
contain any statistical information. We therefore cannot impose a
$\chi^2$ on the models, let us say, based on an assumed statistical
distribution of $\chi^2$ values. As an example, for two degrees of
freedom and at the $95\,\%$~CL, and assuming a Gaussian distribution
such a $\chi^2$ cut would eliminate all models whose best $\chi^2$
value is larger than $6$. Evidently, in this case, none of the models
based on the NLO moments would survive and we would be forced to
reject, at least, the nonrelativistic expansion up to NLO as a
legitimate tool to calculate the moments from QCD for the sets of
$n$'s considered in this work. 
\subsection{Determination of $M_b$ with Constraints on $\alpha_s$}
\label{subsectionconstraint}
We now carry out the fitting procedure for $M_b$ if $\alpha_s$ is
taken as an input, e.g. from the current world average. At
this point one might be tempted to simply cut out of the gray shaded
region in Fig.~\ref{fignnlofull} the part for which $\alpha_s$ is
located in the preferred range. Due to the sizable correlation between
$M_b$ and $\alpha_s$ this would then lead to a much smaller
uncertainty in 
$M_b$ than given in Eq.~(\ref{nnloMbottom}). However, the naive
procedure just described is not the correct way to account for a
constraint on $\alpha_s$. This comes from the fact that for the
unconstraint fit performed in Section~\ref{subsectionunconstraint} the
strong coupling is essentially a function of the model parameters 
${\cal{M}}=\{\mu_{\rm soft},\mu_{\rm hard},\mu_{\rm fac},
r_c,a_{cor},\{n\}\}$, i.e.\ $\alpha_s$ is not independent of the choice
for ${\cal{M}}$. If $\alpha_s$ is taken as an input, however, we have
to treat $\alpha_s$ and ${\cal{M}}$ as independent, because we have to
be able to freely assign values to them. Thus, if we take $\alpha_s$
from the 
world average, we can expect that for a number of models the allowed
range for $M_b$ will be located outside the gray shaded region in
Fig.~\ref{fignnlofull}. As a consequence, the constraint fit will in
general lead to larger uncertainties in $M_b$ than the unconstraint
one. In addition, due to the positive correlation between $M_b$ and
$\alpha_s$ we can also expect that the result for the allowed region
of $M_b$ for the constraint fit will be located at slightly larger
masses than for the unconstraint fit. 

We would like to point out that there are many ways to account for a
constraint on $\alpha_s$ which all might lead to slightly different
results. In this work we account for a constraint on $\alpha_s$ by
treating it in the same way as the parameters in ${\cal{M}}$, i.e.\ we
also scan over the preferred range of $\alpha_s$. The allowed range of
$M_b$ is then obtained in the same way as for the unconstraint fit
carried out in Section~\ref{subsectionunconstraint} with the
difference that now only a one-parameter fit is performed (see also
Section~\ref{sectionfitting}). It should be noted
that this method treats $\alpha_s$ entirely on the same footing as the
theoretical parameters in ${\cal{M}}$, i.e.\ the uncertainties on
$\alpha_s$ are not taken into account as Gaussian or statistical
errors.
\begin{figure}[t!] % figasconstraint
\begin{center}
\leavevmode
\epsfxsize=5cm
\epsffile[220 420 420 550]{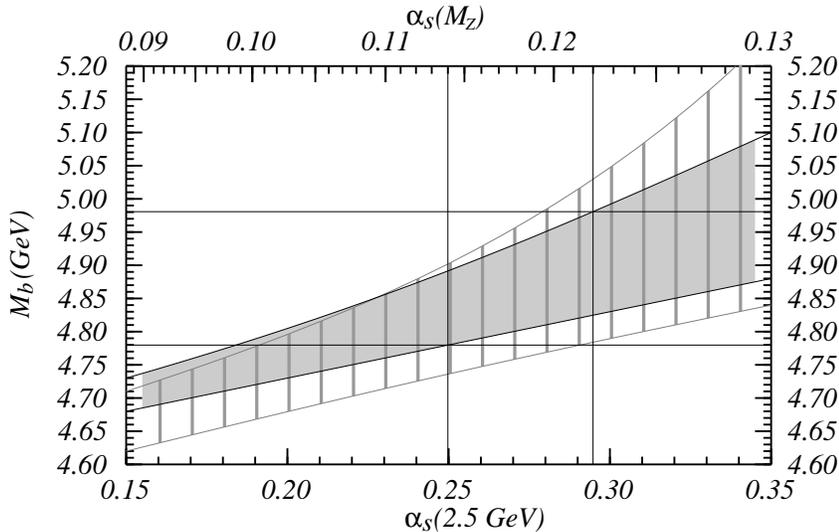}
%
%\centerline{\epsfig{file=got.ps,height=3.5in,width=3.5in}}
%\vspace{10pt}
%
\vskip  4.5cm
 \caption{\label{figasconstraint} 
Result for the allowed $M_b$ values for a given value of $\alpha_s$.
The grey shaded region corresponds to the allowed ranges for the NNLO
analysis and the striped region for the NLO analysis. Experimental
errors are included at the $95\%$ CL level. It is illustrated how
allowed range for $M_b$ at NNLO is obtained if
$0.114\le\alpha_s(M_z)\le 0.122$ is taken as an input.
}
%\label{fignlofull}
 \end{center}
\end{figure}
In Fig.~\ref{figasconstraint} the allowed range for $M_b$ based on the
NNLO moments is presented as a function of $\alpha_s$. For each given
value for $\alpha_s$ the allowed range for $M_b$, which is obtained by
scanning all the parameters in ${\cal{M}}$ over the
ranges~(\ref{parameterranges}), is the projection of the gray shaded
region onto the $M_b$ axis. If a region for $\alpha_s$ is given the
allowed range for $M_b$ is obtained by projecting the gray shaded
region for all the $\alpha_s$ valued in the preferred region onto the
$M_b$ axis. As an example which is also illustrated in
Fig.~\ref{figasconstraint}, staring from the world average for
$\alpha_s$ as given by Stirling~\cite{Stirling1}, $0.114 \le
\alpha_s(M_z) \le 0.122$, we arrive at
\begin{center}
\begin{minipage}{17cm}
\begin{eqnarray}
4.78\,\, \mbox{GeV}\quad  \le & M_b & \le  \quad 4.98\,\,\mbox{GeV}
\label{nnloasconstraintMbottom}
\end{eqnarray} 
\end{minipage}\\[0.3cm]
\mbox{(NNLO analysis, $\alpha_s(M_z)$ taken from the world
average~\cite{Stirling1})}
\end{center}
for the bottom quark pole mass.
The result is consistent with Eq.~(\ref{nnloMbottom}) obtained from
the unconstraint fit. However, as expected, the allowed range for
$M_b$ is wider and, in addition, located at slightly larger masses.
In fact, the uncertainty on $M_b$
for the constraint for is almost a factor of two larger.
We have checked that the result for $M_b$ is insensitive to the
particular choice of the scanning ranges for $\mu_{\rm hard}$,
$\mu_{\rm fac}$, $a_{cor}$ and $r_c$. However, the bottom quark mass
tends toward lower values if $\mu_{\rm soft}$ is chosen larger. We
have also displayed the result for the NLO analysis in
Fig.~\ref{figasconstraint} as the
striped area. As for the unconstraint fit the inclusion of the NNLO
contributions to the moments leads to a smaller uncertainty for $M_b$,
although the improvement is not as dramatic.
We want to mention that the larger uncertainty for $M_b$ obtained from
the constraint fit is partly a consequence of the fact that our
fitting procedure does not treat the error on $\alpha_s$ as Gaussian
or statistical. Therefore one might argue that the uncertainties in
Eq.~(\ref{nnloasconstraintMbottom}) are too conservative. However, from
the way how a world average is gained, it is certain that the error of
$\alpha_s$ constrains a sizable systematic contribution. Because an
accurate quantitative description of such a systematic error is quite
difficult, we take the position that the error on $\alpha_s$ should be
treated in a conservative way.

Using the result in Eq.~(\ref{nnloasconstraintMbottom}) and the
two-loop relation between the running and the pole
mass~\cite{Broadhurst1} we obtain
\begin{eqnarray}
4.08\,\, \mbox{GeV}\quad  \le & m_b(M_{\Upsilon(1S)}/2) & 
  \le  \quad 4.28\,\,\mbox{GeV}
\,,\\
4.16\,\, \mbox{GeV}\quad  \le & m_b(m_b) & 
  \le  \quad 4.33\,\,\mbox{GeV}
\,.
\label{nnlombrunning2}
\end{eqnarray} 
for the running bottom quark mass. It is remarkable that this result
and the result for the running quark mass based on the unconstraint
fit, Eq.~(\ref{nnlombrunning}), are almost identical.
\par
\vspace{0.5cm}
\section{Comments on Previous Analyses}
\label{sectioncomments}
In the past few years there have been three previous analyses by
Voloshin~\cite{Voloshin1}, Jamin and Pich~\cite{Jamin1} and K\"uhn,
Penin and Pivovarov~\cite{Kuhn1} where the bottom quark pole mass and
the strong coupling have been extracted from data on the $\Upsilon$
mesons and using the same sum rule as in our analysis. We would like
to emphasize that in Refs.~\cite{Voloshin1,Jamin1,Kuhn1} no
consistent determination of NNLO corrections has been carried out and
that the results by Voloshin
\begin{equation}
M_b = 4.827\pm 0.007\,\,\mbox{GeV}\,,
\qquad
\alpha_s(M_z) \, = \, 0.109\pm0.001\,,
\qquad\mbox{(Voloshin)}
\label{resultVoloshin}
\end{equation}
Jamin and Pich (JP)
\begin{equation}
M_b = 4.60\pm 0.02\,\,\mbox{GeV}\,,
\qquad
\alpha_s(M_z) \, = \, 0.119\pm0.008\,,
\qquad\mbox{(Jamin, Pich)}
\label{resultJP}
\end{equation}
and K\"uhn, Penin and Pivovarov (KPP)
\begin{equation}
M_b = 4.75\pm 0.04\,\,\mbox{GeV}\,
\qquad
\alpha_s(M_z) \, = \, 0.118\pm0.006\,,
\qquad\mbox{(K\"uhn {\it et al.})}
\label{resultKPP}
\end{equation}
are contradictory to each other and partly to our own results. In
particular, although no NNLO contributions have been included, all
results in Refs.~\cite{Voloshin1,Jamin1,Kuhn1} are claimed to have
much smaller
uncertainties than any of the results obtained in our analyses. In
this section we will explain the origin of those discrepancies and
give some comments on the methods used in
Refs.~\cite{Voloshin1,Jamin1,Kuhn1} from the
point of view of the strategies followed in this work.
To organize the discussion we will analyze the methods used in
Refs.~\cite{Voloshin1,Jamin1,Kuhn1} with respect to three aspects: (i)
theoretical
expression for the moments, (ii) optimization and tuning of the
perturbative series for the moments and (iii) fitting procedure and
error analysis. Because the theoretical uncertainties in the
determination of $M_b$ and $\alpha_s$ are much larger than the
experimental ones, we will neither
focus on the treatment of experimental errors nor on the formulae
used for the experimental moments. Compared to the effects caused by
using different methods to handle the theoretical uncertainties, the
differences in the treatment of the experimental side of the analysis
represent only a minor issue. We also would like to mention that in
the analyses of Voloshin, JP and KPP moments with $n$ as large as $20$
were used. According to the estimates given in
Section~\ref{sectionbasicidea} this means that the effective smearing
range contained in those moments is already of the same size as or
even smaller than $\Lambda_{\mbox{\tiny QCD}}$. This leads to a an additional
source of systematic theoretical errors in the results of Voloshin, JP
and KPP. We have checked, however, that using moments with $10\le
n\le 20$ causes only shifts in the results for $M_b$ (and
$\alpha_s$) which are small compared to the size of the theoretical
uncertainties at the NLO level as we have estimated them in our
analysis. Therefore we will not raise this issue in the following
discussion. For a NNLO analysis, however, where the uncertainties in
$M_b$ are shown to be smaller, the use of values of $n$ which are too
large is an important issue and can lead to considerable errors.\\[5mm]
{\bf Theoretical expressions for the moments:}\\[2mm]
Voloshin's moments are identical to ours at the NLO level.\\[1mm]
The moments used by KPP have been calculated in the same way as
Voloshin's (and ours at the NLO level) with the difference that the
dispersion relation in Eq.~(\ref{momentcrosssectionrelation}) has been
performed numerically in terms of it covariant form, i.e.\ without 
using the asymptotic expansion~(\ref{integrationmeasure}) and the
inverse Laplace transform. We have checked that for the values of
$n$`s considered by Voloshin, KPP and us the difference between both
approaches is negligible. Thus, the moments used by KPP are
equivalent to Voloshin's and ours at the NLO level.\\[1mm]
The moments by JP, on the other hand, were obtained from the Born,
one-loop and two-loop expressions for $R(e^+e^-\to b\bar b)$
supplemented by a resummation of LO Coulomb singularities in form of
the Sommerfeld factor (see
Eq.~(\ref{Rthreshnonrelativistic})). Further, the one-loop corrections
to the Coulomb potential have been implemented by inserting them
directly into the Sommerfeld factor, i.e.\ without using
time-independent perturbation theory. For the dispersion
integration~(\ref{momentcrosssectionrelation}) JP have only taken into
account c.m.\ energies above the threshold point ($s > 4M_b^2$). We
disagree with the moments used by JP in two major points. Most
important, JP did not take into account the bound state poles of the
cross section $R$, which are located below the threshold point ($s<
4M_b^2$). We have demonstrated in Section~\ref{sectionproperties} that
the bound states represent the dominant contribution to the moments
for large values of $n$ (see Tab.~\ref{tab4}). Thus the moments used by
JP are far too small which causes the bottom quark pole mass obtained
from the fits to be too low.\footnote{The same conclusion has been
drawn in Ref.~\cite{Kuhn1}.}
In fact, one can easily see that omitting
the bound state poles for large values of $n$ will always lead to a
bottom quark pole mass $M_b \le M_{\Upsilon(1S)}/2 \approx 4.7$~GeV
regardless whether $\alpha_s$ is determined from the fit or taken as
an input. This explains why the value for $M_b$ in the analysis by JP
is significantly smaller than in the analyses by Voloshin, KPP and
us. In addition, we do not think that the effects of the running of
the Coupling governing the Coulomb potential have been treated
properly. JP simply inserted the one-loop corrections to the Coulomb
potential into the Sommerfeld factor. Whereas this legitimate for the
non-logarithmic corrections, it is not for the logarithmic ones
because the effects arising from virtual momenta below and above the
scale $\sim M_b \alpha_s$ are not taken into account correctly. This
can only be achieved by using time-independent perturbation theory (or
by solving 
the Schr\"odinger equation exactly). We therefore conclude that the
results obtained by JP contain large systematic theoretical errors
which are by far larger than indicated by their error analysis. That
the value for $\alpha_s$ obtained by JP still seems reasonable is a
consequence of the fact that the moments are much less sensitive to
$\alpha_s$ than to $M_b$. \\[5mm]
{\bf Optimization and Tuning:}\\[2mm]
We have shown in Section~\ref{sectionproperties} that the perturbative
corrections to the moments and the resulting scale dependences are
quite large. This behavior is particularly obvious at the NNLO
level. However, already at NLO the corrections are uncomfortably
large. In our analysis this feature has been fully taken into account
during our fitting procedure. In fact, it is the main source of
theoretical uncertainties in our results. In the analyses by Voloshin
and KPP, however, the perturbative expansion for the theoretical
moments has been tuned to improve the convergence. \\[1mm]
In Voloshin's work, at each value of $n$ the soft scale $\mu_{\rm
soft}$ has been fixed such that the NLO corrections caused by
$V^{(1)}_c$, Eq.~(\ref{Vrunning1loop}), vanish exactly and the hard
scale $\mu_{\rm hard}$ has been fixed to the BLM
scale~\cite{Brodsky1}. Thus, Voloshin has eliminated the scale
dependences of the moments. We would like to
emphasize that we consider Voloshin's prescription as one possible
choice for the renormalization scales, which essentially corresponds
to selecting one single model out of the range of models used in our
analysis. We have shown in Section~\ref{sectionresults} (see
e.g.\ Fig.~\ref{fignnlofull})  that the
results for $M_b$ and $\alpha_s$ depend significantly on such a
choice. Because we think that no argument can be found why
Voloshin's choice should be better than others, we have the position
that a scan over all ``reasonable'' models should be carried
out. Because Voloshin has not carried out such a scan we consider the
theoretical uncertainties quoted in his analysis as largely
underestimated. \\[1mm]
In the analysis by KPP, at each value of $n$ a non-logarithmic piece
of $V^{(1)}_c$ has been absorbed into the LO nonrelativistic
Green function, Eq.~(\ref{CoulombGreenfunctionregularized}), such that
the NLO corrections caused by the non-absorbed piece (calculated via
first order time-independent perturbation theory) vanish. This
optimization is quite similar to Voloshin's but leaves the soft scale
unfixed. It should be mentioned that KPP have explicitly identified
soft and hard
scale which has eliminated the possibility to vary both scales
independently. This reveals why the uncertainties quoted by KPP are much
larger than Voloshin's, and partly explains why they are still much
smaller than the uncertainties obtained from our NLO analysis
where no optimization has been performed. (See
Fig.~\ref{fignlofull} for a graphical comparison).\\[1mm]
JP have not carried out any optimization. However, due to their way to
calculate the moments starting from the expressions of the covariant
multi-loop expressions for the cross section, JP implicitly
identified soft and hard scale.\\[5mm]
{\bf Fitting procedure and error estimate:}\\[2mm]
In the analysis by Voloshin and KPP a two parameter fit was carried
out to obtain $M_b$ and $\alpha_s$ for the sets $\{n\}=
\{8,12,16,20\}$ and $\{10,12,14,16,18,20\}$, respectively. Thus, the
results obtained by Voloshin and KPP should be compared with the
results of our unconstraint fit presented in
Section~\ref{subsectionunconstraint}. Because Voloshin has eliminated 
all scale dependences, he has
estimated the theoretical uncertainties in his analysis using the
assumption that the NNLO corrections to the $n$'the moments can be
parameterized by a global factor $(1+c/n)$ where $c$ is a number of
order one. The size of the uncertainties was obtained from the
variation of the best fits for $M_b$ and $\alpha_s$ if $c$ is first
fixed to zero and then obtained from a three parameter fit. 
The theoretical uncertainties gained by this method have been of the
same size as the (small) experimental errors. We have
shown in Section~\ref{sectionproperties} that the NNLO contributions
to the moments have an entirely different structure (large size,
growing with $n$, tremendous dependence on the soft scale) and cannot
be accounted for by the global factor $(1+c/n)$. Thus, Voloshin's
method to estimate the theoretical error is not capable to account for
the true size of theoretical uncertainties inherent to the
perturbative calculation of the moments.\\[1mm]
In the analysis by JP the theoretical uncertainties for $M_b$ and
$\alpha_s$ were essentially obtained from the variation of the best
fit for $M_b$ and $\alpha_s$ (for fixed $\mu_{\rm soft}=\mu_{\rm
hard}=M_b$) when the two-loop corrections to the Coulomb potential,
$V^{(2)}_c$, are included and when the two-loop contributions to the
high energy cross section are removed. No additional uncertainties
(e.g.\ from the renormalization scale dependence) have been taken into
account based on the argument that this would lead to a double
counting of theoretical uncertainties. We disagree with this statement
because the effects of the inclusion or removal of the two-loop
corrections to the high energy cross section or the Coulomb potential
certainly depends on the value of the other parameters (like the
renormalization scales). This and the fact that JP have
neglected the bound state poles, which are the dominant source of large
corrections to the moments (and their scale dependence) for large
values of $n$, have lead to an underestimate of the theoretical
uncertainties (besides the large systematic errors mentioned
above).\\[1mm]
KPP, finally, have determined $M_b$ and $\alpha_s$
separately. For the determination of $M_b$ $\alpha_s(M_z)=0.118$ was
taken as a fixed input. Thus, the result for $M_b$ by KPP should be
compared to the results of our constraint fit presented in
Section~\ref{subsectionconstraint}. 
The method used by KPP to obtain $M_b$ was
based on solving the equation $P_n^{th} = P_n^{ex}$ for $M_b$ while
$n$ and all the other parameters are fixed to specific values. The
mean value and the uncertainty for $M_b$ has then been gained by
calculating the mean and observing the spread of $M_b$ values when
this procedure was carried out, first, for $\mu_{\rm soft}=\mu_{\rm
hard} = M_b$ and $n=10,12,\ldots,20$ and, second, for fixed $n=14$ and
$1.2\,\mbox{GeV}\le\mu_{\rm soft}=\mu_{\rm hard}\le M_b$. This
procedure effectively scans over some fraction of the range of models
used in our fitting procedure but misses e.g. models with $n\neq 14$
and $\mu_{\rm soft}\le M_b$. This the main reason why the
uncertainties quoted by KPP are much smaller than in our NLO
analysis.\\[2mm]

From the discussion presented above we come to the following final
conclusion about the results by Voloshin, JP and KPP in comparison to
our own analysis: The theoretical
moments calculated by Voloshin and KPP are equivalent to our NLO
moments. We therefore consider the results obtained by Voloshin and
KPP consistent with our own results at the NLO level (see
Fig.~\ref{fignlofull} and \ref{figasconstraint}). However, the
theoretical uncertainties are underestimated in both analyses, which
leads to the apparent contradiction between the results by Voloshin
and KPP. In view of the error analysis performed in our analysis,
where we tried to impose as less bias as possible,
the results by Voloshin and KPP are perfectly
consistent. The apparent contradiction essentially corresponds to a
disjunct (and from our point of view biased) choice of models used for
the fitting procedure in both analyses. We want to emphasize that the
choice by Voloshin is not less plausible than the one made by KPP
which illustrates the need that the whole range of models must be
scanned in the fitting procedure. The comparison between the results of
our analysis, where such a complete scan has been performed, and
the results obtained by Voloshin and KPP makes this obvious.
The theoretical moments determined by JP, on the other hand, do not
take into account the bound state poles which represent the dominant
contributions to the moments for large values of $n$. As a consequence
the $M_b$ value obtained by JP is too small and has to be considered
inconsistent with the results by Voloshin, KPP and us and, in
particular, with the nonrelativistic expansion of QCD, where the bound
state poles are predicted. That the value for $\alpha_s$ obtained by
JP still seems reasonable is a consequence of the fact that the
moments are much less sensitive to $\alpha_s$ than to $M_b$.
 
While this paper was in its final stages there appeared a letter by
Penin and Pivovarov (PP)~\cite{Penin1} where the NNLO corrections to
the large $n$ moments have also been included in the sum rule
determination of the bottom quark pole mass. The formulae for the
moments used by PP were based on previous results for the
cross section published in~\cite{Hoang1,Hoang2,Hoang3,Melnikov1} and
are therefore conceptually equivalent to ours.
For the bottom quark pole mass PP quote the result
$M_b=4.78\pm 0.04$~GeV. The result is consistent with ours.
The uncertainty, however, is smaller and the
allowed range for $M_b$ is somewhat lower than for our NNLO results. To
obtain this apparently more precise result PP have used the same
methods as in Ref.~\cite{Kuhn1}
(which we have already criticized above) with the difference that all
the scales (including the factorization scale) were varied in the
range $M_b\pm 1$~GeV. We consider this
range too high for the soft scale. Further, there is a sign
error in the $C_A C_F$ piece of the ${\cal{O}}(\alpha_s^2)$
short-distance coefficient in Eq.~(2) of~\cite{Penin1}, which is not
contained in the original publication~\cite{Hoang3}, and in
Eqs.~(\ref{kappadef}) and (\ref{c12short}) of this paper.
These issues and the fact the PP used
values of $n$ for the moments between $10$ and $20$, which we consider
too large for a NNLO analysis, are the main reasons why the result by PP
is located at lower masses.
Further, we would like to point out two mistakes in~\cite{Penin1}
which, however, do not affect the results for $M_b$. PP state that the
knowledge of the nonrelativistic correlators is sufficient to
completely determine the vacuum polarization function,
Eq.~(\ref{vacpoldef}), in the threshold regime. This is not true
due to UV-divergences in the real parts of the correlators which still
have to be removed by an additional matching calculation. (See 
the discussion at the end of Section~\ref{subsectionmatching}.) PP
state that the factorization scale $\mu_{\rm fac}$ in the
${\cal{O}}(\alpha_s^2)$ short-distance constant and in the correlators
(Eqs.~(2), (5) and (6) in~\cite{Penin1})) is defined in the
$\overline{\mbox{MS}}$ scheme. This statement is not true. The
formulae used by PP were taken from Ref.~\cite{Hoang3}
where the factorization scale is defined in a cutoff scheme. The
corresponding results in the $\overline{\mbox{MS}}$ scheme can be
obtained by an additional redefinition of the factorization scale
$\mu_{\rm fac}$. (See the discussion following
Eq.~(\ref{CoulombGreenfunctionregularized}).)
\par
\vspace{0.5cm}
\section{Conclusions and Outlook}
\label{sectionconclusions}
Based on the argument of global duality and causality one can
relate the  derivatives of the vacuum correlator of two
bottom-antibottom vector currents at zero momentum transfer to an
integral over the total cross section of the production of hadrons
containing a bottom and an antibottom quark in electron-positron
annihilation
\begin{equation}
P_n \, \equiv \,
\frac{4\,\pi^2\,Q_b^2}{n!\,q^2}\,
\bigg(\frac{d}{d q^2}\bigg)^n\,\Pi_\mu^{\,\,\,\mu}(q)\bigg|_{q^2=0}
\, = \,
\int \frac{d s}{s^{n+1}}\,R^{b\bar b}(s)
\,,
\label{duality}
\end{equation}
It is therefore possible to relate a theoretical calculation of the
moments $P_n$ to experimental data for the cross section $R(e^+e^-\to
\mbox{``$b\bar b$ hadrons''})$. The limit of large values of $n$ is of
special interest for relation~(\ref{duality}) because in this limit
the high energy contributions are suppressed. For the theoretical side
this means that the bottom-antibottom pair can be treated
nonrelativistically, and for the experimental side that data for the
production of $\Upsilon$ mesons are already sufficient to saturate
relation~(\ref{duality}). The requirement that the effective range of
integration is larger than $\Lambda_{\mbox{\tiny QCD}}$~\cite{Poggio1}
imposes an upper bound on the values of $n$ for which a perturbative
calculation of the moments can be trusted. Due to the large size of
the bottom quark mass of order five GeV we are in the fortunate
situation that a range of values of $n$ can be found for which the
$b\bar b$ system can be treated nonrelativistically and, at the same
time, the range of integration is still broad enough compared to
$\Lambda_{\mbox{\tiny QCD}}$. We have identified this range of
``large values of $n$'' as $4\lsim
n \lsim 10$. In this work we have used the arguments just given to
determine the bottom quark pole mass $M_b$ and the strong coupling
$\alpha_s$ in the $\overline{\mbox{MS}}$ scheme from experimental data
on the $\Upsilon$ masses and electronic decay widths.

The aim of this paper was twofold:

\noindent
1.) {\bf Calculation of NNLO corrections:} The complete set of NNLO
corrections in relation~(\ref{duality}) for large values of $n$ has
been calculated. This includes corrections to the expressions in the
nonrelativistic limit or order $\alpha_s^2$, $\alpha_s\,v$ and $v^2$,
where $v$ is the velocity of the bottom quarks in the c.m.\ frame. The
conceptual difficulty in those calculations is that the relativistic
corrections, e.g.\ from the kinetic energy or from higher order
interactions like the Darwin or the Breit-Fermi potential lead to
ultra-violet divergent integrations. We have used the concept of
effective field theories formulated in NRQCD~\cite{Caswell1,Bodwin1}
to deal with
this problem. In NRQCD the latter divergences appear as a natural
consequence of the existence of higher dimensional operators which
lead to the renormalization of lower dimensional ones. The exact form
of the renormalization constants is obtained through matching to full
QCD. This automatically provides a separation of all relevant effects
into short-distance (contained in the renormalization constants) and
long-distance ones (contained in matrix elements). In our case this
leads to an expression for the moments (and the cross section
$R(e^+e^-\to\mbox{``$b\bar b$ hadrons''})$ which is a sum of terms
each of which consists of a nonrelativistic current correlator
multiplied with a
short-distance factor (see Eqs.~(\ref{crosssectionexpansion}) and
(\ref{Pntheoryfinal})). For the leading term in this series we have
performed matching at the two-loop level. 
Although the NNLO contributions are quite large, they lead to a
considerable reduction of the theoretical uncertainties in the
extraction of $M_b$. 

\noindent
2.) {\bf Conservative approach for the error estimates:} The
uncertainties in the determination of $M_b$ (and $\alpha_s$) based on
sum rule~(\ref{duality}) are dominated by theory, in particular, by
the remaining large renormalization scale dependences of the
theoretical moments. These scale dependences are caused by large
coefficients which arise in the perturbative calculation of the
corrections to the moments in the nonrelativistic regime. In contrast
to statistical errors, which can be treated in a standardized way, it
is not an easy task to properly account for theoretical uncertainties,
in particular, if a model-independent (in the framework of QCD)
analysis is intended. In fact, in an analysis where theoretical
uncertainties dominate results may easily become biased depending on
personal preferences. For the case of the determination of $M_b$ 
from sum rule~(\ref{duality}) this has lead to the
paradoxical situation that in two earlier
publications~\cite{Voloshin1,Kuhn1}
contradictory results were obtained although equivalent
theoretical expressions for the moments were used. In this paper it
was attempted to include as less personal preference into the analysis
as possible by scanning all theoretical parameters independently over
{\it reasonably large} ranges which were in size and location
motivated from {\it general} considerations. For each set of
parameters, called a ``model'', a standard statistical fitting
procedure was carried out using the method of least squares to
calculate a $95\%$ CL contour. The external envelope of the contours
obtained for all the scanned models was then taken as the ``overall
allowed range'' which, we want to emphasize, does not have any
well defined statistical meaning due to the dominance of theoretical
uncertainties. This makes the scanning method more conservative (and
in our opinion also more honest) than the methods used in
Refs.~\cite{Voloshin1,Kuhn1}. In addition, the scanning method has the
advantage that it automatically accounts for non-linear dependences on
the theoretical parameters and prevents by construction the
Gaussian-like treatment of theoretical uncertainties. Of course, the
results presented in this work are not completely free from personal
preferences either because of the choice of the ranges used for the
scanning. 

In this paper we have performed two different analyses based on the
scanning method and including the new NNLO corrections. First, $M_b$
and $\alpha_s$ were determined simultaneously using the least squares
method for two parameters. We have obtained 
\begin{center}
\begin{minipage}{17cm}
\begin{eqnarray}
4.74\,\, \mbox{GeV}\quad  \le & M_b & \le  \quad 4.87\,\,\mbox{GeV}
\,,
\label{nnloMbottom2}\\
4.09\,\, \mbox{GeV}\quad  \le & m_b(M_{\Upsilon(1S)}/2) & 
  \le  \quad 4.32\,\,\mbox{GeV}
\,,
\label{nnlombrunning3}\\
0.096\quad   \le & \alpha_s(M_z) & \le \quad 0.124
\label{nnloalphasMz2}
\end{eqnarray} 
\end{minipage}\\[0.3cm]
\mbox{(NNLO analysis, $M_b$ and $\alpha_s$ are fitted simultaneously).}
\end{center}
The corresponding result using the NLO expressions for the moments
yielded considerably larger uncertainties (see Figs.~\ref{fignnlofull}
and \ref{fignlofull} and
Eqs.~(\ref{nnloMbottom})--(\ref{nnloalphas25}) and
(\ref{nloMbottom})--(\ref{nloalphas25})). The results show that
relation~(\ref{duality}) allows for a much more precise determination
of the bottom quark mass than for the strong coupling. Second, $M_b$
was determined using the least squares method for one parameter and
taking $\alpha_s$ as a known parameter. We have obtained
\begin{center}
\begin{minipage}{17cm}
\begin{eqnarray}
4.78\,\, \mbox{GeV}\quad  \le & M_b & \le  \quad 4.98\,\,\mbox{GeV}
\,,
\\
4.08\,\, \mbox{GeV}\quad  \le & m_b(M_{\Upsilon(1S)}/2) & 
  \le  \quad 4.28\,\,\mbox{GeV}
\label{nnlombrunning4}
\label{nnloasconstraintMbottom2}
\end{eqnarray} 
\end{minipage}\\[0.3cm]
\mbox{(NNLO analysis, $0.114\le\alpha_s(M_z)\le 0.122$).}
\end{center}
As for the first analysis the NNLO contributions to the theoretical
moments lead to a reduction of the uncertainties (see
Fig.~\ref{figasconstraint}). In our opinion, the sum
rule~(\ref{duality}) can be regarded as a quite precise tool to
determine the bottom quark mass. For the determination of the
strong coupling we are by far less optimistic.

In the past few years there have been three previous analyses by
Voloshin~\cite{Voloshin1}, Jamin and Pich~\cite{Jamin1} and K\"uhn
{\it et al.}~\cite{Kuhn1} where the bottom quark pole mass 
has been determined from experimental data for the masses and the
electronic decay widths for the $\Upsilon$ mesons using the sum
rule~(\ref{duality}). The results
obtained in those three analyses are contradictory to each other, and,
although NNLO corrections have essentially not been included, quote
uncertainties smaller than in our own NNLO analysis. The results
obtained by Voloshin and K\"uhn {\it et al.} are based on moments
which are equivalent to ours at the NLO level. In view of the
uncertainties for $M_b$ (and $\alpha_s$) obtained from our analysis at
NLO, the results by Voloshin and K\"uhn {\it et al.} can therefore be
regarded as consistent with each other (and us), see
Fig.~\ref{fignlofull}. 
The small uncertainties quoted by Voloshin and K\"uhn {\it et al.}
come from too tight, model-like bounds imposed on the theoretical
parameters. The results obtained by Jamin and Pich, on the
other hand, contain a large systematic error due to the negligence of
the bound state contributions in the moments. We consider the result
by Jamin and Pich inconsistent with those by Voloshin,  K\"uhn {\it et
al.} and us, and in particular with the nonrelativistic expansion of
QCD.

It is quite interesting to ask whether and how the results determined
in this work can be further improved in order to arrive at even
smaller uncertainties for the bottom quark pole mass or the strong
coupling. From the technical point of view the answer would simply by
to calculate the NNNLO contributions in relation~(\ref{duality}). Such
a task, however, is highly nontrivial. Apart from the fact that a
three-loop matching would have to be performed also the NNNLO effects
in the bottom-antibottom interactions would have to considered. This
would require a consistent treatment of retardation effects which are
caused by the non-instantaneous exchange of gluons and, as a
prerequisite, a better understanding of higher order Fock
bottom-antibottom-gluon states. In principle a calculation to
determine these effects would be the QCD analogue of the determination
of the Lamb shift contributions to the positronium wave function. So
far no technical instruments have been developed yet to immediately
tackle this challenging problem. We further believe that it is
unlikely that this goal can be achieved entirely in the framework of
perturbation theory because it involves also the bound state energy
$\sim M_b\,v^2 \sim M_b\,\alpha_s^2$ as a relevant scale. For the
bottom quark this scale is already of the same size as the typical
hadronization scale $\Lambda_{\mbox{\tiny QCD}}$, which means that the
bottom-antibottom-gluon propagation is certainly nonperturbative. In
fact, the rather uncomfortably large NNLO corrections in
relation~(\ref{duality}) might be regarded as a first warning sign in
support of this view.

At this point it seems to be just natural to mention the renormalon
ambiguities contained in the definition of the pole
mass~\cite{renormalon} 
which is defined perturbatively as the location of the singularity of
the renormalized quark propagator. This ambiguity indicates that the
pole mass has an intrinsic uncertainty of order
$\Lambda_{\mbox{\tiny QCD}}\sim 200-300$~MeV. It is caused
by the long range sensitivity of the pole mass and reflected in a
factorial growth of the high order coefficients of the perturbation
series connecting the pole mass to other mass definitions like
$\overline{\mbox{MS}}$ which seem to be free from this problem. Our
results
and the rather pessimistic prospect to further improve the results
obtained in this paper certainly support this view.
However, the notion of the renormalons might also give hints toward
a more precise determination of the bottom quark mass because it
implies that with a different mass definition the perturbative series
for the moments might become better behaved. In this work we have not
attempted to make use of this possibility, but we hope to return to
this issue in the near future.
\par
\vspace{.5cm}
\section*{Acknowledgement}
I am grateful to J.~G.~Branson, J.~Kuti, Z.~Ligeti, C.~Morningstar,
V.~A.~Sharma and J.~J.~Thaler for useful conversation.
I thank A.~V.~Manohar for many helpful discussions and for reading the
manuscript, and Z.~Ligeti and M.~B.~Voloshin for their comments to the
manuscript.
This work is supported in part by the U.S.~Department of Energy under
contract No.~DOE~DE-FG03-90ER40546.
\begin{appendix}
\par\vspace{1cm}
\section{NNLO Corrections from $\delta H_{kin}$, $V_{\mbox{\tiny
BF}}$ and $V_{\mbox{\tiny NA}}$}
\label{appendixcrosssection}
In this appendix we present some details about the calculation of the
NNLO corrections to the zero-distance Green function coming from the
kinetic energy $\delta H_{kin}(\vec r)=-\vec\nabla^4/4 M_b^3$, the
Breit-Fermi potential $V_{\mbox{\tiny
BF}}$, Eq.~(\ref{VBreitFermi}), and  the non-Abelian potential
$V_{\mbox{\tiny NA}}$, Eq.~(\ref{Vnonabelian}). 

At NNLO the corrections coming from $\delta H_{kin}$, $V_{\mbox{\tiny
BF}}$ and $V_{\mbox{\tiny NA}}$ are determined from first order
time-independent perturbation theory,
\begin{equation}
[G_c^{(2)}(0,0,E)]^{\mbox{\tiny kin+BF+NA}} \, = \,
-\,\int d^3\vec r \, 
G_c^{(0)}(0,r,E)\,\delta H(\vec r)\,G_c^{(0)}(r,0,E)
\label{app1}
\,,
\end{equation}
where
\begin{equation}
\delta H(\vec r) \, = \,
-\frac{\vec\nabla^4}{4\, M_b^3} + V_{\mbox{\tiny BF}}(\vec r) + 
V_{\mbox{\tiny NA}}(\vec r)
\,.
\label{app2}
\end{equation}
Because the zero-distance Green function only describes
bottom-antibottom pairs in a ${}^3\!S_1$ triplet state, we can take the
angular average and evaluate the spin operators for $\delta H$ in
expression~(\ref{app1}). The form of $\delta H$ then simplifies to
\begin{equation}
\delta H_{3S1} \, = \,
-\frac{\vec\nabla^4}{4 \,M_b^3} -
\frac{C_F\,a_s}{r}\,\frac{\vec\nabla^2}{M_b^2}  + 
\frac{11}{3}\,\frac{C_F\,a_s\,\pi}{M_b^2}\,\delta^{(3)}(\vec r) -
\frac{C_A\,C_F\,a_s^2}{2\,M_b\,r^2}
\,.
\label{app3}
\end{equation}
Using the equation of motion for the Coulomb Green function,
Eq.~(\ref{SchroedingerNR}), we can eliminate the $\vec\nabla^2$ terms
in $\delta H_{3S1}$. For illustration, let us consider the corrections
coming from the term $-C_F\frac{a_s}{r}\frac{\vec\nabla^2}{M_b^2}$ in
$\delta H_{3S1}$. Using the equation of motion we arrive at the
relation
\begin{eqnarray}
\lefteqn{
-\,\int d^3\vec r \, 
G_c^{(0)}(0,r,E)\,\bigg[
-\frac{C_F\,a_s}{r}\,\frac{\vec\nabla^2}{M_b^2}
\,\bigg]\,G_c^{(0)}(r,0,E)
}
\nonumber\\[2mm]
& = &
-\,\int d^3\vec r \, 
G_c^{(0)}(0,r,E)\,\bigg[
-\bigg(\,
\frac{C_F^2\,a_s^2}{M_b\,r^2} +
\frac{C_F\,a_s}{r}\,\frac{E}{M_b}
\,\bigg)\,G_c^{(0)}(r,0,E) -
\frac{C_F\,a_s}{M_b\,r}\,\delta^{(3)}(\vec r)
\,\bigg]
\label{app4}
\end{eqnarray}
The third term in the brackets represents a power divergence which is
dropped in our convention (see the text after
Eq.~(\ref{CoulombGreenfunctionzero})). Using the same arguments for
the kinetic energy term we get
\begin{eqnarray}
\lefteqn{
-\,\int d^3\vec r \, 
G_c^{(0)}(0,r,E)\,\bigg[
-\frac{\vec\nabla^4}{4 \,M_b^3}
\,\bigg]\,G_c^{(0)}(r,0,E)
}
\label{app5}\\[1mm]
& = &
\frac{E}{2\,M_b}\,G_c^{(0)}(0,0,E)
-\,\int d^3\vec r \, 
G_c^{(0)}(0,r,E)\,\bigg[
-\frac{C_F^2\,a_s^2}{4\,M_b\,r^2} - 
\frac{E}{2\,M_b}\,\frac{C_F\,a_s}{r} -
\frac{E^2}{4\,M_b}
\,\bigg]\,G_c^{(0)}(r,0,E)
\,.
\nonumber
\end{eqnarray}
Collecting all terms from Eqs.~(\ref{app3})-(\ref{app5}) we arrive at
\begin{eqnarray}
\lefteqn{
[G_c^{(2)}(0,0,E)]^{\mbox{\tiny kin+BF+NA}}
}
\nonumber\\[1mm] & = &
\frac{E}{2\,M_b}\,G_c^{(0)}(0,0,E)
-\,\int d^3\vec r \, 
G_c^{(0)}(0,r,E)\,\bigg[
-\frac{E^2}{4\,M_b} - \frac{3\,E}{2\,M_b}\,\frac{C_F\,a_s}{r} 
\nonumber\\[1mm] & &  \qquad 
+\frac{11}{3}\,\frac{C_F\,a_s\,\pi}{M_b^2}\,\delta^{(3)}(\vec r) 
-\bigg(\frac{5}{4}+\frac{C_A}{2\,C_F}\bigg)\,\frac{C_F^2\,a_s^2}{M_b\,r^2}
\,\bigg]\,G_c^{(0)}(r,0,E)
\,.
\label{app6}
\end{eqnarray}
The first and the second term in the brackets on the RHS of
Eq.~(\ref{app6}) are handled by redefining the energy, $E\to E+E^2/4
M_b^2$  and the coupling, $a_s\to a_s [1+ 3 E/2 M_b]$, in the
nonrelativistic Coulomb Green function. The calculation of the
$\delta$-function term is trivial. The treatment of the $1/r^2$ term,
on the other hand, is rather awkward. However, we can infer the
correction caused the $1/r^2$ term by using the facts that the
wave functions to the Schr\"odinger equation
\begin{equation}
\bigg(-\frac{\vec\nabla^2}{m^2} - \frac{a}{r} -
\frac{b}{m\,r^2} - E
\,\bigg)\,\Psi(\vec r) \, = \, 0
\label{modeleq}
\end{equation}
can be determined exactly for any energy E (see e.g.\
Ref.~\cite{Landau1})
and that the imaginary part of the Green function $G$ of
Eq.~(\ref{modeleq}) in the continuum, i.e.\ for any positive energy
$E$, is proportional to the modulus square of the scattering wave
function at the energy $E$. From this it is straightforward to derive
for positive energies the relation
\begin{equation}
\mbox{Im}\,G(0,0,E) \, = \,
\lim\limits_{r\to 0}\,\Big[\,(2\,p\,r)^s\,\Big]\,
\frac{m\,p}{4\,\pi}\,\exp\bigg\{\,\frac{a\,\pi\,m}{2\,p}\,\bigg\}\,
\left|\,\frac{\Gamma(1+s-i\frac{a\,m}{2\,p}}{\Gamma(2\,s+2)}\,\right|^2
\,,
\label{app7}
\end{equation}
where $s(s+1)=-b$ and $p=\sqrt{m(E+i\epsilon)}$. Expanding the RHS of
Eq.~(\ref{app7}) in small $b$\footnote{
Because we want to treat the $1/r^2$ potential as a perturbation, the
limit $r\to 0$ has to be taken after the expansion in $b$.
}
and imposing the short-distance cutoff
$\mu_{\rm fac}$ as described in Section~\ref{subsectioncorrelators} we
obtain for positive energies the relation
\begin{equation}
\mbox{Im}\,\bigg[\,
\int d^3\vec r \, 
G_c^{(0)}(0,r,E)\,\bigg(\,
\frac{C_F^2\,a_s^2}{M_b\,r^2}
\,\bigg)\,G_c^{(0)}(r,0,E)
\,\bigg] \, = \, \frac{4\,C_F\,a_s\,\pi}{M_b^2}\,\mbox{Im}\,\Big[\,
\Big(\,G_c^{(0)}(0,0,E)\,\Big)^2\,\Big]
\,.
\label{app8}
\end{equation}
Due to analyticity relation~(\ref{app8}) is then also valid for any
real energy. Up to (irrelevant) constants we can therefore write 
\begin{equation}
\int d^3\vec r \, 
G_c^{(0)}(0,r,E)\,\bigg(\,
\frac{C_F^2\,a_s^2}{M_b\,r^2}
\,\bigg)\,G_c^{(0)}(r,0,E)
\, = \, \frac{4\,C_F\,a_s\,\pi}{M_b^2}\,\Big[\,
G_c^{(0)}(0,0,E)\,\Big]^2
\,.
\label{app9}
\end{equation}
Collecting all terms the final result for the sum of the zero-distance
Coulomb Green function and the NNLO corrections caused by $\delta
H_{kin}$, $V_{\mbox{\tiny BF}}$ and $V_{\mbox{\tiny NA}}$ reads
\begin{eqnarray}
\lefteqn{
G_c^{(0)}(0,0,E,a_s) +
[G_c^{(2)}(0,0,E)]^{\mbox{\tiny kin+BF+NA}} 
}
\nonumber\\[1mm] & = & 
\bigg(\,1+\frac{E}{2\,M_b}\,\bigg)\,
G_c^{(0)}\Big(0,0,E+\frac{E^2}{4\,M_b},
      a_s\,\Big[1+\frac{3\,E}{2\,M_b}\Big]\,\Big)
\nonumber\\[1mm] & & +
\frac{4}{3}\,\bigg(\,1+\frac{3\,C_A}{2\,C_F}\,\bigg)\,
\frac{C_F\,a_s\,\pi}{M_b^2}\,\Big[\,G_c^{(0)}(0,0,E,a_s)\,\Big]^2
\label{app10}
\end{eqnarray}
up to corrections beyond the NNLO level. 
$G_c^{(0)}(0,0,E,a_s)$ is defined as the expression on the RHS of
Eq.~(\ref{CoulombGreenfunctionregularized}).
Rewriting the energy in terms
of $v=\sqrt{(E+i\epsilon)/M_b}$ we arrive at the result displayed in
Eq.~(\ref{GreenfunctionNNLOBF}).

We do not want to leave unmentioned that for the treatment of the
singular $1/r^2$ potential we have ignored the fact that its
coefficient (mainly through the large non-Abelian contribution) is
large enough that the $b\bar b$ system can collapse to a
point~(see e.g.\ \cite{Landau1}). This would lead to the breakdown of
hermiticity. Thus, the result in Eq.~(\ref{app10}) has some heuristic
character. However, we strictly treat the singular $1/r^2$ (and also
the $\delta^{(3)}(\vec r)$) potential as a ``small'' perturbation to
the Coulomb exchange and remove the arising UV singularities through
the matching procedure. No exact treatment of the singular potential
is intended. In this sense the result in Eq.~(\ref{app10}) should be
fine.
\par\vspace{1cm}
\section{Inverse Laplace Transforms}
\label{appendixLaplace}
In this appendix we present the list of inverse Laplace transforms
used to the calculate the theoretical moments at NNLO. In the
following we use the conventions
\begin{eqnarray*}
\Psi(z) & = & \frac{d \ln\Gamma(z)}{dz}
\,,
\\[1mm]
\Psi^{(n)}(z) & = & \frac{d^n}{dz^n}\,\Psi(z)
\,,
\\[1mm]
\Psi^{\prime}(z) & = & \Psi^{(1)}(z) 
\,,
\\[1mm]
\Psi^{\prime\prime}(z) & = & \Psi^{(2)}(z) 
\,,
\\[1mm]
{}_0F_2(a,b;z) & = & \Gamma(a)\,\Gamma(b)\,\sum\limits_{k=0}^{\infty}
\frac{1}{\Gamma(a+k)\,\Gamma(b+k)}\,\frac{z^k}{k!}
\,.
\end{eqnarray*}
\begin{eqnarray}
\frac{1}{2\pi i}\,\int\limits_{\gamma-i\infty}^{\gamma+i\infty}
\frac{1}{x^\nu}\,e^{x\,t}\,dx & = & \frac{t^{\nu-1}}{\Gamma(\nu)}
\,,
\\[2mm]
\frac{1}{2\pi i}\,\int\limits_{\gamma-i\infty}^{\gamma+i\infty}
\frac{\ln x}{x^\nu}\,e^{x\,t}\,dx & = & 
\frac{t^{\nu-1}}{\Gamma(\nu)}\,\Big[\,
\Psi(\nu) - \ln t
\,\Big]
\,,
\\[2mm]
\frac{1}{2\pi i}\,\int\limits_{\gamma-i\infty}^{\gamma+i\infty}
\frac{\ln^2 x}{x^\nu}\,e^{x\,t}\,dx & = & 
\frac{t^{\nu-1}}{\Gamma(\nu)}\,\Big\{\,
\Big[\,\Psi(\nu) - \ln t\,\Big]^2
-\Psi^\prime(\nu)
\,\Big\}
\,,
\\[2mm]
\frac{1}{2\pi i}\,\int\limits_{\gamma-i\infty}^{\gamma+i\infty}
\frac{\ln^3 x}{x^\nu}\,e^{x\,t}\,dx & = & 
\frac{t^{\nu-1}}{\Gamma(\nu)}\,\Big\{\,
\Big[\,\Psi(\nu) - \ln t\,\Big]^3 -
3\,\Big[\,\Psi(\nu) - \ln t\,\Big]\,\Psi^\prime(\nu)
+\Psi^{\prime\prime}(\nu)
\,\Big\}
\,,
\\[2mm]
\frac{1}{2\pi i}\,\int\limits_{\gamma-i\infty}^{\gamma+i\infty}
\frac{1}{x^\nu}\,\sin(\frac{a}{\sqrt{x}})\,e^{x\,t}\,dx & = & 
\frac{a\,t^{\nu-\frac{1}{2}}}{\Gamma(\nu+\frac{1}{2})}\,
{}_0\rm F_2\bigg(\frac{3}{2},\nu+\frac{1}{2},-\frac{a^2}{4}\,t\bigg)
\,,
\\[2mm]
\frac{1}{2\pi i}\,\int\limits_{\gamma-i\infty}^{\gamma+i\infty}
\frac{\ln x}{x^\nu}\,\sin(\frac{a}{\sqrt{x}})\,e^{x\,t}\,dx & = & 
\frac{a\,t^{\nu-\frac{1}{2}}}{\Gamma(\nu+\frac{1}{2})}\,
\bigg\{\,\bigg[\,
\Psi\Big(\nu+\frac{1}{2}\Big)-\ln t \,\bigg]\,
{}_0\rm F_2\bigg(\frac{3}{2},\nu+\frac{1}{2},-\frac{a^2}{4}\,t\bigg)
\nonumber\\[1mm] & & \qquad
- \frac{d}{d\nu}\,
{}_0\rm F_2\bigg(\frac{3}{2},\nu+\frac{1}{2},-\frac{a^2}{4}\,t\bigg)
\,\bigg\}
\,,
\\[2mm]
\frac{1}{2\pi i}\,\int\limits_{\gamma-i\infty}^{\gamma+i\infty}
\frac{\ln^2 x}{x^\nu}\,\sin(\frac{a}{\sqrt{x}})\,e^{x\,t}\,dx & = & 
\frac{a\,t^{\nu-\frac{1}{2}}}{\Gamma(\nu+\frac{1}{2})}
\bigg\{\bigg[\bigg(
\Psi\Big(\nu+\frac{1}{2}\Big)-\ln t\bigg)^2-
  \Psi^\prime\Big(\nu+\frac{1}{2}\Big) \bigg]\,
{}_0\rm F_2\bigg(\frac{3}{2},\nu+\frac{1}{2},-\frac{a^2}{4}t\bigg)
\nonumber\\[1mm] & & \qquad
- 2\,\bigg[\,
\Psi\Big(\nu+\frac{1}{2}\Big)-\ln t \,\bigg]\,
\frac{d}{d\nu}\,
{}_0\rm F_2\bigg(\frac{3}{2},\nu+\frac{1}{2},-\frac{a^2}{4}\,t\bigg)
\nonumber\\[1mm] & & \qquad
+ \frac{d^2}{d\nu^2}\,
{}_0\rm F_2\bigg(\frac{3}{2},\nu+\frac{1}{2},-\frac{a^2}{4}\,t\bigg)
\,\bigg\}
\,.
\end{eqnarray}
\par\vspace{1cm}
\section{The Constants $w^{0,1,2}_p$ and $\tilde w^{0,1,2}_p$}
\label{appendixconstants}
In this appendix the constants $w^{0,1,2}_p$ and $\tilde w^{0,1,2}_p$
from expression~(\ref{rho1NNLOC}) are given. They generically
parameterize the higher order contributions to the Green function of
the Schr\"odinger equation~(\ref{Schroedingerfull}) coming from the
radiative corrections to the Coulomb potential, $V^{(1)}_c$ and
$V^{(2)}_c$, Eqs.~(\ref{Vrunning1loop}) and (\ref{Vrunning2loop}). 
For the constants $w^{0,1,2}_p$ we were able to calculate analytic
expressions. The results read ($p=1,2,3,\ldots$)
\begin{eqnarray}
w^0_p & = &
-\frac{1}{p!\,\Gamma(\frac{p}{2})}\,
\int\limits_0^\infty dt \int\limits_0^\infty du \,
\frac{1}{(1+t+u)^2}\,\ln^p\Big(\frac{(1+t)\,(1+u)}{t\,u}\Big)
\, = \,
-\frac{(p+1)\,\zeta_{p+1}}{\Gamma(\frac{p}{2})}
\,,
\\[2mm]
w^1_p & = &
\frac{1}{p!\,\Gamma(\frac{p}{2})}\,
\int\limits_0^\infty dt \int\limits_0^\infty du \,
\frac{1-\ln(1+t+u)}{(1+t+u)^2}\,\ln^p\Big(\frac{(1+t)\,(1+u)}{t\,u}\Big)
\nonumber\\[1mm] & = &
-\bigg\{\,
\frac{(1+p)}{\Gamma(\frac{p}{2})}\,\bigg[\,\gamma_{\mbox{\tiny E}}\,
  \zeta_{p+1} +
  \sum\limits_{m=0}^\infty\,\frac{\Psi(2+m)}{(1+m)^{p+1}}\,\bigg]
+ \frac{2}{\Gamma(\frac{p}{2})}\,
  \sum\limits_{l=0}^{p-1}\,\sum\limits_{m=0}^\infty\,
  (-1)^{p-l}\,\frac{(1+l)\,\Psi^{(p-l)}(2+m)}{(p-l)!\,(1+m)^{1+l}}
\,\bigg\}
\nonumber\\&&
\\[2mm]
w^2_p & = &
\frac{1}{p!\,\Gamma(\frac{p}{2})}\,
\int\limits_0^\infty dt \int\limits_0^\infty du \,
\frac{\zeta_2-2\,\ln(1+t+u)+\ln^2(1+t+u)}{(1+t+u)^2}\,
\ln^p\Big(\frac{(1+t)\,(1+u)}{t\,u}\Big)
\nonumber\\[1mm] & = &
\frac{(1+p)}{\Gamma(\frac{p}{2})}\,\bigg\{\,
\Big(\,\gamma_{\mbox{\tiny E}}^2+2\,\zeta_2\,\Big)\,\zeta_{1+p} +
  \sum\limits_{m=0}^\infty\,\frac{1}{(1+m)^{1+p}}\,\bigg[\,
2\,\gamma_{\mbox{\tiny E}}\,\Psi(2+m) - 
\Psi^\prime(2+m) + \Big(\Psi(2+m)\Big)^2
\,\bigg]
\,\bigg\}
\nonumber\\[1mm] & &
+ \frac{2}{\Gamma(\frac{p}{2})}\,\sum\limits_{m=0}^\infty\,
\sum\limits_{l=0}^{p-1}\,\frac{(-1)^{p-l}\,(1+l)}{(p-l)!\,(1+m)^{1+l}}\,
\bigg[\,
2\,\gamma_{\mbox{\tiny E}}\,\Psi^{(p-l)}(2+m) - \Psi^{(p-l+1)}(2+m) 
\nonumber\\[1mm] & & \mbox{\hspace{6cm}}
+2\,\Psi^{(p-l)}(2+m)\,\Psi(2+m)
\,\bigg]
\nonumber\\[1mm] & &
+\frac{4}{\Gamma(\frac{p}{2})}\,\sum\limits_{m=0}^\infty\,
\sum\limits_{l=0}^{p-2}\,\sum\limits_{k=1}^{p-l-1}\,
(-1)^{p-l}\,\frac{(1+l)\,\Psi^{(p-l-k)}(2+m)\,\Psi^{(k)}(2+m)}
{(p-l-k)!\,k!\,(1+m)^{1+l}}
\,.
\end{eqnarray}
The constants $\tilde w^{0,1,2}_p$ are calculated
numerically. The corresponding integrals are ($i=0,1,2$)
\begin{eqnarray}
\tilde w^i_p & = & 
\frac{1}{p!\,\Gamma(\frac{p+1}{2})}\,
\int\limits_0^\infty dt 
\int\limits_0^\infty du \,
\int\limits_0^\infty dv \,
\int\limits_0^1 ds \,\,
\omega^i(t,u,v,s)\,
\ln^p\bigg(\,
\frac{(1+t)\,(1+u)\,(1+v)\,(1-s)}{t\,u\,v\,s}
\,\bigg)
\,,
\end{eqnarray}
where
\begin{eqnarray}
\omega^0(t,u,v,s) & = & \frac{3\,x + y}{x^2\,(x + y)^3}
\,,
\\[2mm]
\omega^1(t,u,v,s) & = & 
\frac{x^2-7\,x\,y-2\,y^2}{x^2\,y\,(x + y)^3} 
+\frac{\ln x}{x^2\,y^2} 
+\frac{(y-x)\,(x^2+4\,x\,y+y^2)\,\ln(x+y)}{x^2\,y^2\,(x+y)^3}
\,,
\\[2mm]
\omega^2(t,u,v,s) & = & 
\frac{3\,x+y}{x^2\,(x+y)^3} - \frac{x+3\,y}{y^2\,(x+y)^3}\,\zeta_2 + 
\frac{(x-y)\,(x^2+5\,x\,y+y^2)}{x^2\,y^2\,(x+y)^3}\,\ln(x+y) 
\nonumber\\[1mm] & &
+\frac{3\,x+y}{x^2\,(x+y)^3}\,\ln^2(x+y)
\nonumber\\[1mm] & &
-\frac{1}{x^2\,y^2}\,\bigg[\,\ln x-\Big(\ln x-\ln(x+y)\Big)\,\ln y + 
{\rm Li}_2\Big(\frac{x}{x+y}\Big)\,\bigg]
\,,
\end{eqnarray}
and
\begin{eqnarray*}
x & = & 1+t+u
\,,
\\[2mm]
y & = & 1+v-s
\,.
\end{eqnarray*}
\end{appendix}
\vspace{1.0cm}
%
%\newpage
%%%%%%%%%%%%%%%%%%%%%%%%%%%%%%%%%%%%%%%%%%%%%%%%%%%%%%%%%%%%%%%%%%%%%%%%
\sloppy
\raggedright
\def\app#1#2#3{{\it Act. Phys. Pol. }{\bf B #1} (#2) #3}
\def\apa#1#2#3{{\it Act. Phys. Austr.}{\bf #1} (#2) #3}
\def\lhc{Proc. LHC Workshop, CERN 90-10}
\def\npb#1#2#3{{\it Nucl. Phys. }{\bf B #1} (#2) #3}
\def\nP#1#2#3{{\it Nucl. Phys. }{\bf #1} (#2) #3}
\def\plb#1#2#3{{\it Phys. Lett. }{\bf B #1} (#2) #3}
\def\prd#1#2#3{{\it Phys. Rev. }{\bf D #1} (#2) #3}
\def\pra#1#2#3{{\it Phys. Rev. }{\bf A #1} (#2) #3}
\def\pR#1#2#3{{\it Phys. Rev. }{\bf #1} (#2) #3}
\def\prl#1#2#3{{\it Phys. Rev. Lett. }{\bf #1} (#2) #3}
\def\prc#1#2#3{{\it Phys. Reports }{\bf #1} (#2) #3}
\def\cpc#1#2#3{{\it Comp. Phys. Commun. }{\bf #1} (#2) #3}
\def\nim#1#2#3{{\it Nucl. Inst. Meth. }{\bf #1} (#2) #3}
\def\pr#1#2#3{{\it Phys. Reports }{\bf #1} (#2) #3}
\def\sovnp#1#2#3{{\it Sov. J. Nucl. Phys. }{\bf #1} (#2) #3}
\def\sovpJ#1#2#3{{\it Sov. Phys. LETP }{\bf #1} (#2) #3}
\def\jl#1#2#3{{\it JETP Lett. }{\bf #1} (#2) #3}
\def\jet#1#2#3{{\it JETP Lett. }{\bf #1} (#2) #3}
\def\zpc#1#2#3{{\it Z. Phys. }{\bf C #1} (#2) #3}
\def\ptp#1#2#3{{\it Prog.~Theor.~Phys.~}{\bf #1} (#2) #3}
\def\nca#1#2#3{{\it Nuovo~Cim.~}{\bf #1A} (#2) #3}
\def\ap#1#2#3{{\it Ann. Phys. }{\bf #1} (#2) #3}
\def\hpa#1#2#3{{\it Helv. Phys. Acta }{\bf #1} (#2) #3}
\def\ijmpA#1#2#3{{\it Int. J. Mod. Phys. }{\bf A #1} (#2) #3}
\def\ZETF#1#2#3{{\it Zh. Eksp. Teor. Fiz. }{\bf #1} (#2) #3}
\def\jmp#1#2#3{{\it J. Math. Phys. }{\bf #1} (#2) #3}
\def\yf#1#2#3{{\it Yad. Fiz. }{\bf #1} (#2) #3}
\def\ufn#1#2#3{{\it Usp. Fiz. Nauk }{\bf #1} (#2) #3}
\def\spu#1#2#3{{\it Sov. Phys. Usp.}{\bf #1} (#2) #3}
\def\epjc#1#2#3{{\it Eur. Phys. J. C }{\bf #1} (#2) #3}
%%%%%%%%%%%%%%%%%%%%%%%%%%%%%%%%%%%%%%%%%%%%%%%%%%%%%%%%%%%%%%%%%%%%%%%%


\begin{thebibliography}{99}
%
\bibitem{Stirling1}
W. J. Stirling, Univ. of Durham Report No. DTP-97-80,
hep-ph/9709429.
%
\bibitem{Voloshin1}
M.B. Voloshin, \ijmpA{10}{1995}{2865}.
%
\bibitem{Jamin1}
M. Jamin and A. Pich, \npb{507}{1997}{334}.
%
\bibitem{Kuhn1}
J. H. K\"uhn, A. A. Penin, and A. A. Pivovarov,
Univ. of Karlsruhe Report No. TTP98-01, hep-ph/9801356.
%
\bibitem{Novikov1}
V.A. Novikov {\it et al.}, \pr{41}{1978}{1}.
%
\bibitem{Voloshin2}
M. B. Voloshin and Yu. M. Za\u{i}tsev, 
\ufn{152}{1987}{}361 [\spu{30}{1987}{7}].
%
\bibitem{Hoang1} 
A. H. Hoang, \prd{56}{1997}{5851}.
%
\bibitem{Hoang2}
A. H. Hoang,  proceedings of the ``Workshop on Physics at the First
Muon Collider and the Front End of a Muon Collider'', Fermilab,
November 6--9, 1997, hep-ph/9801273.
%
\bibitem{Hoang3} 
A. H. Hoang and T. Teubner, UCSD Report No. UCSD-PTH 98-01,
hep-ph/9801397.
%
\bibitem{Melnikov1}
K. Melnikov and A. Yelkhovkii, Univ. of Karlsruhe Report
No. TTP98-10, hep-ph/9802379.
%
\bibitem{Caswell1}
W. E. Caswell and G. E. Lepage, \plb{167}{1986}{437}.
%
\bibitem{Bodwin1}
G. T. Bodwin, E. Braaten, and G. P. Lepage, \prd{51}{1995}{1125}.
%
\bibitem{Reinders1}
L. J. Reinders, H. R. Rubinstein, and S. Yazaki, \npb{186}{1981}{109}.
%
\bibitem{Poggio1}
E. C. Poggio, H. R. Quinn, and S. Weinberg, \prd{13}{1976}{1958}.
%
\bibitem{Shifman1}
M. A. Shifman, A. I. Vainshtein, and V. I. Zakharov,
\npb{147}{1979}{385}; \npb{147}{1979}{448}.
%
\bibitem{Hoang4}
A. H. Hoang, \prd{57}{1998}{1615}.
%
\bibitem{Lepage1}
W. E. Caswell, R. R. Horgan, and G. P. Lepage, \pra{18}{1978}{810}.
%
\bibitem{Hoang5}
A. H. Hoang, P. Labelle, and S. M. Zebarjad. \prl{79}{1997}{3387}.
%
\bibitem{Voloshin3}
M. B. Voloshin, \npb{154}{1979}{365}.
%
\bibitem{Wichmann1}
E.H. Wichmann and C.H. Woo, \jmp{2}{1961}{178}.
%
\bibitem{Hostler1}
L. Hostler, \jmp{5}{1964}{591}.
%
\bibitem{Schwinger1}
J. Schwinger, \jmp{5}{1964}{1606}.
%
\bibitem{Fischler1}
W. Fischler, \npb{129}{1977}{157}.
%
\bibitem{Billoire1}
A. Billoire, \plb{92}{1980}{343}.
%
\bibitem{nonabelianpotential}
S. N. Gupta and S. F. Radford, \prd{24}{1981}{2309}, 
{\it ibid.} {\bf 25} (1982) 3430 (Erratum);\\
S. N. Gupta, S. F. Radford, and W. W. Repko, 
\prd{26}{1982}{3305}.
%
\bibitem{Peter1}
M. Peter, \prl{78}{1997}{602}, 
\npb{501}{1997}{471}.
%
\bibitem{Abramowitz1}
M. Abramowitz and I.A. Stegun, eds., 
{\it Handbook of Mathematical Functions}
(Dover Publications, Inc., New York, 1972).
%
\bibitem{Gradshteyn1}
I.S. Gradshteyn and I. M. Ryzhik, {\it Table of Integrals, Series and  
Products}, (Academic Press, Inc., San Diego, 1994).
%
\bibitem{Braun1}
M.A. Braun, \ZETF{54}{1968}{1220} [\sovpJ{27}{1968}{652}].
%
\bibitem{Kallensabry1}
G. K\"allen and A. Sabry, 
{\it K. Dan. Vidensk. Selsk. Mat.-Fys. Medd.}
{\bf 29} (1955) No. 17.
%
\bibitem{Schwinger2}
J. Schwinger, {\it Particles, Sources and Fields},
Vol II, (Addison-Wesley, New York, 1973).
%
\bibitem{Hoang6}
A. H. Hoang, \prd{56}{1997}{7276}.
%
\bibitem{Adkins1}
G. Adkins, R. N. Fell, and P. M. Mitrikov, \prl{79}{1997}{3383}.
%
\bibitem{Czarnecki1}
A. Czarnecki and K. Melnikov, \prl{80}{1998}{2531}.
%
\bibitem{Beneke1}
M. Beneke and V. A. Smirnov, CERN Report No. CERN-TH-97-315, 
hep-ph/9711391;\\
M. Beneke, A. Signer, and V. A. Smirnov, \prl{80}{1998}{2535}.
%
\bibitem{Hoang7}
A. H. Hoang, J. H. K\"uhn, and T. Teubner, \npb{452}{1995}{173}.
%
\bibitem{Karshenboim1}
S.G. Karshenbo\v{\i}m, \yf{56}{1993}{155}.
%
\bibitem{Karplus1}
R. Karplus and A. Klein, \pR{87}{1952}{848}.
%
\bibitem{Barbieri1}
R. Barbieri {\it et al.}, \plb{57}{1975}{455}. 
%
\bibitem{Steinhauser1}
K. G. Chetyrkin, J. H. K\"uhn, and M. Steinhauser, 
\plb{371}{1996}{93}; \npb{482}{1996}{213}.
%
\bibitem{Kuhn2}
K. G. Chetyrkin, J. H. K\"uhn, and A. Kwiatkowski,
\pr{277}{1996}{189}.
%
\bibitem{Chetyrkin1}
K. G. Chetyrkin {\it et al.}, \epjc{2}{1998}{137}.
%
\bibitem{Albrecht1}
H. Albrecht {\it et al}, \zpc{65}{1995}{619}.
%
\bibitem{PDG}
Particle Data Group, R. M. Barnett {\it et al.},
\prd{54}{1996}{77}.
%
\bibitem{Buras1}
A. J. Buras, proceedings of the ``Symposium of Heavy Flavours'',
Santa Barbara, July 7--11, 1997, hep-ph/9711217.
%
\bibitem{BaBar1}
{\it The BaBar Physics Book}, SLAC Report No. SLAC-R-504, in
preparation.
%
\bibitem{Broadhurst1}
N. Gray {\it et al}, \zpc{48}{1990}{673}.
%
\bibitem{DELPHI1}
DELPHI Collaboration, P. Abreu {\it et al.}, CERN Report
No. CERN-PPE-97-141. 
%
\bibitem{Rodrigo1}
W. Bernreuther, A. Brandenburg, and P. Uwer, \prl{79}{1997}{189};\\
G. Rodrigo, M. Bilenky ,and A. Santamaria, \prl{79}{1997}{193};\\
P. Narison and C. Oleari, \plb{407}{1997}{57}.
%
\bibitem{Brodsky1}
S. J. Brodsky, G. P. Lepage, and P. B. Mackenzie, \prd{28}{1983}{228}.
%
\bibitem{Penin1} 
A. A. Penin and A. A. Pivovarov,
Univ. of Karlsruhe Report No. TTP98-13, hep-ph/9803363.
%
\bibitem{renormalon}
M. Beneke and V. M. Braun, \npb{426}{1994}{301};\\
I. I. Bigi, \prd{50}{1994}{2234}.
%
\bibitem{Landau1}
L. D. Landau and E. M. Lifschitz, {\it Quantum Mechanics Vol.3},
(Butterworth-Heinemann, 1981).
%
\end{thebibliography}
\end{document}